\documentclass[aps,prd,10pt,notitlepage,nofootinbib,superscriptaddress,showkeys,showpacs]{revtex4-1}
\pdfoutput=1
\usepackage{amstext,amsmath,amssymb,amsfonts,bbm}
\usepackage[latin1]{inputenc}
\usepackage{epsfig}
\usepackage{hyperref}
\usepackage{amsthm}
\usepackage{subfigure}
\usepackage{color}
\usepackage{multirow}
\usepackage{tikz}

\usetikzlibrary{arrows,snakes,backgrounds}

\usepackage{verbatim}

\usepackage{graphicx}

\usepackage{latexsym}

\newcommand{\bea}{\begin{eqnarray}}	
\newcommand{\eea}{\end{eqnarray}}
\newcommand{\be}{\begin{equation}}	
\newcommand{\ee}{\end{equation}}
\newcommand{\beq}{\begin{equation}}	
\newcommand{\eeq}{\end{equation}}

\newcommand{\cV}{{\mathcal V}}

\newcommand{\Z}{{\mathbb Z}}
\newcommand{\C}{{\mathbb C}}

 \newcommand{\lap}{\bigtriangleup}

\newcommand{\norm}[1]{\left\lVert#1\right\rVert}

\newcommand{\intt}{{\rm int\,}}

\newcommand{\Tr}{{\rm Tr}}

\newcommand{\ov}[1]{\overline{#1}}
\newcommand{\ca}[1]{\mathcal{#1}}

\newcommand{\de}{\partial}

\newcommand{\Sym}{\text{sym}\Big\{1, 2,\dots, d\Big\}}

\def\R{\relax\ifmmode {\mathbb R}  \else${\mathbb R}$\fi}
\def\C{\relax\ifmmode {\mathbb C}  \else${\mathbb C}$\fi}
\def\Z{\relax\ifmmode {\mathbb Z}  \else${\mathbb Z}$\fi}
\def\N{\relax\ifmmode {\mathbb N}  \else${\mathbb N}$\fi}
\def\I{\relax\ifmmode {\mathbb I}  \else${\mathbb I}$\fi}

\begin{document}

\title{Functional Renormalisation Group analysis of Tensorial Group Field Theories on $\mathbb{R}^d$ }


\author{Joseph Ben Geloun}\email{jbengeloun@aei.mpg.de}

\affiliation{Max Planck Institute for Gravitational Physics, Albert Einstein Institute\\
 Am M\"uhlenberg 1, 14476, Potsdam, Germany }

\affiliation{International Chair in Mathematical Physics and Applications, 
ICMPA-UNESCO Chair, 072BP50, Cotonou, Rep. of Benin }

\author{Riccardo Martini}\email{riccardo.martini@studio.unibo.it}

\affiliation{Alma Mater Studiorum, Universit\`a di Bologna, via Zamboni 33,
40126, Bologna, Italy}

\affiliation{Max Planck Institute for Gravitational Physics, Albert Einstein Institute\\
 Am M\"uhlenberg 1, 14476, Potsdam, Germany }

\author{Daniele Oriti}\email{daniele.oriti@aei.mpg.de}

\affiliation{Max Planck Institute for Gravitational Physics, Albert Einstein Institute\\
 Am M\"uhlenberg 1, 14476, Potsdam, Germany }


\begin{abstract}
Rank-d Tensorial Group Field Theories are quantum field theories defined on a group manifold $G^{\times d}$, which represent a non-local generalisation of standard QFT, and a candidate formalism for quantum gravity, since, when endowed with appropriate data, they can be interpreted as defining a field theoretic description of the fundamental building blocks of quantum spacetime. 
Their renormalisation analysis is crucial both for establishing their consistency as quantum field theories, and for studying the emergence of continuum spacetime and geometry from them.
In this paper, we study the renormalisation group flow of two simple classes of TGFTs, defined for the group $G=\mathbb{R}$ for arbitrary rank, both without and with gauge invariance conditions, by means of functional renormalisation group techniques. The issue of IR divergences is tackled by the definition of a proper thermodynamic limit for TGFTs. 
We map the phase diagram of such models, in a simple truncation, and identify both UV and IR fixed points of the RG flow.  
Encouragingly, for all the models we study, we find evidence for the existence of a phase transition of condensation type.

\
 
\noindent Pacs numbers: 11.10.Gh, 05.10.Cc, 04.60.-m, 02.10.Ox\\  
\noindent Key words: Renormalisation, Renormalisation Group Methods,
Group Field Theory, Tensor Models. \\ 
\noindent Report numbers: ICMPA/MPA/2016/xx

\end{abstract}

\maketitle

\tableofcontents

\section{Introduction}

Group field theories \cite{GFTreviews} (GFTs) are a new type of quantum field theories characterised by a peculiar non-local pattern of pairings of field arguments in the interactions. The domain of definition of the fields is, for the most studied models, a (Lie) group manifold, hence the name of the formalism.
The first consequence of the non-locality of the GFT interactions is that their quantum states can be associated to graphs (or networks), while the Feynman diagrams arising in the GFT perturbative expansion are dual to cellular complexes. These graphs and cellular complexes are then decorated by group-theoretic data, corresponding to the degrees of freedom associated to the GFT fields. This implies that a number of standard QFT techniques have to be adapted to this new context, and that a host of new mathematical structures can be explored by such field theoretic means. 
This formalism finds its historic roots, and main applications, at present, as a promising framework for quantum gravity. From this more physical perspective, GFTs are a tentative definition of the microstructure of quantum spacetime and of its fundamental quantum dynamics. The decorated graphs, in this interpretation, are the fundamental quantum structures from which a continuum spacetime and geometry should emerge in the appropriate regime of approximation. 
In fact, group field theories were first proposed \cite{boulatov} as an enrichment, by the addition of group-theoretic data, of tensor models \cite{tensor} (in turn a generalisation of matrix models for 2d quantum gravity \cite{Di Francesco:1993nw} to higher dimensions), with the main goal being to obtain Feynman amplitudes of the form of state sum models of topological field theories. The link with loop quantum gravity \cite{LQG} became quickly clear \cite{carloPR}: group field theories and loop quantum gravity share the same type of quantum states, i.e. spin networks. It is then in the context of loop quantum gravity and state sum models, called spin foam models \cite{SF} and developed as a covariant definition of loop quantum gravity, that most subsequent work has been done, once it was understood \cite{carlomike} that the correspondence between group field theory and spin foam amplitudes is completely general. Finally, the relation between group field theory and lattice quantum gravity, already evident in their origin in tensor models, became stronger because of the appearance of the Regge action in semiclassical analyses of spin foam amplitudes (see for example, \cite{SFasympt}), and, more recently, of the general possibility to recast group field theory amplitudes as (non-commutative) simplicial gravity path integrals \cite{danielearistide}. It is now clear that group field theories sit at the crossroad of several approaches to quantum gravity, as a 2nd quantised framework for loop quantum gravity degrees of freedom \cite{GFT-LQG} as well as an enrichment of tensor models. 
The quantum field theory framework they provide for the candidate fundamental degrees of freedom of quantum spacetime is then crucial for tackling the open issues of these approaches. In particular, it makes possible to take on them a condensed matter-like perspective, making precise the idea of `atoms of space' and to study from this perspective the emergence of continuum spacetime \cite{danieleemergence}, and to use powerful renormalisation group techniques to the analysis of their quantum dynamics.
The renormalisation group analysis of GFT models has two main goals: establishing their perturbative renormalisability and exploring the continuum phase diagram. The first goal is all the more important because these models are initially defined and studied in perturbative expansion around the trivial vacuum, and it is in this expansion that their relation with loop quantum gravity and lattice quantum gravity, as well as their quantum geometric content, is more apparent. Establishing their perturbative renormalisability amounts then to establishing the consistency of this definition, and it also serves the purpose of constraining quantisation ambiguities (the GFT counterpart of those arising in the canonical loop quantum gravity formulation) as well as model building. The second goal is the most important open issue in all these related quantum gravity approaches: their continuum limit, i.e. the macroscopic, collective dynamics of their microscopic degrees of freedom, and the possibility of spacetime and geometry emerging from a phase transition of the same degrees of freedom \cite{danieleemergence}, as it has been proposed also in related approaches  \cite{Tim-DGP,HannoTim, CDTphasetrans, dariojoe,jacub}. It also amounts to controlling the full GFT expansion in terms of sum over cellular complexes and spin foam histories, thus it can be seen as solving, by QFT techniques, the problem of the continuum limit in both dynamical triangulations and spin foam models (for which alternative strategies are also been explored \cite{biancabenny}).

GFT renormalisation is in fact one of the most rapidly developing research directions in this area, and it has benefit greatly from concurrent developments in tensor models \cite{tensor}, which provides analytic tools and many insights concerning the combinatorics and the topology of GFT Feynman diagrams \cite{largeN, doublescal} as well as the possible definitions of the theory space to focus on \cite{universalityTensor,vincentTheorySpace}. Indeed, most of the work in GFT renormalisation has concerned a class of GFTs, called  Tensorial Group Field Theories (TGFTs), in which tensorial structures are prominent.
Several interesting TGFT models have been proven to be renormalisable \cite{GFTrenorm, Carrozza:2013mna} and their RG flow has been also studied, mainly in the vicinity of the UV fixed point  \cite{BenGeloun:2012pu,josephBeta, Samary:2013xla, Sylvain}, showing that asymptotic freedom is a very general feature of TGFT models \cite{Rivasseau:2015ova}. This work has encompassed abelian as well as non-abelian models, and both models with and without the additional gauge invariance properties that characterise GFTs for topological BF theory and 4d gravity, by giving their Feynman amplitudes the structure of lattice gauge theories. The same analysis has also been extended to models defined not on groups but on homogeneous spaces \cite{danielevincent}. 
More recently, non-perturbative GFT renormalisation has been tackled as well. Some work \cite{thomasreiko,Krajewski:2015clk} has been based on the Polchinski equation and on the analysis of the Schwinger-Dyson equations (see also \cite{vincentdanielevincent}). Most work has however been framed in the language of the Functional Renormalization Group approach to QFTs, first adapted to TGFTs in \cite{Benedetti:2014qsa}, after the initial steps taken in \cite{Brezin:1992yc,AstridTim,AstridTim2, Benedetti:2012dx} for matrix models. The first model being studied \cite{Benedetti:2014qsa} was an abelian rank-3 one on $U(1)$, and this analysis was quickly extended to the non-compact case in  \cite{Geloun:2015qfa}.  A model in rank-6 and again based on $U(1)$ was instead analysed in \cite{Benedetti:2015yaa}, this time incorporating gauge invariance. All these models were analysed in a fourth order truncation of in the number of fields. In all these cases, not only it was possible to confirm the asymptotic freedom of the models int he UV, but it was also possible to identify IR fixed points and to provide strong hints of a phase transition. The IR fixed points resemble Wilson-Fisher fixed points for ordinary scalar field theories, and the phase transition appears to separate a symmetric and a broken or condensate phase, with non-zero expectation value for the TGFT field operator. 
 With a different perspective, the existence of phase transition
has been proven for quartic tensor models in  \cite{Delepouve:2015nia,Benedetti:2015ara} 
with a characterisation of the related phases
and also for GFT models related to topological BF theory,
in any dimension \cite{GFTphase}.
In models more directly related to loop quantum gravity and lattice quantum gravity, this type of phase transition was suggested to govern the emergence of an effective cosmological dynamics from such quantum gravity models \cite{GFTcondensate}. 

In this paper, we generalise the analysis of abelian models on $\mathbb{R}$ performed in \cite{Geloun:2015qfa}, in two main ways: we compute and study the RG flow of models of arbitrary rank, and we perform the same analysis also for gauge invariant models, again in arbitrary rank. In both cases, we then specialise the results to rank 3, 4 and 5, identify the UV and IR fixed points and describe the resulting phase diagram. We still work, though, in a fourth order truncation of in the number of fields.

The plan of this paper is as follows. 
Section \ref{sect2:FGRfoGFT}  reviews the Functional Renormalisation Group applied to Group Field Theories following \cite{Benedetti:2014qsa}.
In section \ref{sect3:TGFT}, we describe the analysis of the simplest class of non-compact models, without gauge invariance, for arbitrary rank. We also complete the analysis given in \cite{Geloun:2015qfa} 
by providing the solution of the system of $\beta$-function at second order around the Gaussian fixed point, provide details on the neighbourhood of that trivial fixed point. As in that previous work, the analysis of such non-compact model requires IR regularisation, as we will discuss in detail.
The key point
of our regularisation scheme is the introduction of a new parameter representing the dependence of couplings on the volume
of the direct space. 
In the section \ref{chap:Gauge}, we repeat the analysis  for another interesting class of models obtained introducing an additional gauge invariance in the amplitudes,  by means of suitable projector operators inserted in the GFT action.
After appropriate regularisation, we can again study the RG flow of these models.
In section \ref{concl} we give a summary of our results
and list some important open problems for this approach. Two appendices \ref{app:betaTGFT} and \ref{app:gau} provide more details
of our calculations.

\section{The Functional Renormalisation Group for TGFTs: An overview}
\label{sect2:FGRfoGFT}
In this section we first review the basic ingredients of the (tensorial) group field theory formalism, in its covariant functional integral formulation. Then, we present the 
Functional Renormalisation approach, as it has been adapted and applied to TGFTs in \cite{Benedetti:2014qsa}. 

\subsection{Tensorial Group Field Theories}
\label{subsect:TGFT}
Let us introduce the special class of GFTs we will work with in the following, known as Tensorial Group
Field Theories (TGFT) \cite{GFTrenorm}-\cite{Sylvain}\cite{Gurau:2011tj,
Gurau:2012ix,Bonzom:2012hw}.

Consider a field $\phi$ defined over $d$-copies of a group manifold
$G$, $\phi\,:\,G^{\times d}\,\longrightarrow\,\mathbb{C}$. For the moment, we assume $G$ to be a compact Lie group. Without assuming any symmetry under permutations of field labels and using Peter-Weyl theorem, the field decomposes in group representations as follows:
\begin{eqnarray}
 \phi(g_1,\dots, g_d)=\sum_{\textbf{P}}\phi_{\textbf{P}}\prod_{i=1}^d D^{p_i}(g_i)\,,
\end{eqnarray}
with ${\textbf{P}}=(p_1,\dots,p_d)$, $g_i\in G$ and where the functions $ D^{p_i}(g_i)$ form a complete orthonormal basis of
functions on the group characterised by the labels $p_i$. In a TGFT model, we require fields to have tensorial properties
under basis changes. We define a rank $d$ covariant complex tensor $\phi_{\textbf{P}}$ to transform through the action of the tensor product of unitary
representations of the group $\bigotimes_{i=1}^d U^{(i)}$, each of them acting independently over the indices of field labels:
\begin{eqnarray}
\phi_{{p'}_1,\dots,{p'}_d}=\sum_{\textbf{P}}\,U^{(1)}_{{p'}_1,p_1}\dots U^{(d)}_{{p'}_d,p_d}\;\phi_{p_1,\dots,p_d}\,.
\end{eqnarray}
The complex conjugate field will then be the contravariant tensor transforming as:
\begin{eqnarray}
\ov{\phi}_{{p'}_1,\dots,{p'}_d}=\sum_{\textbf{P}}\,(U^{\dagger})^{(d)}_{{p'}_d,p_d}\dots (U^{\dagger})^{(1)}_{{p'}_1,p_1}\;
   \ov{\phi}_{p_1,\dots,p_d}\,.
\end{eqnarray}

TGFT interactions are defined by `trace invariants' built out of $\phi$ and $\ov{\phi}$, which allow a strong control on the combinatorial structure of field convolutions, and are thus relevant for the construction of renormalisable TGFT actions. Tensorial trace invariants generalise invariant traces over matrices, which indeed are classical unitary invariants. They are obtained contracting pairwise the indices with the same position of covariant and contravariant tensors and saturating all of them.
In this way, they always involve the same number of $\phi$
and $\ov{\phi}$. A simple example is the following:
\beq\label{tr2}
 \Tr(\phi\ov{\phi})=\sum_{\textbf{P},\textbf{Q}}\phi_{\textbf{P}}\ov{\phi}_{\textbf{Q}}
   \prod_{i=1}^d \delta_{p_i,q_i}\,.
\eeq
Considering that $\phi_{\textbf{P}}$ (resp. $\ov{\phi}_{\textbf{P}}$) transforms as a complex vector (resp. 1-form) under
the action of the unitary representations of $G$ on one single index, the fundamental theorem on
classical invariants for $U$ on each index entails that all invariant polynomials in field entries can be written as a
linear combination of trace invariants \cite{abdesselam}. This formulation of tensor models can be adapted to the real field case, where the unitary group is replaced by the orthogonal one \cite{ON}.

An interesting feature, which becomes an important computational tool, is that tensor invariants can be given a graphical
representation as  bipartite coloured graphs, and in fact they are in one to one correspondence with them. A tensor $\phi$
is represented by a (white) node with $d$ labelled half lines outgoing from it. Its complex conjugate is a similar $d$-valent node with a different colour (black). A tensor contraction is represented then by joining the half-lines, equally labelled, of two nodes of different colour.  

Trace invariants can be generalised to convolutions where the contractions are made by operators different from the delta
distribution, i.e. by non-trivial kernels. In this case, the resulting object is not guaranteed to be a unitary invariant. 

We write a generic action for a TGFT model symbolically as:
\begin{align}
 &S[\phi,\ov{\phi}]=\Tr(\ov{\phi}\cdot \ca{K}\cdot\phi)+ S^{\intt}[\phi,\ov{\phi}]\\
\Tr(\ov{\phi}\cdot \ca{K}\cdot\phi)=\sum_{\textbf{P},\textbf{Q}}\ov{\phi}_{\textbf{P}}&\,\ca{K}(\textbf{P};\textbf{Q})\,
     \phi_{\textbf{Q}}\,,
\qquad
  S^{\intt}[\phi,\ov{\phi}]=\sum_{\{n_b\}}\lambda_{n_b}\Tr(\ca{V}_{n_b}
\cdot\phi^n\cdot\ov{\phi}^n)\,.
\nonumber
\end{align}
Here $\ca{K}$ and $\ca{V}_n$ are kernels implementing the convolutions in the kinetic and interaction terms, respectively,
where $n$ indicates the numbers of covariant and contravariant fields appearing in the vertices, $b$ labels the combinatorics
of convolutions (i.e. corresponds to some given bipartite d-coloured graph) and $\lambda_{n_b}$ is a coupling constant for the interaction $n_b$.

The formalism can be easily generalised to a TGFT based on a non-compact group manifold $G$, and in this case the Plancherel decomposition into (unitary) representations replaces the Peter-Weyl one to decompose fields, and the definition of the trace over representation labels involves, in general, also integrals over continuous
variables.

Given an action $S[\phi,\ov{\phi}]$, the partition function is defined as usual:
\begin{eqnarray}
\label{partitionTensors}
\ca{Z}[J,\ov{J}]=e^{W[J, \ov{J}]}=\int d\phi d\ov{\phi}\;e^{-S[\phi,\ov{\phi}]+\Tr(J\cdot \ov{\phi})+\Tr(\ov{J}\cdot\phi)}\,,
\end{eqnarray}
where $J$ is a rank $d$ complex source term and $\Tr(J\cdot \ov{\phi})$
is defined in \eqref{tr2}. 

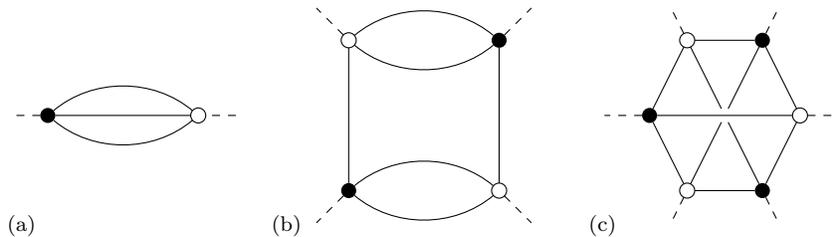
\begin{figure}[h]
\center
\begin{tikzpicture}
\draw[shorten <= 0.1cm, shorten >= 0.1cm] (-5,0) to[out=40, in=140] (-3,0);
\draw[shorten <= 0.1cm, shorten >= 0.1cm] (-5,0) to[out=0, in=180] (-3,0);
\draw[shorten <= 0.1cm, shorten >= 0.1cm] (-5,0) to[out=-40, in=-140] (-3,0);
\draw[shorten <= 0.1cm, dashed] (-5,0) to (-5.5,0);
\draw[shorten >= 0.1cm, dashed] (-2.5,0) to (-3,0);
\fill[fill=black] (-5,0) circle (0.1);
\draw (-3,0) circle (0.1);

\draw[shorten <= 0.1cm, shorten >= 0.1cm] (-1,1) to[out=40, in=140] (1,1);
\draw[shorten <= 0.1cm, shorten >= 0.1cm] (-1,1) to[out=-40, in=-140] (1,1);
\draw[shorten <= 0.1cm, shorten >= 0.1cm] (-1,-1) to[out=40, in=140] (1,-1);
\draw[shorten <= 0.1cm, shorten >= 0.1cm] (-1,-1) to[out=-40, in=-140] (1,-1);
\draw[shorten <= 0.1cm, shorten >= 0.1cm] (-1,1) to (-1,-1);
\draw[shorten <= 0.1cm, shorten >= 0.1cm] (1,1) to (1,-1);
\draw[shorten <= 0.1cm, shorten >= 0.1cm, dashed] (-1,1) to (-1.5,1.5);
\draw[shorten <= 0.1cm, shorten >= 0.1cm, dashed] (-1,-1) to (-1.5,-1.5);
\draw[shorten <= 0.1cm, shorten >= 0.1cm, dashed] (1,1) to (1.5,1.5);
\draw[shorten <= 0.1cm, shorten >= 0.1cm, dashed] (1,-1) to (1.5,-1.5);
\fill[fill=black] (1,1) circle (0.1);
\fill[fill=black] (-1,-1) circle (0.1);
\draw (1,-1) circle (0.1);
\draw (-1,1) circle (0.1);

\draw[shorten <= 0.1cm, shorten >= 0.1cm] (4.5,1) to (5,0);
\draw[shorten <= 0.1cm, shorten >= 0.1cm] (4.5,1) to (3.5,1);
\draw[shorten <= 0.1cm, shorten >= 0.1cm] (4.5,1) to (4,0);
\draw[shorten <= 0.1cm, shorten >= 0.1cm] (4.5,-1) to (5,0);
\draw[shorten <= 0.1cm, shorten >= 0.1cm] (4.5,-1) to (3.5,-1);
\draw[shorten <= 0.1cm, shorten >= 0.1cm] (4.5,-1) to (4,0);
\draw[shorten <= 0.1cm, shorten >= 0.1cm] (3,0) to (3.5,-1);
\draw[shorten <= 0.1cm, shorten >= 0.1cm] (3,0) to (3.5,1);
\draw[shorten <= 0.1cm, shorten >= 0.1cm] (3,0) to (5,0);
\draw[shorten <= 0.1cm, shorten >= 0.1cm] (3.5,1) to (4,0);
\draw[shorten <= 0.1cm, shorten >= 0.1cm] (3.5,-1) to (4,0);
\draw[shorten <= 0.1cm, shorten >= 0.1cm, dashed] (3.5,-1) to (3.25,-1.5);
\draw[shorten <= 0.1cm, shorten >= 0.1cm, dashed] (3.5,1) to (3.25,1.5);
\draw[shorten <= 0.1cm, shorten >= 0.1cm, dashed] (4.5,-1) to (4.75,-1.5);
\draw[shorten <= 0.1cm, shorten >= 0.1cm, dashed] (4.5,1) to (4.75,1.5);
\draw[shorten <= 0.1cm, shorten >= 0.1cm, dashed] (3,0) to (2.3,0);
\draw[shorten <= 0.1cm, shorten >= 0.1cm, dashed] (5,0) to (5.7,0);
\fill[fill=black] (4.5,1) circle (0.1);
\fill[fill=black] (4.5,-1) circle (0.1);
\fill[fill=black] (3,0) circle (0.1);
\draw (5,0) circle (0.1);
\draw (3.5,1) circle (0.1);
\draw (3.5,-1) circle (0.1);
\end{tikzpicture}
\put (-320,-0.7){\footnotesize (a)}
\put(-220,-0.7) {\footnotesize (b)}
\put(-100,-0.7){\footnotesize (c)}
\caption{Three examples of Feynman graphs for a rank 3 TGFT's. The trace invariants used to build the interactions are: figure (a)
$\Tr(\phi\ov{\phi})$, figure (b) an example of $\Tr(\phi\ov{\phi}\phi\ov{\phi})$, figure (c) an example of
$\Tr(\phi\ov{\phi}\phi\ov{\phi}\phi\ov{\phi})$.}
\label{traceinvex}
\end{figure}

The partition function can be expanded in perturbation theory around a Gaussian distribution, and expressed as a (formal) sum over Feynman diagrams.
Feynman diagrams of a rank-$d$ TGFT are obtained by attaching, to the bipartite graph corresponding to a trace invariant defining each interaction vertex, a propagator (dashed line) for each field obtaining a ($d+1$)-coloured graph (some examples are depicted in 
Fig.\ref{traceinvex}).

\subsection{FRG formulation for TGFTs}
\label{FRGfotgft}
The generalisation of the FRG formalism \cite{polchinski,Wetterich:1992yh,Delamotte-review,Morris:1993qb,Berges:2000ew} to TGFTs is straightforward and was first provided in \cite{Benedetti:2014qsa}. Given a partition
function of the type \eqref{partitionTensors}, we choose a UV cut-off $M$ and a IR cut-off $N$\footnote{We adopt a standard QFT terminology for field modes, even if no spacetime interpretation should be attached to them, at this stage.}. Adding to the
action a regulator term of the form:
\begin{eqnarray}
\Delta S_N[\phi,\ov{\phi}]=\Tr(\ov{\phi}\cdot R_N\cdot \phi)=
   \sum_{\textbf{P},\textbf{P}'}\,\ov{\phi}_{\textbf{P}}\,R_N(\textbf{P};\textbf{P}')\,\phi_{\textbf{P}'}\,\,,
\end{eqnarray}
we can perform the usual splitting in high and low modes. In particular, given an action with a generic kernel depending on the 
derivative of the fields $\ca{K}(\nabla\phi)$ and a generalised Fourier transform $\ca{F}$, if we choose $R_N$ to be of the 
specific form
\beq
\label{generalformR}
R_N(\textbf{P};\textbf{P}')=N\delta_{\textbf{P},\textbf{P}'}R\biggl(\frac{\ca{F}(\ca{K}_{\textbf{P}})}{N}\biggr)\,\,,
\eeq
we need to impose on the profile function $R(z)$ the following conditions:

- positivity $R(z)\geq 0$, to indeed suppress and not enhance modes outside of the domain of the regulator function;

- monotonicity $\frac{d}{dz}R(z)\leq 0$, so that high modes will not be suppressed more that low modes;

- $R(0)>0$ and $\lim_{z\to+\infty}R(z)=0$ to exclude functions with constant profile.

The last requirement, together with the form \eqref{generalformR}, guarantees that the regulator is removed for
$Z\to 0$. In accordance with the usual FRG procedure, we define the scale dependent partition function as:
\begin{eqnarray}
\ca{Z}_N[J,\ov{J}]=e^{W_N[J,\ov{J}]}=\int d\phi d\ov{\phi}\;
   e^{-S[\phi,\ov{\phi}]-\Delta S_N[\phi,\ov{\phi}]+\Tr(J\cdot\ov{\phi})+\Tr(\ov{J}\cdot\phi)}
\end{eqnarray}
and the generating functional of 1PI correlation functions after 
Legendre transform are given in terms of  the average field
$\varphi=\langle\phi\rangle$ as
\begin{eqnarray}
\Gamma_N[\varphi,\ov{\varphi}]=\sup_{J,\ov{J}}\biggl\{\Tr(J\cdot\ov{\varphi})+\Tr(\ov{J}\cdot\varphi)
   -W_N[J,\ov{J}]-\Delta S_N[\varphi,\ov{\varphi}]\biggr\}\,\,.
\end{eqnarray}
Given the above definitions, the Wetterich equation takes the form:
\begin{eqnarray}
\label{wetttensor}
\de_t\Gamma_N[\varphi,\ov{\varphi}]=\ov{\Tr}\Big(\de_tR_N\cdot[\Gamma_N^{(2)}+R_N]^{-1}\Big)\,,
\end{eqnarray}
where $t=\log N$, so that $\de_t=N\de_N$, and the ``super''-trace symbol $\ov{\Tr}$ means that we are summing over all 
mode labels. More explicitly, the functional trace reads:
\begin{eqnarray}\label{matlikeEq}
\sum_{\textbf{P},\textbf{P}'}\de_tR_N(\textbf{P};\textbf{P}')
   [\Gamma_N^{(2)}+R_N]^{-1}(\textbf{P}';\textbf{P})\,.
\end{eqnarray}

The presence of the $\de_tR_N$ in the Wetterich equation for TGFT's, enforces the trace to be UV-finite if the profile
function and its derivative go fast enough to $0$,  as $z\to +\infty$. In this way, we can basically forget about the
UV cut-off $M$. In any case, as in any resolution of 
differential equation, we need an initial condition of the type
\begin{equation}\label{gammacondition2}
\Gamma_{N=M}[\varphi,\ov{\varphi}]=S[\varphi,\ov{\varphi}]\,, 
\end{equation}
for some scale $M$. The problem of solving the full quantum theory is now phrased in the one of pushing the initial condition to infinity, which usually requires the existence of a UV fixed point, and solving the Wetterich equation with such initial condition. The full quantum field theory will then be defined by the corresponding solution, i.e. by the full RG trajectory.

The Wetterich equation has a 1-loop structure, and since no (perturbative) approximation is required to obtain it, it is an exact functional equation.
However, although we have expressed the problem of extracting the flow of the theory in terms of a partial differential equation in
one single variable, we still have the issue that all possible (i.e. compatible with symmetry requirements and field content) couplings
are allowed in $\Gamma_k$, which is thus expressible as an infinite sum of monomials in the field (and its conjugate). If we want to perform practical computations, we need some approximation scheme for the form of the
free energy. Usually, this is obtained by truncating $\Gamma_k$ to a maximal power in the fields and in their derivative. It is then a truncation in theory space, which maintains the non-perturbative character of the RG equation.

What is peculiar, and interesting, about the application of FRG to TGFTs, is that $\Gamma_N^{(2)}$ carries inside the Wetterich
equation information about the combinatorial non-locality of the theory, i.e. the intricate combinatorics of TGFT interactions. 
In the case we consider here, that of a non-compact group manifold, this will also back-react  at the level of the $\beta$-functions, in the fact that, depending on the combinatorics of the interaction, the volume contributions appearing in \eqref{wetttensor} will  be not homogeneous and, in general, a natural definition of an effective local potential does not exist. Let us explain this key point, which we will deal with in detail in the following.

In its usual form, namely when applied to a standard, local quantum field theory (see for instance, 
in \cite{Delamotte-review}), the Wetterich equation shows pathological IR divergences due to the presence of
$\delta(0)$ arising from the two-point Green's function computed at a single point $G_k^{(2)}(q,q)$. In the local field theory case, these divergent delta functions are homogeneous and proportional to the total volume of the system,
namely, the domain manifold of the fields. A particular approximation procedure allows to cure this problem and it is called the local potential approximation (LPA) \cite{Delamotte-review}. This procedure 
cannot be applied, at least not in the same straightforward way, to combinatorially non-local
theories as TGFTs. One reason is that, in such non-local theories, the same type of IR divergence arise, in general, in a non-homogeneous combination
of $\delta(0)$ which are strictly dependent on the combinatorics of the interaction. We will discuss this and several other issues characterising TGFTs as QFTs of an interesting new kind.

\

\section{Rank-$d$ Tensorial Group Field Theory on \R}
\label{sect3:TGFT}
As discussed in the introduction, the first model studied within the FRG framework for TGFTs, already in \cite{Benedetti:2014qsa}, was a rank-3 model with compact group manifold $U(1)$, and subsequently, we have studied a non-compact counterpart of the same model, i.e. a rank-3 TGFT on $\mathbb{R}$ \cite{Geloun:2015qfa}. New issues concerning the thermodynamic limit but also more compelling hints for the existence of UV and IR fixed points, and of a condensation phase transitions, were found.

We now extend the analysis and results of the latter work to arbitrary rank (as well as analysing in more detail in the rank-3 model), showing how those intriguing hints are actually confirmed in a more general case. In the following section, we will analyse a modification of the same type of TGFT models which includes a gauge invariance property of fields and amplitudes, thus moving closer to full-fledged TGFT models for quantum geometry and discrete quantum gravity, and related to loop quantum gravity. 

We start by introducing the class of TGFT models we will analyse.

\subsection{The model}
\label{themod}

The TGFTs we work with have  ``melonic'' interactions (in correspondence with $d$-colored graphs called ``melons'') \cite{coloured, tensorNew, criticalTensor}. 
Such melons are dual to special triangulations of the $d$-ball \cite{Bonzom:2012hw} 
and of course correspond  also to trace invariants of the type introduced in section \ref{subsect:TGFT}.

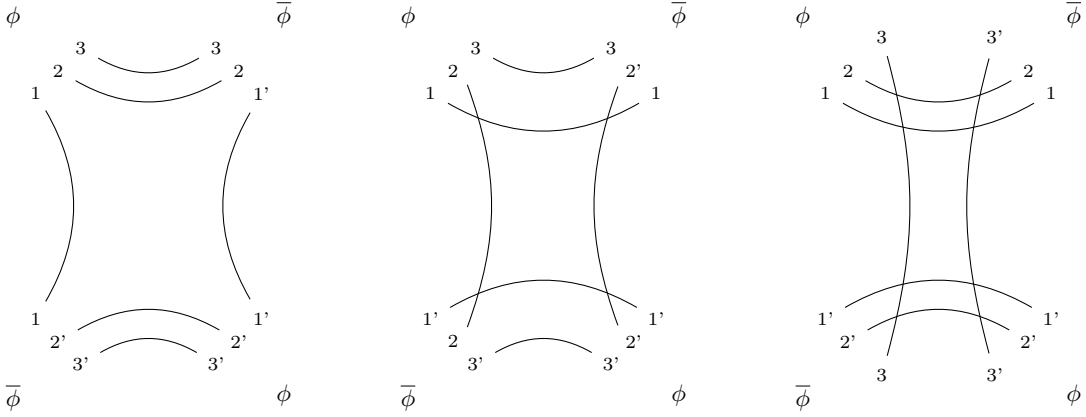
\begin{figure}
\centering
\begin{tikzpicture}[scale=1.5]
\path (0,1) node (1rs1) [shape=circle ]{\scriptsize 1}
      (0,-1) node (1cs1) [shape=circle ]{\scriptsize 1}
      (0.2,1.2) node (2rs1) [shape=circle ]{\scriptsize 2}
      (1.8,1.2) node (2cs1) [shape=circle ]{\scriptsize 2}
      (0.4,1.4) node (3rs1) [shape=circle]{\scriptsize 3}
      (1.6,1.4) node (3cs1) [shape=circle]{\scriptsize 3}
      (2,-1) node (11rs1) [shape=circle]{\scriptsize 1'}
      (2,1) node (11cs1) [shape=circle]{\scriptsize 1'}
      (1.8,-1.2) node (21rs1) [shape=circle]{\scriptsize 2'}
      (0.2,-1.2) node (21cs1) [shape=circle]{\scriptsize 2'}
      (1.6,-1.4) node (31rs1) [shape=circle]{\scriptsize 3'}
      (0.4,-1.4) node (31cs1) [shape=circle]{\scriptsize 3'}
      (-0.2,1.4) node (phi11) [shape=circle,  label=above:$\phi$]{}
      (-0.2,-1.4) node (ovphi11) [shape=circle,  label=below:$\ov{\phi}$]{}
      (2.2,1.4) node (ovphi21) [shape=circle,  label=above:$\ov{\phi}$]{}
      (2.2,-1.4) node (phi21) [shape=circle,  label=below:$\phi$]{};
      \draw (1rs1) to [bend left=30]  (1cs1);
      \draw (2rs1) to [bend right=30]  (2cs1);
      \draw (3rs1) to [bend right=30]  (3cs1);
      \draw (11rs1) to [bend left=30] (11cs1);
      \draw (21rs1) to [bend right=30] (21cs1);
      \draw (31rs1) to [bend right=30] (31cs1);

\path[shift={(3.5,0)}] (0,1) node (1rs2) [shape=circle]{\scriptsize 1}
      (0,-1) node (11cs2) [shape=circle]{\scriptsize 1'}
      (0.2,1.2) node (2rs2) [shape=circle]{\scriptsize 2}
      (1.8,1.2) node (21cs2) [shape=circle]{\scriptsize 2'}
      (0.4,1.4) node (3rs2) [shape=circle]{\scriptsize 3}
      (1.6,1.4) node (3cs2) [shape=circle]{\scriptsize 3}
      (2,-1) node (11rs2) [shape=circle]{\scriptsize1'}
      (2,1) node (1cs2) [shape=circle]{\scriptsize1}
      (1.8,-1.2) node (21rs2) [shape=circle]{\scriptsize 2'}
      (0.2,-1.2) node (2cs2) [shape=circle]{\scriptsize2}
      (1.6,-1.4) node (31rs2) [shape=circle]{\scriptsize 3'}
      (0.4,-1.4) node (31cs2) [shape=circle]{\scriptsize 3'}
      (-0.2,1.4) node (phi12) [shape=circle,  label=above:$\phi$]{}
      (-0.2,-1.4) node (ovphi12) [shape=circle,  label=below:$\ov{\phi}$]{}
      (2.2,1.4) node (ovphi22) [shape=circle,  label=above:$\ov{\phi}$]{}
      (2.2,-1.4) node (phi22) [shape=circle,  label=below:$\phi$]{};
      \draw (1rs2) to [bend right=30]  (1cs2);
      \draw (2rs2.south east) to [bend left=20]  (2cs2.north east);
      \draw (3rs2) to [bend right=30]  (3cs2);
      \draw (11rs2) to [bend right=30] (11cs2);
      \draw (21rs2.north west) to [bend left=20] (21cs2.south west);
      \draw (31rs2) to [bend right=30] (31cs2);
      
\path[shift={(7,0)}] (0,1) node (1rs3) [shape=circle]{\scriptsize 1}
      (0,-1) node (11cs3) [shape=circle]{\scriptsize 1'}
      (0.2,1.2) node (2rs3) [shape=circle]{\scriptsize2}
      (1.8,1.2) node (2cs3) [shape=circle]{\scriptsize2}
      (0.5,1.5) node (3rs3) [shape=circle]{\scriptsize3}
      (1.5,1.5) node (31cs3) [shape=circle]{\scriptsize3'}
      (2,-1) node (11rs3) [shape=circle]{\scriptsize1'}
      (2,1) node (1cs3) [shape=circle]{\scriptsize1}
      (1.8,-1.2) node (21rs3) [shape=circle]{\scriptsize2'}
      (0.2,-1.2) node (21cs3) [shape=circle]{\scriptsize2'}
      (1.5,-1.5) node (31rs3) [shape=circle]{\scriptsize3'}
      (0.5,-1.5) node (3cs3) [shape=circle]{\scriptsize3}
      (-0.2,1.4) node (phi13) [shape=circle,  label=above:$\phi$]{}
      (-0.2,-1.4) node (ovphi13) [shape=circle,  label=below:$\ov{\phi}$]{}
      (2.2,1.4) node (ovphi23) [shape=circle,  label=above:$\ov{\phi}$]{}
      (2.2,-1.4) node (phi23) [shape=circle,  label=below:$\phi$]{};
      \draw (1rs3) to [bend right=30]  (1cs3);
      \draw (2rs3) to [bend right=30]  (2cs3);
      \draw (3rs3) to [bend left=15]  (3cs3);
      \draw (11rs3) to [bend right=30] (11cs3);
      \draw (21rs3) to [bend right=30] (21cs3);
      \draw (31rs3) to [bend left=15] (31cs3);
\end{tikzpicture}
\caption{Colored symmetric interaction terms in rank $d=3$.}
\label{Sym}
\end{figure}

We consider a rank-$d$ model with complex field, $\phi:\R^d\to \C,$ defined by the following action:
\bea
\label{actiondirectspace}
&&
S[\phi,\ov{\phi}]=(2\pi)^d\int_{\mathbb{R}^{\times d}}[dx_i]_{i=1}^d\;
    \ov{\phi}(x_1,\dots,x_d)\biggl(-\sum_{s=1}^d\lap_s+\mu\biggr)\phi(x_1,\dots,x_d)\cr\cr
&&+\frac{\lambda}{2}(2\pi)^{2d}\int_{\mathbb{R}^{\times 2d}}[dx_i]_{i=1}^d[dx'_j]_{j=1}^d\biggl[
    \phi(x_1,x_2,\dots,x_d)\ov{\phi}(x'_1,x_2,\dots,x_d)\phi(x'_1,x'_2,\dots,x'_d)\ov{\phi}(x_1,x_2',\dots,x'_d)\cr\cr
&&\qquad \qquad +\Sym\biggr]\,,
\eea
where $2\pi$ factors  have been conveniently introduced in the definition of the Fourier transform,
 the symbol sym$\{\cdot\}$ represents the rest of the coloured symmetric terms in the interaction (see Fig.\ref{Sym} for a graphical representation\footnote{As a remark, in the following subsections,
illustrations and figures are made in the case $d=3$ because the general
case can be easily recovered from that case.} 
 in rank $d=3$); $\mu$
and $\lambda$ are coupling constants. As it is easy to see, due to the structure of the interaction kernels,
the interaction fully depends on all the six coordinates and this makes it 
non-local from the combinatorial point of view. 

After Fourier transform, we  write the action in momentum space as:
\bea
&&S[\phi,\ov{\phi}]=\int_{\mathbb{R}^{\times d}}[dp_i]_{i=1}^d\;
    \ov{\phi}_{12\dots d}\biggl(\sum_{s=1}^d p_s^2 +\mu\biggr)\phi_{12\dots d}\\
&&+\frac{\lambda}{2}\int_{\mathbb{R}^{\times 2d}}[dp_i]_{i=1}^d[dp'_j]_{j=1}^d\;\biggl[
    \phi_{12\dots d}\ov{\phi}_{1'2\dots d}\phi_{1'2' \dots d'}\ov{\phi}_{12'\dots d'}
    +\Sym\biggr]\,,
\nonumber
\eea
where we use the conventions
\begin{align}
\phi_{12 \dots d}=\phi_{p_1,p_2,\dots, p_d}=\phi(\textbf{p})&=\int_{\mathbb{R}^{\times d}}[dx_i]_{i=1}^d\;
     \phi(x_1,x_2,\dots, x_d)\,e^{-i\sum_ip_ix_i}\,,\\
\phi(x_1,x_2,\dots, x_d)&=\frac{1}{(2\pi)^d}\int_{\R^{\times d}}[dp_i]_{i=1}^d\;\phi_{12 \dots d}\,e^{i\sum_ip_ix_i}\,.
\end{align}
We represent the propagator  as a stranded line made with $d$ segments (strands). See Fig.\ref{FeynPropa} for the case $d=3$. The combinatorics of the interaction is preserved by the Fourier transform.

\begin{figure}[h]
\centering
\begin{tikzpicture}
\path coordinate (a) at (0.3,0)
      coordinate (a1) at (-2.7,0)
      coordinate (b) at (0.3,0.2)
      coordinate (b1) at (-2.7,0.2)
      coordinate (c) at (0.3,-0.2)
      coordinate (c1) at (-2.7,-0.2)
      node (1) at (0.6,0) [right]{$\phi$}
      node (2) at (-3,0) [left]{$\ov{\phi}$}
      node (propa) at (3,0) []{$=\biggl(\sum_sp_s^2+\mu\biggr)^{-1}$};
      \draw (a) to (a1);
      \draw (b) to (b1);
      \draw (c) to (c1);
\end{tikzpicture}
\caption{Feynman rule for the propagator at $d=3$.}
\label{FeynPropa}
\end{figure}

We can now proceed with the dimensional analysis to fix the dimensions of the coupling constants. In order to make sense of the exponentiation of the action in the partition function, we must set $[S]=0$. Furthermore, we fix the dimensions to be in unit of the momentum, i.e., $[p]=[dp]=1$ \footnote{Notice that the physical dimension of such momentum variables, if any, is not especially relevant in this context; what matters is the relative dimension of the various ingredients entering the TGFT action.}. Now, for consistency we must have $[\mu]=2$.
This leads us to the following equations:
\begin{align}
&3+2[\phi]+2=0\;\Rightarrow\;[\phi]=-\frac{d+2}{2}\,,\\
&[\lambda]+2d+4[\phi]= 0\;\Rightarrow\;[\lambda]=4\, ,
\end{align}
which fix the dimension of the TGFT fields depending on the rank $d$ of the model.

\subsection{Effective action and Wetterich equation}
\label{Wettgftapp}

In order to proceed with the Functional Renormalisation Group analysis, following the general template described in the previous section, we introduce an IR cut-off $k$ and a UV cut-off $\Lambda$. We need to perform a truncation on the form of the effective action. A natural choice, compatible with the condition \eqref{gammacondition2}, is to truncate the effective action to be of the same form of the action itself for any value of the cut-offs, that is:
\begin{align}
\label{gammaModel}
\Gamma_k[\varphi,\ov{\varphi}]=&\int_{\R^{\times d}} [dp_i]_{i=1}^d\;\ov{\varphi}_{12\dots d}(Z_k\sum_sp_s^2+\mu_k)\varphi_{12\dots d}\\
&+\frac{\lambda_k}{2}\int_{\R^{\times 2d}} [dp_i]_{i=1}^d[dp'_j]_{j=1}^d\;\biggl[
    \varphi_{12\dots d}\ov{\varphi}_{1'2\dots d}\varphi_{1'2' \dots d'}\ov{\varphi}_{12' \dots d'}
    +\Sym  \biggr]\,,
\nonumber
\end{align}
where $\varphi=\langle\phi\rangle$. As we have already stressed, this is a non-perturbative truncation of the theory, and any of the ensuing results should then be tested by extending this truncation, including more invariants (including other types of $\Tr(\phi^4)$ invariants, i.e. with different combinatorics, as well as higher order terms $\Tr(\phi^{2n})$, $n\geq 3$; in general, one should include also disconnected invariants such as multi-traces, $\Tr(\phi^{2n})\Tr(\phi^{2m})\dots$) and checking for (qualitative) convergence. 
Enlarging the theory space  is postponed for future
investigations, but it should be obvious that, even in the truncation given by \eqref{gammaModel}, the calculations and the outcome of the
present analysis remain highly non-trivial.

From the dimensional analysis of the previous section and from the fact that
$[\Gamma_k]=0$ and $[\varphi]=[\phi]$, one infers $[Z_k]=0$, $[\mu_k]=[\mu]=2$, $[\lambda_k]=[\lambda]=4$.

We introduce a regulator kernel of the following form \cite{Litim:2011cp, Litim:2001up}
\begin{eqnarray}\label{reR}
R_k(\textbf{p},\textbf{p}')=\delta(\textbf{p}-\textbf{p}')Z_k(k^2-\sum_sp_s^2)\theta(k^2-\sum_sp_s^2)\,,
\end{eqnarray}
where $\theta$ stands for the Heaviside step function. This form of the regulator is convenient because it allows to solve analytically many spectral sums. It is easy to show that $R_k$ satisfies the minimal requirements for a regulator kernel:
\begin{itemize}
\item 
as a consequence of the fact that $\theta(-|x|)=0$, we have
\begin{eqnarray}
R_{k=0}(\textbf{p},\textbf{p}')=\delta(\textbf{p}-\textbf{p}')Z_k(-\sum_sp_s^2)\theta(-\sum_sp_s^2)=0\,;
\end{eqnarray}
\item 
at the scale $k=\Lambda$, the regulator takes the form:
\begin{eqnarray}
R_{k=\Lambda}(\textbf{p},\textbf{p}')=\delta(\textbf{p}-\textbf{p}')
   Z_{\Lambda}(\Lambda^2-\sum_sp_s^2)\theta(\Lambda^2-\sum_sp_s^2)\,,
\end{eqnarray}
which at the first order gives: $R_{k=\Lambda}\simeq Z_{\Lambda}\Lambda^2$;
\item
for $k\in[0,\Lambda]$, we have also:
\begin{eqnarray}
&&R_k(\textbf{p},\textbf{p}')=0\,, \qquad \forall\textbf{p},\textbf{p}',
\textrm{ such that }\; |\textbf{p}|,\, |\textbf{p}'|>k\,,\\
&&R_k(\textbf{p},\textbf{p}')\simeq Z_kk^2\,, \qquad \forall\,\textbf{p},\textbf{p}', \textrm{ such that }\;
  |\textbf{p}|,|\textbf{p}'|<k\,.
\end{eqnarray}
\end{itemize}
The derivative of the regulator kernel with respect to the logarithmic scale $t=\log k$, entering in the Wetterich equation, evaluates as:
\begin{eqnarray}
\de_tR_k(\textbf{p},\textbf{p}')=\theta(k^2-\Sigma_sp_s^2)[\de_tZ_k(k^2-\Sigma_sp_s^2)+2k^2Z_k]\delta(\textbf{p}-\textbf{p}')\,.
\end{eqnarray}
One notes that $R_k$ and $\de_t R_k$ are both symmetric
kernels, which is important in evaluating the convolutions induced by the 
Wetterich equation. 

\begin{figure}[h]
\centering
\begin{tikzpicture}
\path coordinate (a) at (0.2,0)
      coordinate (a1) at (-2.6,0)
      coordinate (b) at (0.3,0.3)
      coordinate (b1) at (-2.7,0.3)
      coordinate (c) at (0.2,-0.3)
      coordinate (c1) at (-2.6,-0.3);
      \draw (b) -- (b1) ;
\fill[fill=black] (a) circle (0.1);
\fill[fill=black] (a1) circle (0.1);
\fill[fill=black] (c) circle (0.1);
\fill[fill=black] (c1) circle (0.1);
\path coordinate (A) at (6,0)
      coordinate (A1) at (3,0)
      coordinate (B) at (5.9,0.3)
      coordinate (B1) at (3.1,0.3)
      coordinate (C) at (6,-0.3)
      coordinate (C1) at (3,-0.3);
      \draw (A) -- (A1) ;
  \draw (C) -- (C1) ;
\fill[fill=black] (a) circle (0.1);
\fill[fill=black] (a1) circle (0.1);
\fill[fill=black] (c) circle (0.1);
\fill[fill=black] (c1) circle (0.1);
\fill[fill=black] (B) circle (0.1);
\fill[fill=black] (B1) circle (0.1);
\end{tikzpicture}
\put(5,20){\footnotesize $q'_1$}
\put(-103,19){\footnotesize $q_1$}
\put(5,9){\footnotesize $p_2$}
\put(5,-3){\footnotesize $p_3$}
\put(-103,9){\footnotesize $p_2$}
\put(-103,-3){\footnotesize $p_3$}
\put(-155,20){\footnotesize $p_1$}
\put(-262,20){\footnotesize $p_1$}
\put(-155,9){\footnotesize $q_2'$}
\put(-262,-3){\footnotesize $q_3$}
\put(-262,9){\footnotesize $q_2$}
\put(-155,-3){\footnotesize $q_3'$}
\caption{Terms of the second variation of $\Gamma_k$ at rank $d=3$.}
\label{SecondVar}
\end{figure}
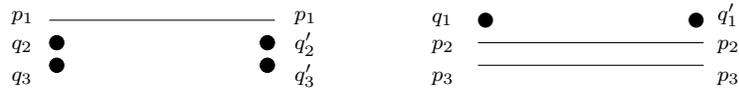

Computing the 1PI 2-point function yields:
\begin{align}\label{eqq}
\Gamma_k^{(2)}(\textbf{q},\textbf{q}')&
=(Z_k\sum_sq_s^2+\mu_k)\delta(\textbf{q}-\textbf{q}')
   +\lambda_k\biggl[\int_{\R} dp_1\;\varphi_{p_1q'_2\dots q'_d}\ov{\varphi}_{p_1q_2\dots q_d}\delta(q_1-q'_1)\crcr
&+\int_{\R^{\times (d-1)}} [dp_i]_{i=2}^{d}\;\varphi_{q'_1p_2\dots p_d}\ov{\varphi}_{q_1p_2\dots p_d}
  [\prod_{i=2}^{d}\delta(q_i-q_i')]+\Sym\biggr]\crcr
&=(Z_k\sum_sq_s^2+\mu_k)\delta(\textbf{q}-\textbf{q}')+F_k(\textbf{q},\textbf{q'})\,.
\end{align}
There is a simple graphical way to picture the various 
terms contributing to $F_k$. Each summed index can be represented by a segment
and each fixed index (not summed) by a dot. As an example in rank $d=3$, Fig.\ref{SecondVar} displays two terms coming from the second variation of the interaction
labeled by colour 1 (the ones which appear explicitly in \eqref{eqq}). The other terms appearing in sym$\{\cdot\}$
can be  inferred by colour permutation. 

Defining the operator $P_k$ with kernel 
\be\label{Pterm}
P_k(\textbf{p},\textbf{p'})=R_k(\textbf{p},\textbf{p'})+(Z_k\sum_sp_s^2+\mu_k)\delta(\textbf{p}-\textbf{p'})\,,
\ee
 the Wetterich equation can be recast as:
\begin{eqnarray}
\label{WettEqModel}
\de_t\Gamma_k=\Tr[\de_tR_k\cdot(P_k+F_k)^{-1}]\,.
\end{eqnarray}
The r.h.s. of \eqref{WettEqModel} generates an infinite series of
terms with convolutions involving an arbitrary number of fields. 
In order to compare the two sides of \eqref{WettEqModel}, we must
therefore perform a truncation in this series to match with
the l.h.s. of that equation. This may be achieved expanding \eqref{WettEqModel} in powers of $F_k\cdot(P_k)^{-1}$, that is, 
in powers of $\varphi\ov{\varphi}$, and considering only the terms up to the power 2:
\begin{align}
\label{wettexpansionPHI4}
\de_t\Gamma_k&=\Tr[\de_tR_k\cdot(P_k)^{-1}
\cdot(1+F_k\cdot(P_k)^{-1})^{-1}]\\
&=\Tr[\de_tR_k\cdot(P_k)^{-1}\cdot(1-F_k\cdot(P_k)^{-1}+F_k\cdot(P_k)^{-1}\cdot F_k\cdot(P_k)^{-1})
 +o((\varphi\ov{\varphi})^3)]\,.
\nonumber
\end{align}
The vacuum term proportional to the 0-th order in the above expansion will be discarded because it does not reflect
 any term in the l.h.s. of \eqref{WettEqModel}. As an explicit example, the trace at linear order
takes the form:
\beq
\de_t\Gamma_k^{\textrm{kin}}=
   \int_{\R^{\times 12}}
   \de_tR_k(\textbf{p},\textbf{p}')(P_k)^{-1}(\textbf{p}',\textbf{q})
      F_k(\textbf{q},\textbf{q}')(P_k)^{-1}(\textbf{q}',\textbf{p})\,. 
\eeq
Already, from the structure of the operators, $\de_t R_k$, $P_k$ 
and $F_k$,  we expect the presence of singular $\delta$-functions which need to be regularised. Indeed, the appearance of $\delta(0)$-terms reflects the fact that 
we have infinite volume effects which have to be treated. 
The presence of such infinities, as we have anticipated above, is not a specific feature of  TGFTs, as it also arises
in standard QFT. What is peculiar in TGFTs is the fact that, 
due to the combinatorics of the vertex operators, these divergences cannot be addressed
by projection on the constant fields. Roughly speaking, in ordinary (local) field theories
projecting on constant fields allows to factorise out the full
volume of the space entering some given power of $\delta(0)$, and depending only on the order of the field interaction. Such a procedure cannot be applied
in the present setting, the main reason being that the order of volume divergences depends not only on the order of field interactions but also on their precise convolution pattern. This would be entirely lost in a constant field projection, and must instead be checked term-by-term in the expansion of \eqref{WettEqModel}. The best way to tackle these divergences is to resort to a compactification of configuration space, corresponding to a discretisation in the conjugate space, and define an appropriate thermodynamic limit. This is explained in the next section. 

\subsection{IR divergences and thermodynamic limit}
\label{IRthermo}

In order to regularise volume divergences, we perform a compactification of the direct space and a lattice regularisation
in the conjugate space, following the conventions of \cite{salm}, and generalising to arbitrary rank the procedure adopted in \cite{Geloun:2015qfa}. 
Defining the model \eqref{actiondirectspace} over a compact set $D\subset\R^{\times d}$ with volume $L^d=(2\pi r)^d$, and taking a Fourier transform, the domain of integration (actually, summation) of the effective action, in momentum space, becomes the lattice
\beq
D^*=\biggl(\frac{2\pi}{L}\Z\biggr)^{\times d}= \biggl(\frac{1}{r}\Z\biggr)^{\times d}
:=\biggl(l\Z\biggr)^{\times d}\,,
\eeq
so that we have, for any function $F(\textbf{p})$,
\begin{eqnarray}
 \int_{D^*}[dp_i]_{i=1}^d\;F(\textbf{p})=l^d\sum_{p_1,p_2,\dots, p_d\in D^*}F(\textbf{p})\,\,.
\end{eqnarray}
We define the delta distribution in ${D^*}$ in momentum space as:
\begin{eqnarray}
 \delta_{D^*}(\textbf{p},\textbf{q})=l^{-d}\delta_{\textbf{p},\textbf{q}}\,.
\end{eqnarray}
with $\delta_{\textbf{p},\textbf{q}}= \prod_s \delta_{p_s,q_s}$, the Kronecker delta.
Choosing an orthonormal basis $(e_\textbf{p})_{\textbf{p}\in D^*}$ for the space of fields such that 
$e_\textbf{p}(\textbf{q})=\delta_{D^*}(\textbf{p},\textbf{q})$, we have:
\begin{eqnarray}
 \phi(\textbf{p})=\langle e_{\textbf{p}},\phi\rangle_{D^*}\,. 
\end{eqnarray}
For a generic observable $A$, we then have 
\begin{eqnarray}
 (A\phi)(\textbf{p})=\int_{D^*}[dq_i]_{i=1}^d\;A(\textbf{q},\textbf{p})\phi(\textbf{p})=
   \int_{D^*}[dq_i]_{i=1}^d\;\langle e_\textbf{q},A e_\textbf{p}\rangle_{D^*}\phi(\textbf{p})\,\,.
\end{eqnarray}
Whenever $A$ is invertible, then the inverse operator satisfies
\bea
 \label{inverseOperator}
 \int_{D^*}[dr_i]_{i=1}^d\;A(\textbf{p},\textbf{r})A^{-1}(\textbf{r},\textbf{q})=
  \delta_{D^*}(\textbf{p},\textbf{q})\,\, .
\eea
We also define the regularised functional derivative as:
\bea
\label{functionalderivativediscrete}
 \frac{\delta}{\delta\phi(\textbf{p})}=l^{-d}\frac{\de}{\de\phi(\textbf{p})}\,,
\eea
so that the following relations hold:
\beq
 \frac{\delta}{\delta\phi({\textbf{p}})}\,\phi({\textbf{q}})=\delta_{D^*}({\textbf{p}},{\textbf{q}})\,, \qquad \quad
 \frac{\delta}{\delta J(\textbf{p})}\,e^{\langle J,\phi\rangle_{D^*}}=J(\textbf{p})\,e^{\langle J,\phi\rangle_{D^*}}\,.
\eeq
This set of conventions is of course consistent with the continuous version of field theory, where $\delta_{D^*}$
becomes the Dirac $\delta$-distribution and the derivative \eqref{functionalderivativediscrete} becomes the standard
functional derivative.

Using this regularization prescription, the effective action of the model takes the form: 
\begin{align}
 \Gamma_k[\varphi,\ov{\varphi};l]&=
   l^d\sum_{\textbf{p}\in D^*}\ov{\varphi}_{12\dots d}\biggl(Z_k\sum_sp_s^2+\mu_k\biggr)\varphi_{12\dots d}\nonumber\\
&+l^{2d}\,\frac{\lambda_k}{2}\sum_{\textbf{p},\textbf{p}'\in D^*}\biggl[
    \varphi_{12\dots d}\ov{\varphi}_{1'2\dots d}\varphi_{1'2'\dots d'}\ov{\varphi}_{12'\dots d'}
    +\Sym\biggr],
\end{align}
where, using the same notation $\varphi$ for the field and its Fourier
transform, one has:  
\bea
\label{fourier}
 &&\varphi(x_1,x_2,\dots, x_d)=(2\pi)^{-d}l^{d}\sum_{\textbf{p}\in D^*}e^{i\sum_ip_ix_i}\varphi(\textbf{p})\,, \cr\cr
&&\varphi(\textbf{p})=\int_D[dx_i]_{i=1}^d\;e^{-i\sum_ip_ix_i}\varphi(x_1,x_2,\dots, x_d)\,. 
\eea
Now we use the relations \eqref{fourier} to transform $\delta_{D^*}$ and
obtain
\beq
 (2\pi)^{-d}l^d\sum_{\textbf{p}\in D^*}\delta_{D^*}(\textbf{p},\textbf{q})e^{i\sum_ip_ix_i}=(2\pi)^{-d}e^{i\sum_iq_ix_i}\,. 
\eeq
Thus, an integral representation of the delta distribution over $D^*$ can be consistently defined as 
\beq
 \delta_{D^*}(\textbf{p},\textbf{q})=(2\pi)^{-d}\int_{D}[dx_i]_{i=1}^d\;e^{-i\sum_i(p_i-q_i)x_i}\,.
\eeq
As a final result, we have:
\beq
  \delta_{D^*}(\textbf{p},\textbf{p})=\frac{(2\pi r)^d}{(2\pi)^d}=r^d=\frac{1}{l^d}\,.
\eeq
From these formulae, the continuum description will be recovered in the thermodynamic limit $l\to 0$. 

This procedure makes the dependence on the volume of the direct space  explicit. We can then rescale also the coupling constants of the model so to incorporate in their definition a dependence on the same volume. Then, we can use this dependence in such a way that the non-compact (thermodynamic) limit of the theory becomes well defined and all divergences are consistently removed.

\subsection{$\beta$-functions and RG flows}
\label{betastgft}

We introduce a regularisation as outlined in section \ref{IRthermo}, 
and write the regularised effective action as: 
\begin{align}
\label{gammaModelDiscrete}
\Gamma_k[\varphi,\ov{\varphi}]=&\int_{D^*} [dp_i]_{i=1}^d\;\ov{\varphi}_{12\dots d}(Z_k\sum_sp_s^2+\mu_k)\varphi_{12 \dots d}\\
+&\frac{\lambda_k}{2}\int_{{D^*}^{\times 2}} [dp_i]_{i=1}^d[dp'_j]_{j=1}^d\;\biggl[
    \varphi_{12\dots d}\ov{\varphi}_{1'2\dots d}\varphi_{1'2'\dots d'}\ov{\varphi}_{12' \dots d'}
    +\Sym\biggl]\,.
\nonumber
\end{align}
We can study the Wetterich equation corresponding to the action \eqref{gammaModelDiscrete}, incorporating a dependence on the volume in the coupling constants, and perform a thermodynamic limit at the end of the
computation to extract the coefficients valid in the non-compact case.

The set of $\beta$-functions that we obtain from the discretised model is\footnote{Important steps of the calculation are detailed in appendix
\ref{app:betaTGFT}}:
\bea\label{betadimensionful}
&&\hspace{-0.8cm}
\left\{
\begin{aligned}
\beta(Z_k)&=\frac{\lambda_k}{(Z_kk^2+\mu_k)^2}\Big\{\de_tZ_k\Big[2(d-1)\frac{k}{l}
   +\frac{\pi^{\frac{d-1}{2}}}{\Gamma_E\Big(\frac{d+1}{2}\Big)}\frac{k^{d-1}}{l^{d-1}}\Big]
   +2Z_k\Big[(d-1)\frac{k}{l}
   +\frac{\pi^{\frac{d-1}{2}}}{\Gamma_E\Big(\frac{d-1}{2}\Big)}\frac{k^{d-1}}{l^{d-1}}\Big]\Big\}\crcr
\beta(\mu_k)&=-\frac{d\,\lambda_k}{(Z_kk^2+\mu_k)^2}\Big\{\de_tZ_k\Big[\frac{4}{3}\frac{k^3}{l}
   +\frac{\pi^{\frac{d-1}{2}}}{\Gamma_E\Big(\frac{d+3}{2}\Big)}\frac{k^{d+1}}{l^{d-1}}\Big]
   +2Z_k\Big[2\frac{k^3}{l}
   +\frac{\pi^{\frac{d-1}{2}}}{\Gamma_E\Big(\frac{d+1}{2}\Big)}\frac{k^{d+1}}{l^{d-1}}\Big]\Big\}\\
\beta(\lambda_k)&=\frac{2\lambda_k^2}{(Z_kk^2+\mu_k)^3}
    \Big\{\de_tZ_k\Big[
\frac{\pi^{\frac{d-1}{2}}}{\Gamma_E\Big(\frac{d+3}{2}\Big)}\frac{k^{d+1}}{l^{d-1}}
+ 
\frac{4(2d-1)}{3}\frac{k^3}{l} + 2\delta_{d,3} k^2
       \Big]\crcr
&+
2Z_k\Big[\frac{\pi^{\frac{d-1}{2}}}{\Gamma_E\Big(\frac{d+1}{2}\Big)}\frac{k^{d+1}}{l^{d-1}}
+ 2(2d-1)\frac{k^3}{l}
   + 2\delta_{d,3} k^2    \Big]\Big\}
\end{aligned}
\right. \crcr
&&
\eea
It must be stressed that the coefficients appearing in \eqref{betadimensionful} are computed with integrals like in the continuous setup. This is however not an issue, once
the volume dependence has been factored out, the order of taking the limit
and performing the integral does not matter.

Some interesting features of the system \eqref{betadimensionful} must be stressed.  
At this intermediate step (the limit $\lim_{l\to 0}$ still has to be taken), this is a non-autonomous system and it involves terms of different powers in the
cut-off $k$ (we refer to this feature as ``non-homogeneity'' in $k$).
Non-autonomous systems
are known to occur in other contexts, for example quantum
field theory at finite temperature \cite{Berges:2000ew}, or on a curved \cite{Benedetti:2014gja} and non-commutative spacetime \cite{Gurau:2009ni}.  The non-homogeneity in $k$ of the system signals the presence of an external scale, for the system; here, the radius of the compactified configuration space. The specific form of the terms appearing in this case is an effect of the particular combinatorics of the vertices of 
the theory which, after differentiation, yields 1PI 2-point
function with terms  with different volume
contributions. If the $l$ parameter is kept finite, we see two different system arising in the UV and IR limits, coming 
from different leading terms. Such a feature has been found
in previous work \cite{Benedetti:2014qsa} and both the two limits
and the intermediate regime investigated. In the two limits one can compute the analogue of fixed points, which however cannot be straightforwardly interpreted as such. 

On the other hand, if one tries to proceed in the usual way, extracting the dimensions of the coupling constants using one parameter ($k$ or $l$), one obtains a
set of $\beta$-functions which are either trivial or still divergent
in the limit. Hence, in the end the
non-local combinatorics of the TGFT interactions
requires a drastic revision of conventional procedures of local
QFTs.  As we now show the correct way of proceeding in the TGFT case requires taking advantage of the presence of both the two parameters $(k,l)$, when defining the scaling of the couplings.

To make sense of the above system,  consider the following ansatz:
\beq\label{dimA}
Z_k=\ov{Z}_kl^{\chi}k^{-\chi},\quad\mu_k=\ov{\mu}_k\ov{Z}_kl^{\chi}k^{2-\chi},
   \quad\lambda_k=\ov{\lambda}_k\ov{Z}_k^2l^{\xi}k^{\sigma}\,,
\eeq
where $[\ov{Z}_k]=[\ov{\mu}_k]=[\ov{\lambda}_k]=0$, $[\varphi]=-\frac{d+2}{2}$ and $\xi+\sigma=4$.
We look for the scaling of dimensionless coupling constants, i.e. for dimensionless $\beta$-functions. 
From \eqref{dimA}, and using the convention $\eta_k = \de_t \ln \ov{Z}_k$, one finds:
\begin{eqnarray}
\eta_k & =& \frac{1}{\ov{Z}_k}\beta(\ov{Z}_k)=\frac{1}{Z_k}\beta(Z_k)+\chi\,,\crcr
 \beta(\ov{\mu}_k) &=&\frac{1}{\ov{Z}_kl^{\chi}k^{2-\chi}}\beta(\mu_k)-\eta_k\ov{\mu}_k-(2-\chi)\ov{\mu}_k\,,\crcr
\beta(\ov{\lambda}_k)&=&\frac{1}{l^{\xi}k^{\sigma}\ov{Z}_k^2}\beta(\lambda_k)-2\eta_k\ov{\lambda}_k
   -\sigma\ov{\lambda}_k\,,
\end{eqnarray}
and inserts this in \eqref{betadimensionful} to reach the following expressions: 
\bea\label{dimlessbetanongauge}
&&\hspace{-0.4cm}
\left\{
\begin{aligned}
\eta_k  &=
\frac{\ov{\lambda}_kl^{\xi}k^{\sigma}}{l^{2\chi}k^{2(2-\chi)} (1+\ov{\mu}_k)^2}
\Big\{(\eta_k -\chi)\Big[
   \frac{\pi^{\frac{d-1}{2}}}{\Gamma_E\Big(\frac{d+1}{2}\Big)}\frac{k^{d-1}}{l^{d-1}} + 2(d-1)\frac{k}{l}\Big]
   +2\Big[(d-1)\frac{k}{l}
   +\frac{\pi^{\frac{d-1}{2}}}{\Gamma_E\Big(\frac{d-1}{2}\Big)}\frac{k^{d-1}}{l^{d-1}}\Big]\Big\} + \chi 
\cr\cr
\beta(\ov{\mu}_k)&=
-\frac{d\,\ov{\lambda}_kl^{\xi}k^{\sigma}}{l^{2\chi}k^{6-2\chi}(1+\ov{\mu}_k)^2}
\Big\{(\eta - \chi)\Big[
\frac{\pi^{\frac{d-1}{2}}}{\Gamma_E\Big(\frac{d+3}{2}\Big)}\frac{k^{d+1}}{l^{d-1}} + \frac{4}{3}\frac{k^3}{l}\Big]
   +2\Big[2\frac{k^3}{l}
   +\frac{\pi^{\frac{d-1}{2}}}{\Gamma_E\Big(\frac{d+1}{2}\Big)}\frac{k^{d+1}}{l^{d-1}}\Big]\Big\} -\eta_k\ov{\mu}_k-(2-\chi)\ov{\mu}_k \cr\cr 
\beta(\ov{\lambda}_k)&=
 \frac{2\ov{\lambda}_k^2 l^{\xi}k^{\sigma}}{l^{2\chi}k^{6-2\chi}(1+\ov{\mu}_k)^3}
  \Big\{ (\eta - \chi)\Big[
\frac{\pi^{\frac{d-1}{2}}}{\Gamma_E\Big(\frac{d+3}{2}\Big)}\frac{k^{d+1}}{l^{d-1}}
+ 
\frac{4(2d-1)}{3}\frac{k^3}{l} + 2\delta_{d,3} k^2
       \Big]\crcr
&+
2\Big[\frac{\pi^{\frac{d-1}{2}}}{\Gamma_E\Big(\frac{d+1}{2}\Big)}\frac{k^{d+1}}{l^{d-1}}
+ 2(2d-1)\frac{k^3}{l}
   + 2\delta_{d,3} k^2    \Big]  \Big\}
-2\eta_k\ov{\lambda}_k-\sigma\ov{\lambda}_k 
\end{aligned}
\right. \crcr
&&
\eea
In order to make the non-compact limit regular,  we must solve the system in the variables $\xi$
and $\chi$ by requiring that the highest degree of divergence (highest negative power of $l$) is regularised and all the
sub-leading infinities sent to zero. 
This is achieved by solving, for any $d\geq 3$,  
\beq
\xi-2\chi-(d-1)=0\,. 
\eeq
We make a natural choice $\chi=0$ (thus implying that $Z_k$ is dimensionless),
and obtain 
\bea
(\chi=0,\; \xi=d-1) \qquad\Rightarrow\qquad
\sigma=5-d\,.
\eea
The resulting system of equations for the theory is:
\bea\label{dimessd}
&&\hspace{-0.4cm}
\left\{
\begin{aligned}
\eta_k  &=
 \frac{2\pi^{\frac{d-1}{2}}}{\Gamma_E\Big(\frac{d-1}{2}\Big)} \frac{\ov{\lambda}_k}{ (1+\ov{\mu}_k)^2}
\Big[\frac{\eta_k}{d-1} +1\Big]
\crcr
\beta(\ov{\mu}_k)&=
\frac{-2d\,\pi^{\frac{d-1}{2}}}{\Gamma_E\Big(\frac{d+1}{2}\Big)}
\frac{\ov{\lambda}_k }{(1+\ov{\mu}_k)^2}
\Big[ \frac{\eta_k}{d+1} +1\Big] -\eta_k\ov{\mu}_k-2\ov{\mu}_k 
\crcr 
\beta(\ov{\lambda}_k)&
=
 \frac{4\pi^{\frac{d-1}{2}}}{\Gamma_E\Big(\frac{d+1}{2}\Big)} \frac{\ov{\lambda}_k^2  }{ (1+\ov{\mu}_k)^3 }
 \Big[\frac{\eta_k}{d+1} + 1 \Big]
-2\eta_k\ov{\lambda}_k-(5-d)\ov{\lambda}_k 
\end{aligned}
\right. \crcr
&&
\eea 
which defines an autonomous system of coupled differential equations describing the flow of dimensionless couplings constants.

These equations hold for generic rank $d$. They could be solved at the same level of generality, in principle, but we find more useful to specialise the analysis for various interesting choices of rank, so that the results can be reported in more explicit terms.
Specifically, we study the above  system of equations when restricted to the first non-trivial 
rank situations at $d=3,4, 5$. We will analyse the rank $d=3$ in all details,
and, will simply report the key results 
in higher ranks $d=4,5$. 

\subsection{Rank $d=3$}\label{rank3}
At rank $d=3$, the system \eqref{dimessd} reduces to 
\begin{equation}
\label{betaad}
\left\{
\begin{aligned}
\eta_k=&\frac{\pi\ov{\lambda}_k}{(1+\ov{\mu}_k)^2}(\eta_k+2)\\
\beta(\ov{\mu}_k)=&-\frac{3\pi\ov{\lambda}_k}{(1+\ov{\mu}_k)^2}(\frac{\eta_k}{2}+2)
     -\eta_k\ov{\mu}_k-2\ov{\mu}_k\\
\beta(\ov{\lambda}_k)=&\frac{\pi\ov{\lambda}_k^2}{(1+\ov{\mu}_k)^3}(\eta_k+4)
     -2\eta_k\ov{\lambda}_k-2\ov{\lambda}_k
\end{aligned}
\right.
\end{equation}

Before proceeding with the standard analysis, which consists in finding fixed points of the flow and studying the linearised
equations around them, we point out that, because of the non-linear nature of the $\beta$-functions, we have a singularity  at 
$\ov{\mu}=-1$ and $\ov{\lambda}=(1+\ov{\mu})^2/\pi$. This is a common feature in dealing with a truncated Wetterich
equation.
In a neighbourhood of those singularities, we do not trust the linear approximation, and being interested in the part of the RG flow
connected with the Gaussian fixed point, we will not study the flow beyond the mentioned divergence of the $\beta$-functions.

By numerical integration, we find a Gaussian fixed point and three non-Gaussian fixed points in the plane
$(\ov{\mu},\ov{\lambda})$ at:
\beq
 _{d=3}P_1=(8.619,-47.049)\,,     \quad \;   _3P_2=10^{-1}(-6.518,0.096)\,,     \quad \;   _3P_3=10^{-1}(-8.010,0.212)\,.
\eeq
A quick inspection proves that $_3P_3$ lies in the sector disconnected from the origin, so we will not perform any analysis around it. 

We linearise the system of equations by evaluating the stability matrix around the other three fixed points:
\begin{equation}
 \left(
\begin{aligned}
 &\beta(\ov{\mu}_k)\\
&\beta(\ov{\lambda}_k)
\end{aligned}
\right)
=
\left(
\begin{aligned}
 &\de_{\ov{\mu}_k}\beta(\ov{\mu}_k)  &\de_{\ov{\lambda}_k}\beta(\ov{\mu}_k)\\
&\de_{\ov{\mu}_k}\beta(\ov{\lambda}_k) &\de_{\ov{\lambda}_k}\beta(\ov{\lambda}_k)
\end{aligned}
\right)_{\text{F.P.}}
\left(
\begin{aligned}
&\ov{\mu}_k\\
&\ov{\lambda}_k 
\end{aligned}
\right)\,.
\end{equation}
In a neighbourhood of the Gaussian fixed point, the stability 
matrix is of the form 
\beq
(\beta^*_{ij})\Big|_{GFP}:= \left(
\begin{array}{cc}
 -2 & -6 \pi  \\
 0 & -2 \\
\end{array}
\right)
\eeq 
which has an eigenvalue with algebraic multiplicity 2, corresponding to the
canonical scaling dimensions of the couplings $\lambda_k$ and $\mu_k$: $_{d=3}\theta_0=-2$.
The geometric multiplicity of $_{3}\theta_0$ is 1, hence, the matrix of the linearised system turns out to be not diagonalisable
and has a single eigenvector $_3\textbf{v}_{0}=(1,0)$.

\begin{figure}[h]
 \centering
\includegraphics[scale=0.238]{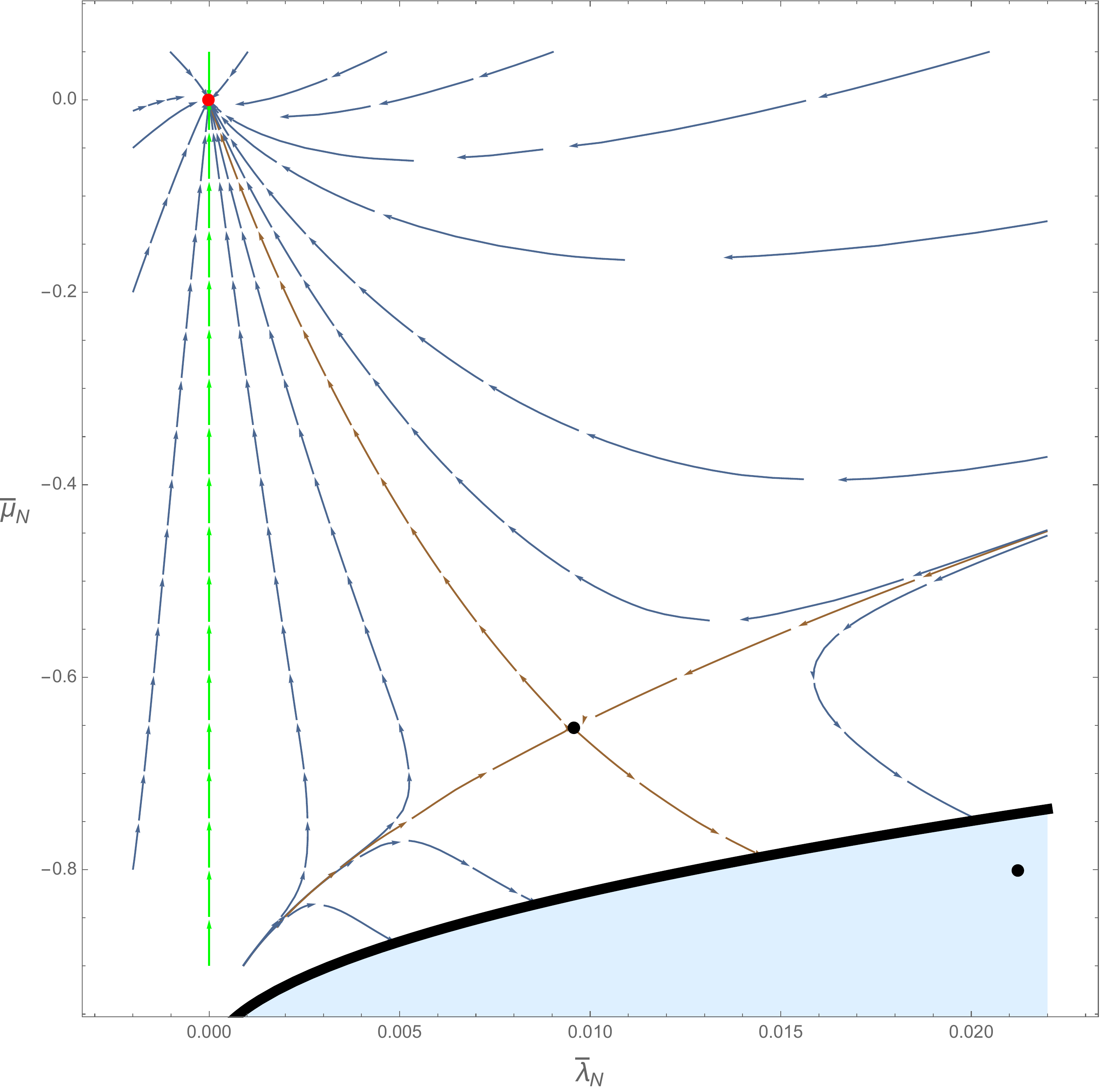}
\qquad 
\includegraphics[scale=0.25]{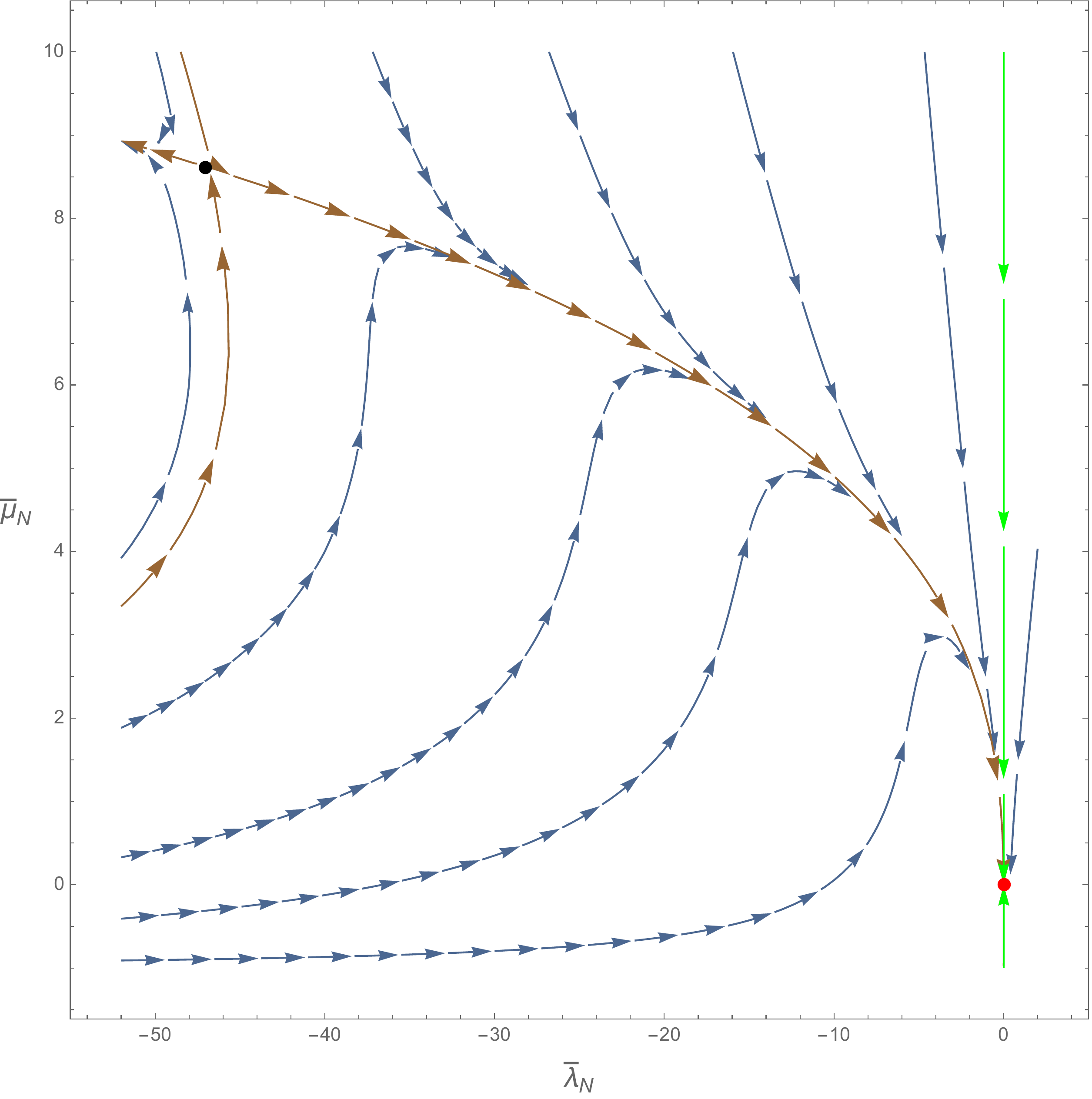}
\caption{Flow of the theory. The red and blue lines represent respectively the zeros of $\beta(\ov{\mu}_k)$ and
$\beta(\ov{\lambda}_k)$, the brown arrows are the eigenperturbations of the non-Gaussian fixed points (represented in black),
and the green ones those of the Gaussian fixed point (in red). Arrows point in the UV direction. The thick black line is the
singularity of the flow.}
\label{plots}
\end{figure}

In a neighbourhood of the non-Gaussian fixed points (NGFP)  we have:
\begin{align}
_3P_{1} \quad  &_3\theta_{11}\sim0.351\; \text{ for }\; _3\textbf{v}_{11}\sim10^{-1}(0.65,-9.98),\\
_3P_{1} \quad   &_3\theta_{12}\sim-2.548\; \text{ for }\; _3\textbf{v}_{12}\sim10^{-1}(-6.88,7.26),\\
_3P_{2} \quad &_3\theta_{21}\sim10.066\; \text{ for }\; _3\textbf{v}_{21}\sim10^{-1}(9.996,-0.269),\\
_3P_{2} \quad   &_3\theta_{22}\sim-1.988 \text{ for }\; _3\textbf{v}_{22}\sim10^{-1}(9.987,0.506).
\end{align}
Because of the difference in their magnitudes (distance from the origin), it becomes difficult to plot the two NGFP's simultaneously with enough precision in their vicinity. We plot
two sectors of the RG flow in the plane $(\ov{\mu}_k, \ov{\lambda}_k)$
(see  Fig.\ref{plots}).

In the vicinity of a fixed point, we define as relevant directions those eigendirections that are UV attractive with respect to
the cut-off, while we call irrelevant the UV repulsive eigendirections. Marginal directions can be attractive or
repulsive depending on the initial condition of the trajectory.
The origin is a {\it great} attractor and has one relevant direction connecting it to the other two fixed points.
The absence of a second eigenvector for the stability matrix around the Gaussian fixed point requires an approximation
beyond the linear order, when the flow is studied analytically, and is a signal of the presence of a marginal perturbation. We can instead integrate numerically the flow, and we find that this marginal direction will still be UV attractive, which means that it corresponds to a marginally relevant direction.
The fact that the GFP is a sink for the flow, means that this model is asymptotically free with respect to the cut-off.
Both non-Gaussian fixed points have one relevant  and one irrelevant directions.
The eigendirections connecting the three fixed
points turn out to be stable under RG transformations and they are characterised by an effect known as
{\it large river effect} \cite{Delamotte-review}.  
This signifies that all the RG trajectories in a neighbourhood of these eigendirections get closer and closer to them while
pointing in the UV. This effect shows a splitting of the space of coupling in two regions not connected by any RG 
trajectory. Thus, the relevant directions for the Gaussian fixed point reflect the properties of a critical surface and suggest
the presence of phase transitions in the model.
In the $\ov{\lambda}_k>0$ plane, the flow is similar to the one of standard local scalar field theory in a neighbourhood of the
Wilson-Fisher fixed point,
but the presence of a second non-Gaussian fixed point in the $\ov{\lambda}_k<0$ plane makes the theory quite different.
Nevertheless, the properties of this second NGFP are basically the same
as the former one. 

In this context, therefore,
we do have strong hint of a phase transition with 
two phases: a symmetric and a broken one. The spontaneous symmetry breaking would happen while crossing the critical
surface, generating a condensed state of the TGFT field (non-zero expectation value of the field operator). This is interesting from a physical perspective, because, in more involved models defined in a simplicial gravity or LQG context, this kind of phase transition has been suggested to relate to the emergence of a geometric spacetime from the theory \cite{danieleemergence}, and the corresponding condensate states have been shown to admit a cosmological interpretation \cite{GFTcondensate}.
To confirm this condensate interpretation of the broken phase, one should change parametrisation for the effective potential and study
the theory around the new (degenerate) ground state solving the equation of motion in the saddle point approximation. This (complicated)
analysis of our TGFT model is left for future work. Here we only notice that, in the constant modes approximation, which forgets about the peculiar combinatorial
non-locality of our interactions, and whose results should therefore be taken with great care, we find and algebraic equation of Ginsburg-Landau type for a $\phi^4$ scalar complex theory, which indeed describe this type of condensate phase transitions. 

\subsection{Rank $d=4,5$}
We now give a streamlined analysis of the flow in the case of rank $d=4$,
which is very similar to the case $d=3$, and the rank $d=5$
which share similarities but also a few differences that we will 
list. 

Writing the system in rank $d=4$ as 
\bea\label{dimessd4}
&&\hspace{-0.4cm}
\left\{
\begin{aligned}
\eta_k  &=
 \frac{4\pi}{3} \frac{\ov{\lambda}_k}{ (1+\ov{\mu}_k)^2}
(\eta_k +2)
\crcr
\beta(\ov{\mu}_k)&=
\frac{-32\pi }{3}
\frac{\ov{\lambda}_k }{(1+\ov{\mu}_k)^2}
\Big[ \frac{\eta_k}{5} +1\Big] -\eta_k\ov{\mu}_k-2\ov{\mu}_k 
\crcr 
\beta(\ov{\lambda}_k)&
=
 \frac{16\pi}{3} \frac{\ov{\lambda}_k^2  }{ (1+\ov{\mu}_k)^3 }
 \Big[\frac{\eta_k}{5} + 1 \Big]
-2\eta_k\ov{\lambda}_k-\ov{\lambda}_k 
\end{aligned}
\right. \crcr
&&
\eea
we find, in addition to the Gaussian fixed-point, 
 the following NGFPs: 
\beq
  _4P_1 =10^{-1}(-6.402,0.058)\,, \quad 
_4P_2 =(1.612,-0.496482)\,,    \quad 
 _4P_3=10^{-1}(-8.452,0.112)\,.
\eeq
As in the case $d=3$, the fixed point $_4P_3$ lies beyond the singularity. 
The eigenvalues and eigenvectors in the vicinity of the GFP and of
$_4P_1$
 and $_4P_2$ are given in the following table
\begin{align}
\text{ GFP}_4\quad  & _4\theta_0^+ = -2  \;
\text{ for }\;  _4\textbf{v}_0^+ = (1,0) \\ 
\text{ GFP}_4\quad  & _4\theta_0^- = -1  \; 
\text{ for }\;  _4\textbf{v}_0^- = (-\frac{32\pi}{3},1) \\ 
_4P_1\quad   &_4\theta_{11}\sim 7.899\;  \text{ for } \; _4\textbf{v}_{11}\sim  10^{-1}(10,-0.106) \\
_4P_1  \quad   &_4\theta_{12}\sim -1.570\;  \text{ for } \;  _4\textbf{v}_{12}\sim  10^{-1}(10,0.279),
\\ 
_4P_2\quad   &_4\theta_{21}\sim -3.082\;  \text{ for } \; _4\textbf{v}_{21}\sim  10^{-1}(-10,0.521) \\
_4P_2  \quad   &_4\theta_{22}\sim 0.439\;  \text{ for } \;  _4\textbf{v}_{22}\sim  10^{-1}(8.193,-5.733). 
\end{align}
Negative eigenvalues at the vicinity of the GFP shows
that its eigendirections are all relevant. The NGFPs
have a relevant and an irrelevant direction. 

\begin{figure}[h]
 \centering
\includegraphics[scale=0.3]{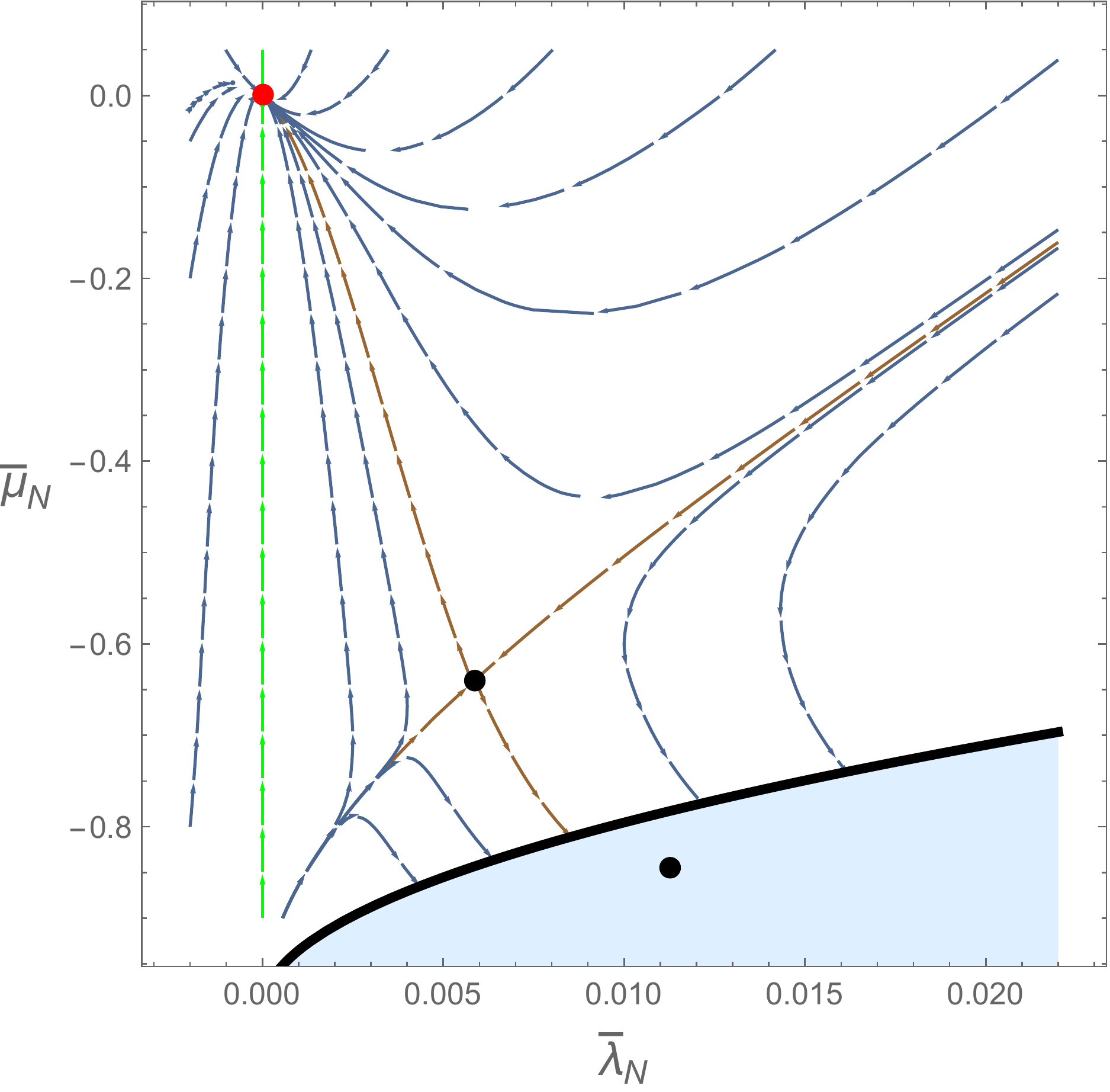}
\qquad 
\includegraphics[scale=0.3]{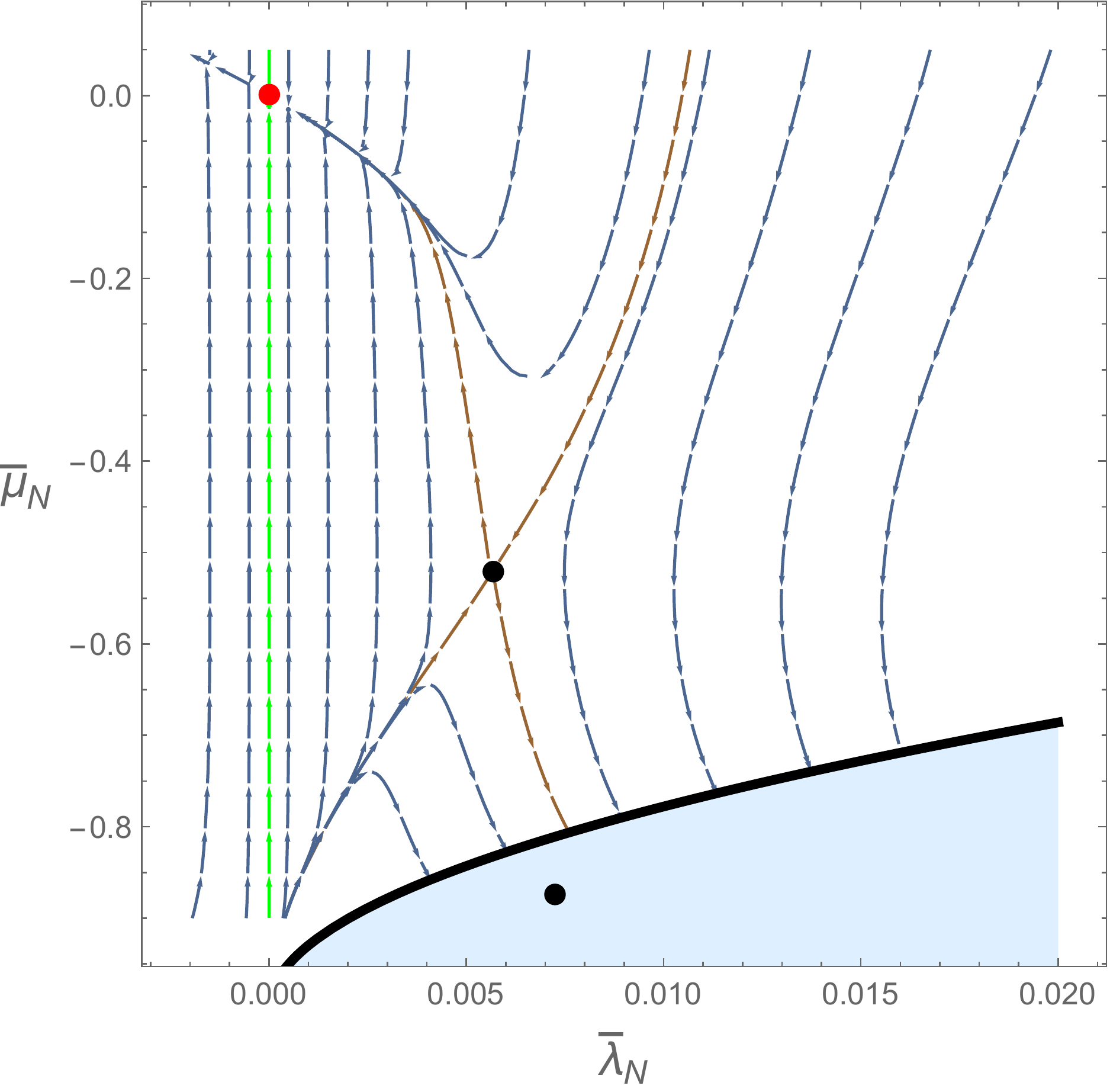}
\caption{Flow at rank $d=4$ (left) and
$5$ (right).}
\label{R4plot}
\end{figure} 

In rank $d=5$, on the other hand, the system \eqref{dimessd} specialises
as 
\bea\label{dimessd5}
&&\hspace{-0.4cm}
\left\{
\begin{aligned}
\eta_k  &=
 \frac{\pi^{2}}{2} \frac{\ov{\lambda}_k}{ (1+\ov{\mu}_k)^2}
(\eta_k +2)
\crcr
\beta(\ov{\mu}_k)&=
-5\pi^{2}
\frac{\ov{\lambda}_k }{(1+\ov{\mu}_k)^2}
\Big[ \frac{\eta_k}{6} +1\Big] -\eta_k\ov{\mu}_k-2\ov{\mu}_k 
\crcr 
\beta(\ov{\lambda}_k)&
=
2\pi^{2} \frac{\ov{\lambda}_k^2  }{ (1+\ov{\mu}_k)^3 }
 \Big[\frac{\eta_k}{6} + 1 \Big]
-2\eta_k\ov{\lambda}_k  
\end{aligned}
\right. \crcr
&&
\eea
Here, along with the GFP, we identify two NGFPs as 
\bea
  _5P_1 =\Big(\frac{-23 + \sqrt{34}}{33}, \frac{4 (191 - 4 \sqrt{34})}{
 11979\pi^2}\Big)= 10^{-1}(-5.202,0.056)\,, \qquad
 _5P_2=10^{-1}(-8.736,0.072)\,.
\eea
Again, one of them, $_5P_2$, is beyond the singularity so we will skip 
its analysis.  
We list eigenvalues and eigenvectors in the vicinity of the GFP and $_5P_1$
as follows:  
\begin{align}
\text{ GFP}_5\quad  & _5\theta_0^+ = -2  \;
\text{ for }\;  _5\textbf{v}_0^+ = (1,0) \\ 
\text{ GFP}_5\quad  & _5\theta_0^- = 0  \; 
\text{ for }\;  _5\textbf{v}_0^- = (-\frac{5\pi^2}{2},1) \\ 
_5P_1\quad   &_5\theta_1\sim 2.947\;  \text{ for } \; _5\textbf{v}_1\sim  (-249.652, 1) \\
_5P_1  \quad   &_5\theta_2\sim -0.843\;  \text{ for } \;  _5\textbf{v}_2\sim  (66.431, 1).
\end{align}
The GFP has one relevant eigendirection (corresponding to 
the mass coupling) and a marginal one. The numerical integration of 
the flow shows that this direction is marginally relevant for positive $\lambda$ but become irrelevant for a negative $\lambda$. 
Similar to the previous case, the NGFP $_5P_1$ has one relevant
and one irrelevant eigendirections.

The models at rank 4 and 5 are very similar to the previous rank 3 case. Hence, similar conclusions concerning the
analysis of their flow hold, in particular the separation of the space of couplings in regions which are not connected by 
any RG trajectories which again suggests of phase transition. 
The numerical flow of the rank 4 and 5 models have been 
given in Figure \ref{R4plot}. Note that we did not display 
the NGFP $_4P_2$ of the rank 4 model which should be 
similar to the second fixed point of the rank 3.

\section{Gauge invariant Rank-$d$ Tensorial Group Field Theory on \R}
\label{chap:Gauge}

We now proceed to analyse a modified version of the TGFT models studied in the previous section, in which an additional gauge invariance condition is included in the model. 
These models define topological lattice gauge theories of BF type for the gauge group $G$ at the level of their Feynman amplitudes.
The first model of this type has been studied in \cite{Benedetti:2015yaa} in rank-6 for the group $G=U(1)$. 
 We therefore extend these first results by working with a non-compact group manifold, albeit still abelian, keeping the rank arbitrary. As for the previous model, we first introduce the gauge-invariant model, then proceed with the FRG analysis in the general case, and finally specialise to interesting choices of rank to show explicitly the results of our analysis.

\subsection{The gauge projection}
\label{5.1gaugeproj}

We work with rank-$d$ fields over the group manifold $G$ satisfying the gauge invariance condition 
\beq
\label{gaugecondition}
\phi(g_1,g_2,\dots, g_d)=\phi(g_1h,g_2h, \dots, g_dh)\,, \quad \forall h \in G\,.
\eeq  
This invariance condition can be imposed directly at the level of the space of fields or as a condition on the dynamics, which then restricts indirectly the field degrees of freedom. In both cases, this translates into a modification of the action \eqref{actiondirectspace}. This modification can take different forms and should be implemented with some care. A possible (formal) way to implement it would be to allow only the propagation of modes satisfying \eqref{gaugecondition} by inserting
in the kinetic kernel a projector on the space of these modes.
Defining the projector $P$ and a kinetic kernel $\mathcal{k}$, one may encounters some ambiguity.
A proper inspection shows that in our case, where the kinetic term has the form of a Laplacian plus constant, $\mathcal{K}$ and $P$ commute.
We choose to implement
the kinetic term of the action in the following form:
\bea
S[\phi,\ov{\phi}]=\int_{G^{\times d}} [dg_i]_{i=1}^d[dg'_i]_{i=1}^d\;
   \ov{\phi}(g_1,g_2, \dots, g_d)(P\cdot\ca{K})(\{g_i\}_{i=1}^d;\{g'_i\}_{i=1}^d)\phi(g'_1,g'_2,\dots, g'_d)
+\ca{V}[\phi,\ov{\phi}]\,,
\eea
where $\ca{V}$ is the interaction term.
The main issue of this formulation is that a projector is by definition not invertible, thus, a kinetic kernel built out of such an operator cannot, in general, define a covariance of a field theory measure.
We partially avoid this problem by inverting the kinetic kernel in the operatorial sense, in such a way that the same constraint will
define the covariance itself. In other words, also the propagator is defined as $P \cdot \ca{K}^{-1}$.

Now we restrict our description to the case of the Abelian additive group $\R$ and consider $\ca{V}$ with same combinatorics
used in section \ref{sect3:TGFT}
\bea
\label{gaugeinpropaNC}
S_1[\phi,\ov{\phi}]
&=&(2\pi)^{d-1}\int_{\R^{\times (2d+1)}} d\textbf{x}d\textbf{y}dh\;
   \ov{\phi}(\textbf{x})\prod_{i=1}^d\delta(x_i+h-y_i)(-\sum_{s=1}^d\Delta_{y_s}+\mu)\phi(\textbf{y})\crcr
&+&\frac{\lambda}{2}(2\pi)^{2d} \int_{\R^{\times 2d}}d\textbf{x}d\textbf{x}'\biggl[
   \phi(x_1,x_2,\dots,x_d)\ov{\phi}(x'_1,x_2,\dots,x_d)\phi(x'_1,x'_2,\dots,x'_d)\ov{\phi}(x_1,x_2',\dots,x'_d)\crcr
&+&\Sym\biggr]\,,
\eea
where $\textbf{x}=(x_i),\textbf{x}'=(x_i')$ and $\textbf{y}=(y_i)$ are vectors in $\R^d$, and $h \in \R$. 

We expect that the Wetterich equation will exhibit IR divergences of the same type encountered in the non-projected
model, although the gauge invariance conditions relate in a non-trivial way the arguments of the fields entering the interactions, and therefore modifies the combinatorics of the same; as a result, we expect a different degree of IR divergences with respect to the case we have treated in the previous section. In any case, we introduce again a regularisation scheme. 
We consider a compact subset $D$ of $\R$ homeomorphic to $S^1$ and write a regularised action as: 
\bea
\label{gaugeinpropa}
S_1[\phi,\ov{\phi}]
&=&(2\pi)^{d-1} \int_{D^{\times (2d+1)}} d\textbf{x}d\textbf{y}dh\;
   \ov{\phi}(\textbf{x})\prod_{i=1}^d\delta(x_i+h-y_i)(-\sum_{s=1}^d\Delta_{y_s}+\mu)\phi(\textbf{y})\crcr
&+&\frac{\lambda}{2}(2\pi)^{2d} \int_{D^{\times 2d}}d\textbf{x}d\textbf{x}'
\Big[ 
   \phi(x_1,x_2,\dots,x_d)\ov{\phi}(x'_1,x_2,\dots,x_d)\phi(x'_1,x'_2,\dots,x'_d)\ov{\phi}(x_1,x_2',\dots,x'_d)\crcr
&+&\Sym\Big] \,,
\eea
where we used the same notations introduced in section \ref{IRthermo}.

The computation will be performed in momentum space.
Using again the same notation for the lattice as
$D^*=\ca{D}^{\times d}$, and denoting the gauge invariance constraint on the corresponding lattice as $\delta_{\ca{D}}(X):=\delta_{\ca{D}}(X,0)$,
the Fourier series of the model \eqref{gaugeinpropa}  reads:
\beq
S_1[\phi,\ov{\phi}]
=l^d\sum_{\textbf{p}\in D^*}\ov{\phi}(\textbf{p})\Big[\Sigma_sp_s^2+\mu\Big]\phi(\textbf{p})
      \delta_{\ca{D}}(\Sigma p)+
      \frac{\lambda}{2}\, l^{2d}\sum_{\textbf{p},\textbf{p}'\in D^*}\Big[
 \phi_{12\dots d}\ov{\phi}_{1'2\dots d}\phi_{1'2' \dots d'}\ov{\phi}_{12'\dots d'}
+\Sym\Big].
\eeq
The general FRG formalism introduced in section \ref{FRGfotgft} applies to this model as to the one in the previous section. In particular, the regulator kernel will incorporate the same gauge constraint appearing in the
kinetic term. The Wetterich equation has the same structure as well and expands again as \eqref{wettexpansionPHI4}.

We choose to truncate the effective action as:
\bea
\Gamma_k^1[\varphi,\ov{\varphi}]
&=&l^d\sum_{\textbf{p}\in D^*}\ov{\varphi}(\textbf{p})\Big[Z_k\Sigma_sp_s^2+\mu_k\Big]\varphi(\textbf{p})
      \delta_{\ca{D}}(\Sigma p)\crcr
&+&\frac{\lambda_k}{2}l^{2d}\sum_{\textbf{p},\textbf{p}'\in D^*}\biggl[
      \varphi_{12\dots d}\ov{\varphi}_{1'2\dots d}\varphi_{1'2' \dots d'}\ov{\varphi}_{12' \dots d'}+\Sym\biggr]\,,
\eea
and, then, we introduce the kernels (using the same notation as \eqref{wettexpansionPHI4}):
\bea
\label{RGaugenongood}
R_k(\textbf{q},\textbf{q'})&=&\Theta(k^2-\Sigma_sq_s^2)Z_k(k^2-\Sigma_sq_s^2)
    \delta_{\ca{D}}(\Sigma q)
    \prod\delta_{\ca{D}}(\textbf{q},\textbf{q'})\,,\\
\label{FGaugenongood}
 F^{1}_k(\textbf{q},\textbf{q}')&=&\frac{\delta^2}{\delta\ov{\varphi}_{\textbf{q}'}\delta\varphi_{\textbf{q}}}
   \ca{V}^1[\varphi,\ov{\varphi}]\,,
\eea
where $\cV^1_k$ refers to the interaction part of $\Gamma^1_k$.
This is a natural choice following directly from a straightforward FRG formulation of \eqref{gaugeinpropa}. Performing the
computation of the Wetterich equation, however, one realises that  
this proposal drastically fails: the delta's 
enforcing the gauge constraints do not convolute properly with the TGFT fields.
 This is due to the fact that, if one evaluates \eqref{wettexpansionPHI4} using
\eqref{RGaugenongood} and \eqref{FGaugenongood}, the fields appearing in the r.h.s. come from the $F_k^1$ operator, while
the constraints always come from the mass-like terms. The comparison of the two sides of the Wetterich equation for
this model, then would lead to all $\beta$-functions being trivial.

A moment of reflection shows that another way of choosing the interaction term produces a more sensible result.  We simply insert gauge
projections also in all fields in the interaction.
An interaction satisfying this requirement expresses as:
\begin{align}
\label{vertexprojected}
\ca{V}[\phi,\ov{\phi}]=&\frac{\lambda_k}{2}
(2\pi)^{2d-4} 
\int_{D^{\times (6d+4)}} \{d\textbf{w}^i\}_{i=1}^4 d\textbf{x}d\textbf{x}'\{dh_j\}_{j=1}^4
      \phi(\textbf{w}^1)\ov{\phi}(\textbf{w}^2)\phi(\textbf{w}^3)\ov{\phi}(\textbf{w}^4)\crcr
&\times\delta(x_1+h_1-w_1^1)\delta(x_2+h_1-w_2^1)
\dots \delta(x_d+h_1-w_d^1)\crcr
&\times\delta(x'_1+h_2-w_1^2)\delta(x_2+h_2-w_2^2)
\dots \delta(x_d+h_2-w_d^2)
\crcr
&\times\delta(x'_1+h_3-w_1^3)\delta(x'_2+h_3-w_2^3)
\dots \delta(x'_d+h_3-w_d^3)\crcr
&\times\delta(x_1+h_4-w_1^4)\delta(x'_2+h_4-w_2^4)
\dots \delta(x'_d+h_4-w_d^4)\crcr
&+\Sym\crcr
=&\frac{\lambda_k}{2}l^{2d}\sum_{\textbf{p},\textbf{p}'}
    \phi_{12\dots d}\ov{\phi}_{1'2\dots d}\phi_{1'2'\dots d'}\ov{\phi}_{12' \dots d'}\delta_{\ca{D}}(\Sigma p)
    \delta_{\ca{D}}(\Sigma p')\crcr
&\times\delta_{\ca{D}}(p'_1+p_2+\dots +p_d)\delta_{\ca{D}}(p_1+p'_2+\dots +p'_d)+\Sym\,.
\end{align}
Hence, re-starting the analysis from the beginning, we define a model with gauge constraints on both the kinetic and interaction kernels via the
action: 
\bea
\label{gaugemodeldiscrete}
&&
S[\phi,\ov{\phi}]
=
l^{2d}\sum_{\textbf{p}}\ov{\phi}(\textbf{p})\Big[\Sigma_sp_s^2+\mu\Big]\phi(\textbf{p})
      \delta_{\ca{D}}(\Sigma p)\crcr
&&+\frac{\lambda_k}{2}l^{2d}\sum_{\textbf{p},\textbf{p}'}
    \phi_{12\dots d}\ov{\phi}_{1'2\dots d}\phi_{1'2'\dots d'}\ov{\phi}_{12' \dots d'}\delta_{\ca{D}}(\Sigma p)
    \delta_{\ca{D}}(\Sigma p')
\delta_{\ca{D}}(p'_1+p_2+\dots +p_d)\delta_{\ca{D}}(p_1+p'_2+\dots+p'_d)
\crcr
&&+  
\Sym\,,
\eea
with corresponding continuous model defined by
\bea
\label{gaugemodelcontinuous}
&&
S[\phi,\ov{\phi}]=\int d\textbf{p}\;\ov{\phi}(\textbf{p})\Big[\Sigma_sp_s^2+\mu\Big]\phi(\textbf{p})
      \delta(\Sigma p)\crcr
&&+\frac{\lambda_k}{2}\int d\textbf{p}d\textbf{p}'\;
    \phi_{12\dots d}\ov{\phi}_{1'2\dots d}\phi_{1'2'\dots d'}\ov{\phi}_{12' \dots d'}\delta_{\ca{D}}(\Sigma p)
    \delta_{\ca{D}}(\Sigma p')
\delta_{\ca{D}}(p'_1+p_2+\dots +p_d)\delta_{\ca{D}}(p_1+p'_2+\dots+p'_d)
\crcr 
&&
+\Sym\,.
\eea
In fact, with hindsight, one realises that this result could  have been guessed from a more general consideration. Even if the perturbative quantum amplitudes of the theory do not depend on whether the gauge projection appears in the kinetic term, in the interaction or in both, and only gauge invariant degrees of freedom have non trivial Feynman amplitudes (spin foam models) the non-perturbative analysis is of course radically different. From a non-perturbative point of view one is suggested to simply project the model to the space of gauge invariant fields, and thus insert projections in all elements of the TGFT action. From this point of view, a model which presents this constraint in only one of the two terms  cannot be consistent. This directly reflects in the analysis we just presented.

At the same time, notice that inserting gauge projections on all fields in the action, both in kinetic and interaction terms, results in a  trivial overall divergence equal to the volume of the domain, due to the fact that the combinatorics of field pairings is such that imposing gauge invariance on all but one field in each polynomial automatically implies the gauge invariance of the last one. We can easily remove this trivial divergence, therefore, by removing one gauge projection from one of the fields in each polynomial term in the action.
 The above prescription of the effective action together with \eqref{WettEqModel}, coincides with the Wetterich equation 
as formulated in \cite{Benedetti:2015yaa} (albeit the formalism differs by the nature of the field background).

We can now proceed further using the model \eqref{gaugemodelcontinuous}.

\subsection{Effective action and Wetterich equation}
\label{sec5.2}

Having defined the main ingredients of the model, 
we are  in position to analyse its FRG equation. 
We shall again restrict to a simple truncation of the effective action for the model \eqref{gaugemodeldiscrete},
which reads: 
\bea
\label{GammaGauge}
\Gamma_k[\varphi,\ov{\varphi}]&=&l^d\sum_{\textbf{p}}\ov{\varphi}(\textbf{p})\Big[Z_k\Sigma_sp_s^2+\mu_k\Big]\varphi(\textbf{p})
      \delta_{\ca{D}}(\Sigma p)\crcr
&+&\frac{\lambda_k}{2}l^{2d}\sum_{\textbf{p},\textbf{p}'}
    \varphi_{12\dots d}\,\ov{\varphi}_{1'2\dots d}\,\varphi_{1'2' \dots d'}\,\ov{\varphi}_{12'\dots d'}\,\delta_{\ca{D}}(\Sigma p)
    \delta_{\ca{D}}(\Sigma p')\crcr
&\times &\delta_{\ca{D}}(p'_1+p_2+\dots+p_d)\delta_{\ca{D}}(p_1+p'_2+\dots+p'_d)+\Sym\,.
\eea
Considering that $[\delta_{\ca{D}}(p)]=-1$, the dimensional analysis for the coupling constants gives different results from the model of section \ref{sect3:TGFT}. We have:
\bea
[Z_k]=0\;&\Rightarrow &\;[\mu_k]=2\crcr
2[\varphi]+d+2-1=0\;&\Rightarrow &\;[\varphi]=-\frac{d+1}{2}\crcr
[\lambda_k]+2d+4[\varphi]-4=0\;&\Rightarrow &\;[\lambda_k]=6\,,
\eea
where, again, we set the canonical dimensions by requiring $[S]=[\Gamma_k]=0$ and $[dp]=1$.

We introduce:
\bea
R_k(\textbf{q},\textbf{q'})&=&
\Theta(k^2-\Sigma_sq_s^2)Z_k(k^2-\Sigma_sq_s^2)\delta_{\ca{D}}(\Sigma q)
    \delta_{D^*}(\textbf{q},\textbf{q}')\,,\label{Rgaug}\\\cr
\de_tR_k(\textbf{q},\textbf{q'})&=&\Theta(k^2-\Sigma_s q_s^2)[\de_t Z_k(k^2-\Sigma_sq_s^2)+2k^2Z_k]
    \delta_{\ca{D}}(\Sigma q)\delta_{D^*}(\textbf{q},\textbf{q}')\,,\label{deRgaug}\crcr
&& \\ 
F_k(\textbf{q},\textbf{q}')&=&\lambda_k\biggl[l^{d-1}\sum_{m_i}
    \varphi_{q'_1 m_2\dots m_d}\ov{\varphi}_{q_1m_2\dots m_d}
\delta_{\ca{D}}(\Sigma q)
    \delta_{\ca{D}}(q'_1+m_2+\dots+m_d)\cr\cr
&\times&\delta_{\ca{D}}(q'_1+q_2+\dots+q_d)\delta_{\ca{D}}(q_1+m_2+\dots+m_d)
    \delta_{\ca{D}}(q_2-q'_2)\dots \delta_{\ca{D}}(q_d-q'_d)\cr\cr
&+&l\sum_{m_1}\varphi_{m_1q'_2 \dots q'_d}\ov{\varphi}_{m_1q_2 \dots q_d}
\delta_{\ca{D}}(\Sigma q)
    \delta_{\ca{D}}(m_1+q'_2+\dots+q'_d)\crcr
&\times&\delta_{\ca{D}}(m_1+q_2+\dots+q_d)\delta_{\ca{D}}(q_1+q'_2+\dots+q'_d)\delta_{\ca{D}}(q_1-q'_1)\cr\cr
&+&\Sym\biggr]\,,\label{Fgaug}\\
P_k(\textbf{q},\textbf{q}')&=&
R_k(\textbf{q},\textbf{q'}) +  \Big(Z_k\sum_sq_s^2+\mu_k\Big)
\delta_{\ca{D}}(\Sigma q)
    \delta_{D^*}(\textbf{q},\textbf{q}')  \,.
\label{Pgaug}
\eea
This leads to the Wetterich equation:
\bea
\label{wettgaugeTr}
\de_t\Gamma_k&=&\Tr[\de_tR_k\cdot(P_k+F_k)^{-1}]\crcr
&=&l^{2d}\sum_{\textbf{p},\textbf{p}'}\de_tR_k(\textbf{p},\textbf{p}')
    \Big(P_k+F_k\Big)^{-1}(\textbf{p}',\textbf{p})\,.
\eea
On the left hand side, as in section \ref{sect3:TGFT}, we truncate at the level
of the quartic interactions. This gives then, for the r.h.s. of the Wetterich equation, the same expansion shown in
\eqref{wettexpansionPHI4}, where now the operators involved are given by \eqref{deRgaug}, \eqref{Fgaug} and \eqref{Pgaug}.

An extra subtlety must be paid attention to, however, in extracting the $\beta$-functions of this model. The $\delta$'s implementing the convolutions which appear in the $P_k$ operators can be inverted using \eqref{inverseOperator}, and summing over their indices we do not modify the dimensions of the whole expression. This is, however, not true for the $\delta$'s coming from the gauge constraints because they are not summed, so we need to keep them in the denominator. In any case, these constraints, turn out to be redundant with other delta functions coming from the $F_k$ and $\de_tR_k$ operators, in such a way that their contribution, 
because of the regularization, is equivalent to some power of $l$, and it is naturally well defined.

\subsection{$\beta$-functions and RG flows}
\label{betfuncGaug}

Expanding the FRG equation \eqref{wettgaugeTr}, we find the following system of dimensionful $\beta$-functions (the main steps of the calculations
are given in appendix \ref{app:gau}): 
\beq\label{dimensfulGaug}
\left\{
\begin{aligned}
\beta_{d\ne 4}(Z_k)&=\frac{d\lambda_k}{(Z_kk^2+\mu_k)^2}\Big\{\de_tZ_k\Big[
    \frac{\pi^{\frac{d-2}{2}}}{(d-1)^{\frac{3}{2}}\Gamma_E\Big(\frac{d}{2}\Big)}\frac{k^{d-2}}{l^d}
    +\frac{1}{l^2}\Big]+\frac{2\,\pi^{\frac{d-2}{2}}Z_k}
    {(d-1)^{\frac{3}{2}}\Gamma_E\Big(\frac{d-2}{2}\Big)}\frac{k^{d-2}}{l^d}\Big\}\crcr
\beta_{d\ne 4}(\mu_k)&=-\frac{d\lambda_k}{(Z_kk^2+\mu_k)^2}
    \Big\{\de_tZ_k\Big[\frac{k^d}{l^d}\frac{\pi^{\frac{d-2}{2}}}
    {\sqrt{d-1}\Gamma_E\Big(\frac{d+2}{2}\Big)}+\frac{k^2}{l^2}\Big]
    +2Z_k\Big[\frac{k^d}{l^d}\frac{\pi^{\frac{d-2}{2}}}
    {\sqrt{d-1}\Gamma_E\Big(\frac{d}{2}\Big)}+\frac{k^2}{l^2}\Big]\Big\}
\crcr
\beta_{d\ne 4}(\lambda_k)&=
\frac{2\lambda_k^2}{(Z_kk^2+\mu_k)^3}
   \Big\{\crcr
& \de_tZ_k\Big[\frac{2\pi^{\frac{d-2}{2}}}{d\sqrt{d-1}\Gamma\Big(\frac{d}{2}\Big)}
   \frac{k^d}{l^d}+(2[d+\delta_{d,3}]-1)\frac{k^2}{l^2}\Big]
+2Z_k\Big[\frac{\pi^{\frac{d-2}{2}}}{\sqrt{d-1}\Gamma\Big(\frac{d}{2}\Big)}
   \frac{k^d}{l^d}+(2[d+\delta_{d,3}]-1)\frac{k^2}{l^2}\Big]\Big\}
\end{aligned}
\right. 
\eeq
and, at $d=4$, we have
\beq\label{dimensfulG4}
\left\{
\begin{aligned}
&
\beta_{d=4}(Z_k)=\frac{\lambda_k}{(Z_kk^2+\mu_k)^2}
\Big\{ 
\de_t Z_k \Big[ \frac{\pi}{\sqrt{3}} \frac{k^2}{l^4}  + \frac{4}{l^2 } \Big] 
+ \frac{2\pi}{\sqrt{3}}\frac{k^2}{l^4} 
Z_k 
\Big\} \crcr
&
\beta_{d=4}(\mu_k)=
-\frac{4\lambda_k}{(Z_kk^2+\mu_k)^2}
\Big\{ 
\de_tZ_k \Big[ \frac{\pi}{2\sqrt{3}} \frac{ k^4  }{l^4 }   +  \frac{ k^2}{l^2 }  \Big] 
+ 
2Z_k \Big[ \frac{\pi}{\sqrt{3}} \frac{ k^4  }{l^4 }  + \frac{k^2}{l^2 }   \Big] 
\Big\} \crcr
&\beta_{d=4}(\lambda_k)=
\frac{2\lambda_k^2}{(Z_kk^2+\mu_k)^3}
   \Big\{\de_tZ_k\Big[\frac{2\pi}{4\sqrt{3}}
   \frac{k^4}{l^4}+7\frac{k^2}{l^2}\Big]
+2Z_k\Big[\frac{\pi}{\sqrt{3}}
   \frac{k^4}{l^4}+7\frac{k^2}{l^2}\Big]\Big\}
\end{aligned}
\right. 
\eeq
In order to obtain a well defined non-compact limit of the model, 
we use a modified ansatz (different from the one of section \ref{betastgft}):
\beq
Z_k=\ov{Z}_kk^{-\chi}l^{\chi}\,,\quad\mu_k=\ov{\mu}_k\ov{Z}_kk^{2-\chi}l^{\chi}\,,
\quad\lambda_k=\ov{\lambda}_k\ov{Z}_k^2k^{6-\xi}l^{\xi}\,,
\eeq 
from which we obtain the dimensionless $\beta$-functions 
according to the following calculation: 
\bea
\eta_k&=&\frac{1}{\ov{Z}_k}\beta(\ov{Z}_k)=\frac{k^{\chi}l^{-\chi}}{\ov{Z}_k}\beta(Z_k)+\chi\,,\crcr
\beta(\ov{\mu}_k)&=&\frac{k^{\chi-2}l^{-\chi}}{\ov{Z}_k}\beta(\mu_k)-\eta_k\ov{\mu}_k+(\chi-2)\ov{\mu}_k\,,\\
\beta(\ov{\lambda}_k)&=&\frac{k^{\xi-6}l^{-\xi}}{\ov{Z}_k^2}\beta(\lambda_k)-2\eta_k\ov{\lambda}_k
   +(\xi-6)\ov{\lambda}_k\,.\nonumber
\eea
Inserting the above in \eqref{dimensfulGaug}, we 
deduce the equations for the dimensionless coupling constants: 
\bea
\label{dimenslssGaug0}
\eta_k&=&
\frac{d\ov{\lambda}_k k^{2-\xi +2\chi}l^{\xi - 2\chi} }{(1+\ov{\mu}_k)^2}
\Big\{ (\eta_k -\chi)\Big[
    \frac{\pi^{\frac{d-2}{2}}}{(d-1)^{\frac{3}{2}}\Gamma_E\Big(\frac{d}{2}\Big)}\frac{k^{d-2}}{l^d}
    +\frac{1}{l^2}\Big]+\frac{2\,\pi^{\frac{d-2}{2}}}
    {(d-1)^{\frac{3}{2}}\Gamma_E\Big(\frac{d-2}{2}\Big)}\frac{k^{d-2}}{l^d}\Big\} + \chi 
\cr\cr
\beta_{d \ne 4}(\ov{\mu}_k)&=&-
\frac{d \ov{\lambda}_kk^{-\xi+2\chi}l^{\xi-2\chi} }{(1+\ov{\mu}_k)^2}
    \Big\{(\eta_k -\chi)\Big[\frac{k^d}{l^d}\frac{\pi^{\frac{d-2}{2}}}
    {\sqrt{d-1}\Gamma_E\Big(\frac{d+2}{2}\Big)}+\frac{k^2}{l^2}\Big]
    +2\Big[\frac{k^d}{l^d}\frac{\pi^{\frac{d-2}{2}}}
    {\sqrt{d-1}\Gamma_E\Big(\frac{d}{2}\Big)}+\frac{k^2}{l^2}\Big]\Big\}
\crcr
&& -\eta_k\ov{\mu}_k+(\chi-2)\ov{\mu}_k 
\cr\cr 
\beta_{d \ne 4}(\ov{\lambda}_k)&=&
\frac{2\ov{\lambda}_k^2 k^{2\chi-\xi}l^{\xi-2\chi}}{  
(1+\ov{\mu}_k)^3}
   \Big\{
(\eta_k -\chi)\Big[\frac{2\pi^{\frac{d-2}{2}}}{d\sqrt{d-1}\Gamma\Big(\frac{d}{2}\Big)}
   \frac{k^d}{l^d}+(2[d+\delta_{d,3}]-1)\frac{k^2}{l^2}\Big]
\cr\cr
&& +2\Big[\frac{\pi^{\frac{d-2}{2}}}{\sqrt{d-1}\Gamma\Big(\frac{d}{2}\Big)}
   \frac{k^d}{l^d}+(2[d+\delta_{d,3}]-1)\frac{k^2}{l^2}\Big]\Big\}
 -2\eta_k\ov{\lambda}_k  +(\xi-6)\ov{\lambda}_k
\eea

As in section \ref{betastgft}, the system of $\beta$-functions is non-autonomous in the IR cut-off $k$, as long as $l$ is kept
finite. We also notice a different dependence on the parameters $k$ and $l$ with respect to \eqref{dimlessbetanongauge}. The difference is of course a consequence of the presence
of the delta functions which, having non-trivial dimensions, change both the canonical and scaling dimensions of couplings
and fields, and remove degrees of freedom from the space of dynamical fields by imposing the gauge invariance constraints. Concerning this,
we point out that, had we introduced one delta for each field appearing in both the kinetic and interaction
kernels, this operation would have caused some extra divergences, but it would have also allowed us to absorb, from the point of view of the dimensions, the contribution of deltas inside a redefinition of the fields. In that case we would expect the couplings to have the same (canonical) dimensions of those appearing in the
previous model. Finally, we can also note that the system might be re-expressed in terms of a shifted anomalous dimension $\eta_k \to \eta_k -\chi$, thus it could be defined up to constant $\chi$. In the following, we have set $\chi = 0$. 

To get an autonomous system in the limit of the regulator being removed, we set
\beq
\xi - 2\chi - d = 0 \,, 
\eeq
and fixing $\chi = 0$, we come to $\xi =d$. 
 In the thermodynamic limit, for $d\ne 4$, we obtain the autonomous
system, 
\beq
\label{dimenslssGaug}
\left\{
\begin{aligned}
&\eta_{k}=
\frac{d\ov{\lambda}_k  }{(1+\ov{\mu}_k)^2}
\frac{\pi^{\frac{d-2}{2}}}{(d-1)^{\frac{3}{2}}}\Big\{ \eta_k
    \frac{1}{\Gamma_E\Big(\frac{d}{2}\Big)}
    +\frac{2}
    {\Gamma_E\Big(\frac{d-2}{2}\Big)}\Big\}  \crcr
&
\beta_{d\ne 4}(\ov{\mu}_k)=-
\frac{d \ov{\lambda}_k }{(1+\ov{\mu}_k)^2}
    \frac{\pi^{\frac{d-2}{2}}}
    {\sqrt{d-1}}\Big\{\eta_k \frac{1}
    {\Gamma_E\Big(\frac{d+2}{2}\Big)}
    +\frac{2}
    {\Gamma_E\Big(\frac{d}{2}\Big)}\Big\}
 -(\eta_k+2)\ov{\mu}_k 
\crcr 
&\beta_{d\ne 4}(\ov{\lambda}_k)=
\frac{2\ov{\lambda}_k^2 }{ 
(1+\ov{\mu}_k)^3}  \frac{\pi^{\frac{d-2}{2}}}
    {\sqrt{d-1}}
   \Big\{\eta_k 
\frac{1}{\Gamma_E\Big(\frac{d+2}{2}\Big)}
+\frac{2}{\Gamma_E\Big(\frac{d}{2}\Big)}\Big\} 
 -2\eta_k\ov{\lambda}_k +(d-6)\ov{\lambda}_k
\end{aligned}
\right. 
\eeq

In passing, we observe that at $d=6=\xi$, the coupling $\lambda_k$
becomes marginal.

The same analysis performed at $d=4$ yields 
\beq\label{dimenslssG4}
\left\{
\begin{aligned}
&
\eta_k=\frac{\ov{\lambda}_k}{(1+\ov{\mu}_k)^2}
\frac{\pi}{\sqrt{3}} (\eta_k +2 )
  \crcr
&
\beta_{d=4}(\mu_k)=
-\frac{4\ov{\lambda}_k}{(1+\ov{\mu}_k)^2}
\frac{\pi}{\sqrt{3}} \Big(\frac12\eta_k +  2\Big) 
 -(\eta_k+2)\ov{\mu}_k\crcr
&\beta_{d=4}(\lambda_k)=
\frac{2\ov{\lambda}^2_k}{(1+\mu_k)^3}
  \frac{\pi}{\sqrt{3}}\Big(\frac{1}{2}\eta_k
+2 \Big) -2(\eta_k+1)\ov{\lambda}_k
\end{aligned}
\right. 
\eeq

\subsection{Rank $d=3,4$}
\label{betfuncGaug34}
We can now fix the rank $d$, to be able to explicitly 
compute the flow. 

We start with the case $d=3$.
The dependence in $\chi$ can be re-absorbed by  a 
redefinition $\eta_k \to \eta_k-\chi$ (and the resulting variable
is called again $\eta_k$). 
We therefore have finally a system of dimensionless
 $\beta$-functions given by 
\beq\label{dimenlessGaug}
\left\{
\begin{aligned}
\eta_k&=\frac{3\ov{\lambda}_k}{\sqrt{2}(1+\ov{\mu}_k)^2-3\ov{\lambda}_k}\crcr
\beta(\ov{\mu}_k)&=-\frac{6\ov{\lambda}_k\sqrt{2}}{(1+\ov{\mu}_k)^2}
    \Big(\frac{\eta_k}{3}+1\Big)-\eta_k\ov{\mu}_k-2\ov{\mu}_k\\
\beta(\ov{\lambda}_k)&=\frac{4\ov{\lambda}_k^2\sqrt{2}}{(1+\ov{\mu}_k)^3}
    \Big(\frac{\eta_k}{3}+1\Big)-2\eta_k\ov{\lambda}_k-3\ov{\lambda}_k
\end{aligned}
\right.
\eeq
Like in the model without gauge projection, the system presents a divergence in the flow due to the truncation scheme. Here the
singularity occurs at $\ov{\mu}=-1$ and $\ov{\lambda}=\frac{\sqrt{2}}{3}(1+\ov{\mu})^2$. In the plane $(\ov{\mu},\ov{\lambda})$, we find four fixed points, the Gaussian (GFP) and three non-Gaussian fixed points (NGFP) at:
\beq
  _3P_1=(10)^{-1}(-7.083,0.154)\,, \quad \, 
_3P_2=10^{-1}(-7.935,0.273)\,, \quad  \, 
 _3P_3=(-12.809,169.635)\,.
\eeq
\begin{figure}[h]
\centering
 \includegraphics[scale=0.3]{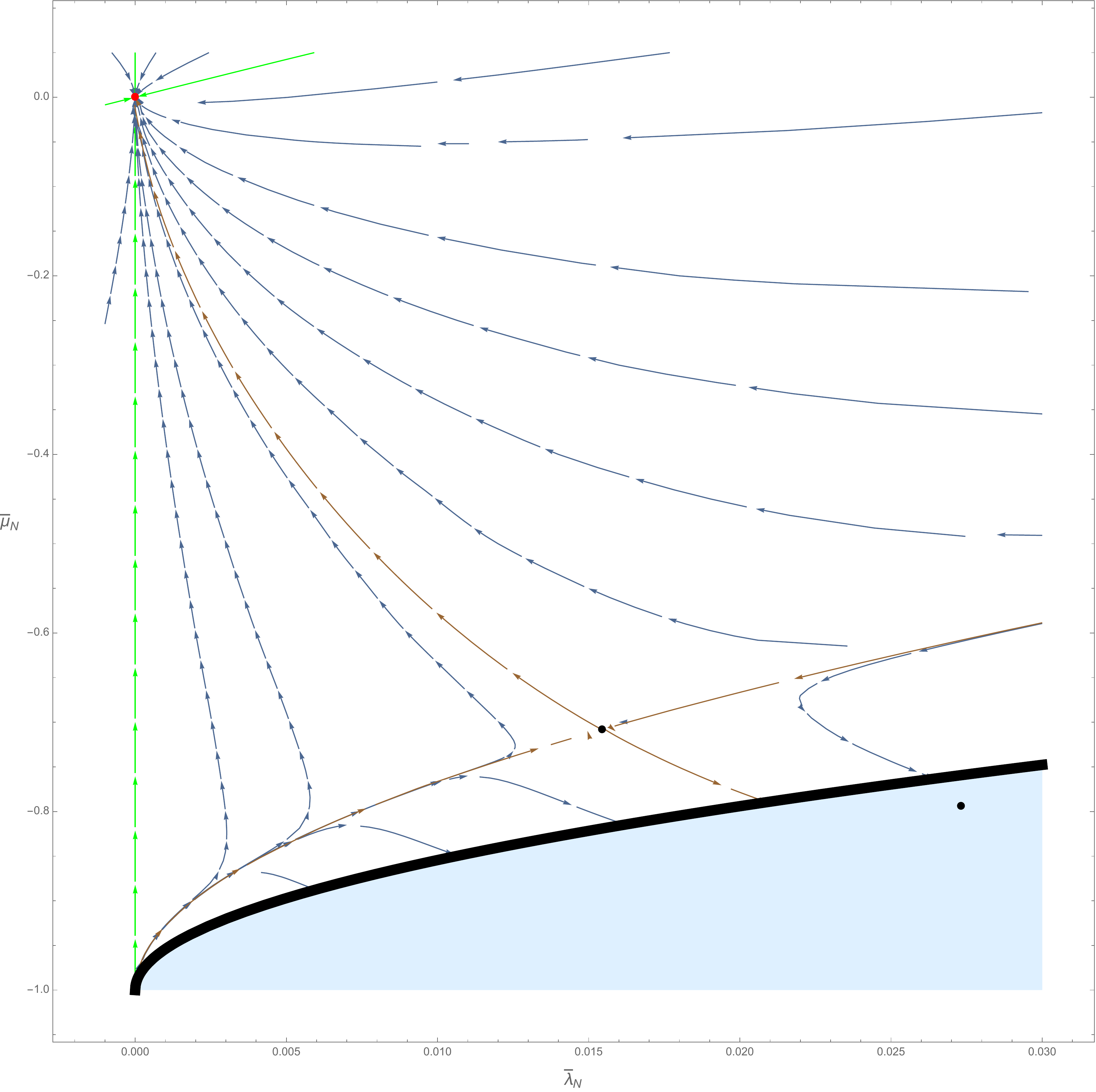}
\caption{Flow for the rank-3 gauge invariant model. Brown arrows represent the eigendirections of the NGFP (in black),
while green arrows are the eigendirections of GFP (in red). The thick black line indicates the singularity of the system.}
\label{Fig:plotGauge}
\end{figure}
Both $_3P_2$ and $_3P_3$ lie in the sector disconnected from the origin, therefore we restrict the analysis and linearize the system only around $_3P_1$ and
the Gaussian fixed point. 
The following eigenvalues and eigenvectors can be found by calculation from the stability matrix: 
\begin{align}
\text{GFP} _3 \quad   & _3\theta_0^+=-2\; \text{ for } \; _3 \textbf{v}_0^+=(1,0),\\
\text{GFP}_3\quad  & _3\theta_0^-=-3\; \text{ for } \; _3\textbf{v}_0^-=(6\sqrt{2},1),\\
_3P_1 \quad  & _3\theta_1\sim14.47\; \text{ for }\; _3\textbf{v}_1 \sim 10^{-1}(9.986,-0.529),\\
_3P_1  \quad  & _3\theta_2\sim-2.29\; \text{ for }\; _3\textbf{v}_2\sim 10^{-1}(9.948,1.022).
\end{align}
Negative eigenvalues represents UV-attractive eigendirections, while positive eigenvalues correspond to UV-repulsive
eigendirections.
From the plot in Fig.\ref{Fig:plotGauge}, we see that the Gaussian fixed point, where we have two negative eigenvalues
corresponding to the scaling dimensions of the couplings, is a UV-attractor and has two relevant directions. Thus, we infer that 
the model is asymptotically free in the UV.
Meanwhile, the NGFP has one relevant direction and one irrelevant direction.
In this model, there are no marginal directions in the flow and,
qualitatively, the structure of the plot is again reminiscent of the Wilson-Fisher fixed point in standard scalar field
theory in three dimensions. This is again a strong hint 
to a phase transition between a symmetric 
and a broken phase, interpreted as a condensate phase labeled by a non-zero expectation value of the TGFT field operator. 

Comparing this model with the one studied in section \ref{rank3}, we can list some similarities, as well as the differences that follow then directly from the new gauge invariance imposition. 

From the computational point of view, there are no fundamental differences. The presence of the gauge constraints influences the end result for what concerns the exact dependence of the FRG equations on the parameters $k$ and $l$. The way the thermodynamic limit turns the regularized system of RG equations into an autonomous one is similar, but resulting from different canonical dimensions attributed to the various elements of the theory. For example, the canonical dimension of the $\phi^4$-coupling changes from one model to the other.
We claim  that these models are not in the same universality  class.

From a qualitative point of view, we find in both models the same number of non-Gaussian fixed points, but their distribution in the plane $(\ov{\mu},\ov{\lambda})$ is different. 
The TGFT model without gauge projection has two interesting NGFPs in the region of the plane $(\ov{\mu},\ov{\lambda})$ connected to the origin, whereas the gauge projected model has a unique NGFP lying in the same region. 
Also, the linearised theory around the Gaussian fixed point turns out to be slightly different. While in the
previous section we have found a non-diagonalizable stability matrix with only one strictly relevant direction, for the gauge invariant model 
we have two relevant directions and the eigenperturbations form indeed a basis for the linearised system.  On the other hand, the GFPs of both models are sinks, and so both models are asymptotically free.

In rank $d=4$, the results are very similar to the above rank $d=3$. 
We obtain, in addition to the Gaussian fixed point, the fixed points
\beq
_4{P_1}=(10)^{-1}(-7.05,0.093)\,, \quad
_4{P_2}=10^{-1}(-8.465,0.228)\,, \quad 
_4{P_3}=(10.051,-97.962)\,.
\eeq
$_4{P_2}$ which stands below the singularity will be not further analysed. 
We will focus on the rest of the fixed points and perform a linearisation around those.  

Around the Gaussian fixed point the stability matrix becomes
\beq
(\beta^*_{ij})\Big|_{GFP}:= \left(
\begin{array}{cc}
 -2 & -\frac{8 \pi }{\sqrt{3}} \\
 0 & -2 \\
\end{array}
\right)
\eeq 
which has an eigenvalue $_4\theta_0=-2$ with multiplicity 2
with a single eigenvector $_4\textbf{v}_0=(1,0)$. We cannot diagonalise
it and will integrate numerically the flow around this point. 

We have the following critical exponents:
\begin{align}
\text{ GFP}_4 \quad   & _4\theta_0=-2\, \text{ for } \, _4\textbf{v}_0=(1,0),
\\
_4P_1 \quad  &  _4\theta_{11}\sim 11.819\, \text{ for }\, _4\textbf{v}_{11}\sim10^{-1}(10,-0.225),   \\
 _4P_1  \quad  & _4\theta_{12}\sim -2.158\, 
\text{ for }\, _4\textbf{v}_{12}\sim10^{-1}(10,0.624), 
\\
_4P_3 \quad  &  _4\theta_{31}\sim -2.654\,  \text{ for }\, _4\textbf{v}_{31}\sim10^{-1}(-3.891, 9.211),\\
 _4P_3 \quad  & _4\theta_{32}\sim 0.624\, \text{ for }\, _4\textbf{v}_{32}\sim10^{-1}(0.316, -10). 
\end{align}

Both NGFPs have one relevant and one irrelevant directions. 
The analysis of perturbations around the fixed points leads to the phase diagram and RG flow presented in Fig\ref{R4plot}. From the numerical integration, we observe
that the second eigendirection of the GFP is marginally relevant. We represent the phase diagram in Fig.\ref{R4plot}. 
\begin{figure}[h]
 \centering
\includegraphics[scale=0.3]{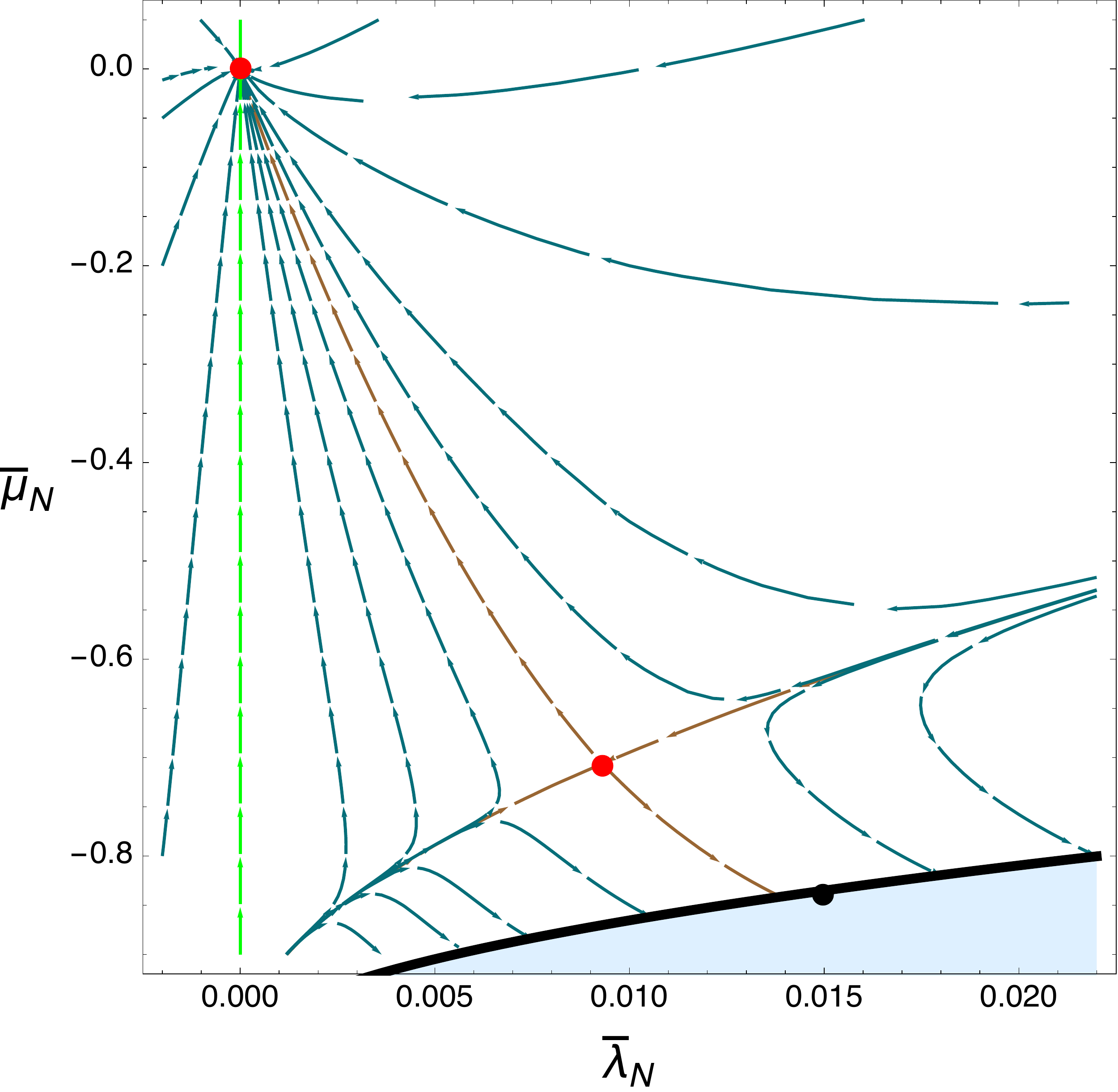}
\caption{Flow of the gauged model at rank $d=4$.}
\label{R4gplot}
\end{figure} 

We see once more RG trajectories indicating asymptotic freedom in the UV, and the presence of a phase transition between a symmetric and a broken phase in the IR. 

\subsection{Rank $d=6$}
Another interesting case to look at in more detail is the one for $d=6$.
For this rank, the model has one marginal direction around the GFP as
the scaling dimension of the coupling $\lambda$ vanishes. 
 In this case, in fact, we can compare our results directly with the ones obtained in \cite{Benedetti:2015yaa}. This comparison has two aspects. At the regularised level, with the system restricted to (six copies of) the compact domain $S^1$, we expect our RG equations to match the ones found in  \cite{Benedetti:2015yaa}, up to normalisations. This can indeed be verified, but we do not report on it. On the other hand, by studying the RG flow in the thermodynamic limit, we will then be able to check how the phase diagram we obtain compares with the limiting cases studied for the compact model, expecting a qualitative agreement with the  results found there in the UV approximation.

\begin{figure}[h]
 \centering
\includegraphics[scale=0.3]{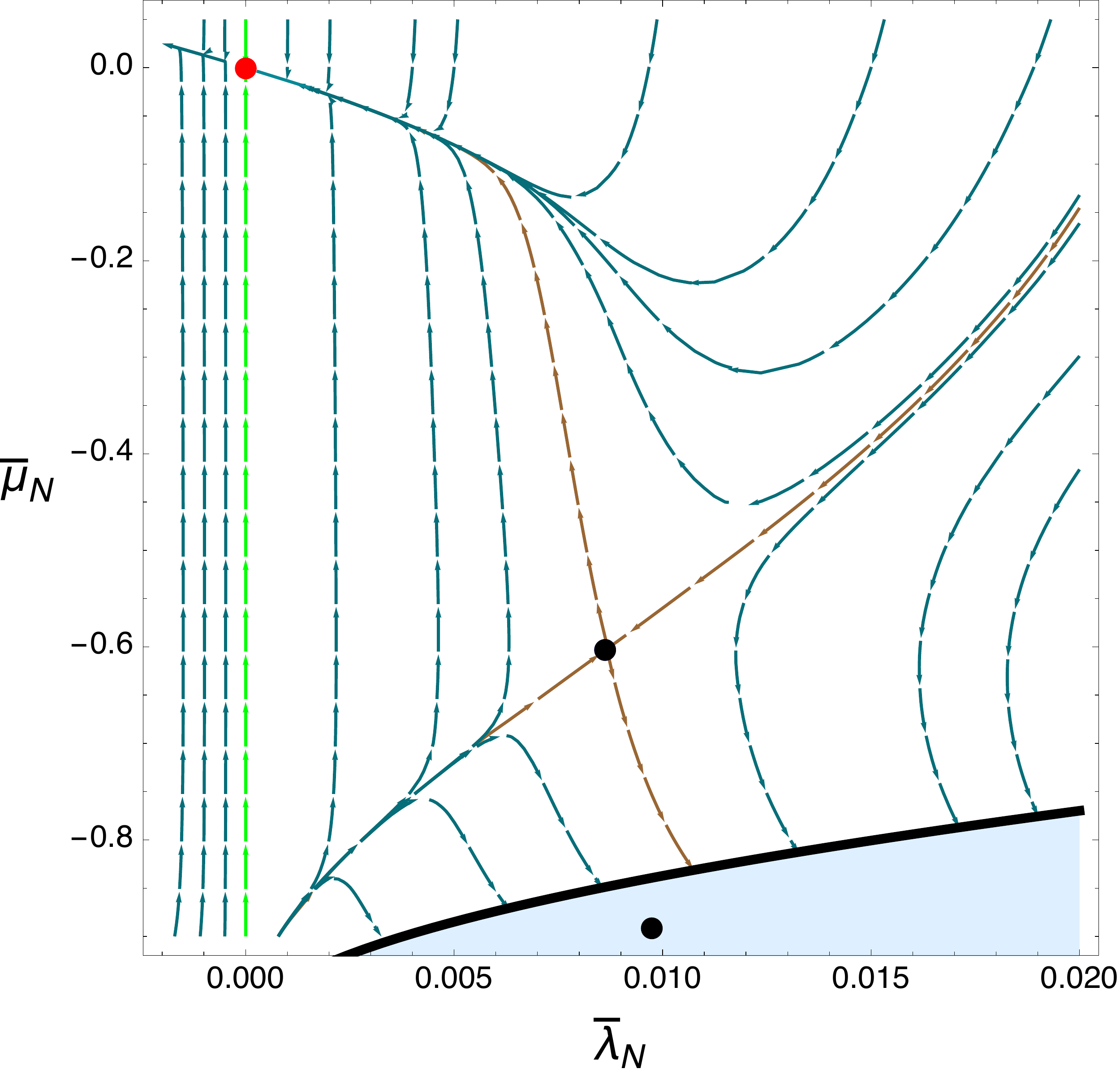}
\caption{Flow of the gauged model at rank $6$.}
\label{R6gplot}
\end{figure}

In rank $d=6$, we have the following fixed points alongside the Gaussian fixed point: 
\begin{align}
 & _6{P_\pm}= \Big( \frac{1}{234}(-175  \pm \sqrt{1141}) , \, \frac{\sqrt{5} \big(43309 \mp 79\sqrt{1141}\big) }{1067742\pi^2} \Big)
\end{align}
The NGFP $_6{P_-}$ is below the singularity. We focus on the Gaussian FP and $_6P_+$ which gives
\begin{align}
\text{ GFP}_6 \quad   & _6\theta_0=-2\,\text{ for }\,_6\textbf{v}_0^+=(1,0),\\
\text{ GFP}_6 \quad   & _6\theta_0=-2\,\text{ for }\,_6\textbf{v}_0^-=(-\frac{3\pi^2}{\sqrt{5}},1),\\
_6P_+ \quad  &  _6\theta_1\sim 4.859\text{ for }_6\textbf{v}_1\sim (-185.549,1),\\
 _6P_+  \quad  & _6\theta_2\sim-0.9\text{ for }_6\textbf{v}_2\sim10^{-1}(31.289,1).
\end{align}
The GFP has one relevant (mass) direction, and one marginally relevant direction for positive  $\lambda$, which signals asymptotic freedom. Notice that for negative $\lambda$ we do not expect the theory to be non-perturbatively well-defined.
On the other hand, the NGFP has a relevant and irrelevant direction
and share a similar structure as the Wilson-Fisher FP.  
The analysis of perturbations around the fixed points in this case, then, leads to the phase diagram and RG flow presented in Fig.\ref{R6gplot}.
Same conclusions discussed so far hold again in the present rank 6. 

After the following change of normalisation
$\lambda\to 2\lambda$, the NGFP $_6P_+$, its critical exponents 
and those of the GFP match with 
the results in rank $d=6$ in the large mode limit of \cite{Benedetti:2015yaa}. 
The RG flow lines are also very similar. 
Interestingly, at least for this model at rank 6, this coincidence means that the large radius sphere limit of the TGFT corresponds to our thermodynamic limit
with our particular choice of scaling the coupling  including both 
IR cut-off-scaling  and lattice spacing scaling. In fact, we expect this to be true more generally (for instance at any rank 
$d$ or for any background of the fields).

\

As pointed out in \cite{Benedetti:2015yaa}, the presence of both an attractive UV fixed point and an IR Wilson-Fisher fixed point seems to be a general feature of TGFT's. While many other (local) Quantum Field Theories present just one of these results (this is the case of QCD for just asymptotic freedom and of scalar field theory for the IR fixed point), the non-local models that we studied appear to always have a well defined behaviour in both the limits.  Moreover, there is another important property which might be interesting and fertile for future developments. All the models that we studied also present a second IR fixed point lying beyond the singularity of the flow. Even if we said that the presence of the anomalous dimension as a parameter in our effective action generates a divergence which prevents us from trusting in the flow across itself, we should remember that far from the infinite values of the flow the computation is probably correct. In other words, given initial conditions in the sector connected with the origin, we are not able to integrate the RG equation beyond the singularity but, had we given initial conditions in the other sector, the situation would be the opposite. Even if we cannot reconnect the flows over all the space of couplings, there are hints that other fixed points could arise and, with them, there is the possibility to find new (non-trivial) UV attractors. If this is confirmed by further investigations, TGFTs would also show asymptotic safety in the UV, in some regions of parameter space, and for specific models at least. If reproduced for 4d gravity models with more quantum geometric structure, this result would be in agreement with the hypothesis of asymptotic safety proposed by Weinberg and Reuter for quantum gravity theories \cite{AS}. However, it is not immediate to match TGFT results of this type with the asymptotic safety programme for quantum gravity, since this is based directly on quantum Einstein gravity, thus quantum field theories on spacetime involving directly a metric field, while TGFTs aim to be models of the microscopic constituents of spacetime and geometry itself. Still, it may taken to suggest a nice convergence of results from different directions.

\section{Conclusion}
\label{concl}

We have undertaken the Functional Renormalisation Group analysis of two classes of Tensorial Group Field Theories, as a further application of the formalism first studied in \cite{Benedetti:2014qsa}. 

The models are defined on the non-compact group manifold $\mathbb{R}$ and for arbitrary tensor rank. They are endowed with melonic combinatorial interactions and distinguished from the presence (or absence) of a projection on the gauge invariant dynamics under the diagonal group action on the field arguments. 

Both classes of models are simplified with respect to full-fledged TGFT models for quantum gravity, usually based on the group manifolds $SU(2)$ or $SL(2,\mathbb{C})$, and characterised by additional condition on the dynamics, in addition to the gauge invariance models. However, they may captures many of their relevant features, and they are in any case of great interest from a more technical/mathematical point of view, and the FRG analysis is a further step towards controlling and understanding this new type of quantum field theories. More generally, any GFT defines a sum over cellular complexes, which can be interpreted as a discrete definition of the covariant path integral for quantum gravity (with the details of the interpretation depending of course on the details of the amplitudes of the model), of the same type as those defining the dynamical triangulations approach to quantum gravity. The FRG analysis has the main objective of probing their continuum limit and phase structure, which would be, for quantum gravity models, a continuum limit for the pre-geometric, discrete and quantum building blocks of spacetime. The search for a continuum geometric phase governed by a general relativistic dynamics is in fact the main outstanding open issue of these quantum  gravity theories.

At a more technical level, the specific aim of our study was to obtain a picture of the fixed points and phase diagram, while enlightening the peculiarities coming from the
non-compactness of the underlying group manifold, and thus comparing these results to previous work on TGFTs based on abelian compact groups \cite{Benedetti:2014qsa, Benedetti:2015yaa}.  

The main new issue posed by the non-compactness of the group manifold is the presence of IR divergences in the expansion of the Wetterich equation, which cannot be dealt with in the same way in which one removes simple infinite volume factors in local field theories, due to the particular combinatorics of TGFT interaction terms.
We have shown, generalising the previous work \cite{Geloun:2015qfa}, how to regularise, first, and then remove these divergences using the appropriate thermodynamic limit. 
In particular, a comparison with \cite{Benedetti:2015yaa} and the verified matching of critical exponents and scaling dimensions, suggests a new concept of scaling dimension for this class of theories. While in the previous work the dimensional analysis leading to scaling dimensions was based on a perturbative approach and on the analysis of $n$-loops greens functions, at this non-perturbative level we find more appropriate to rely on the order of divergences that need to be regularised to make the theory consistent in the non-compact limit.
In this limit,  all the models we study define a well-posed autonomous system of RG equations for the coupling constants,  we then proceed to solve numerically for various interesting values of the rank, in a simple truncation of the effective action. 

In this simple truncation, and for all models considered, we identify UV and IR fixed points, study the perturbations around them, and obtain the corresponding phase diagram. 
In all these models, we find strong indications of: 1) asymptotic freedom in the UV; 2) a number of non-Gaussian fixed points in the IR; 3) a phase transition similar to the Wilson-Fisher type, between a symmetric and a broken (or condensate) phase with a non-zero expectation value of the TGFT field operator. 

The first point is interesting because it confirms, by different means, the apparently generic asymptotic freedom of TGFT models, due to the dominance of wave function renormalisation over coupling constant renormalisation \cite{Rivasseau:2015ova}. The last point, on the other hand, is important because phase transitions (in particular, of condensation type) have been suggested to mark the emergence of spacetime and geometry in GFT models of 4d quantum gravity \cite{danieleemergence, Rivasseau:2011hm}, and because GFT condensate states have in fact been used to extract effective cosmological dynamics directly from the microscopic GFT quantum dynamics \cite{GFTcondensate}. 

However, more work is certainly needed to further corroborate these findings. 

Even for this simple class of TGFT models, one would need to improve the truncation scheme to include more terms in the effective action entering the Wetterich equation. And, concerning the study of the phase transition, a clear understanding of the different phases require at least solving the equations of motion (thus a mean field analysis), which is highly non-trivial due to the combinatorial structure of the TGFT interactions and the integro-differential nature of the equations, and a change of parametrisation for the effective potential (see the discussion in \cite{Benedetti:2014qsa}). 

And of course, we need to proceed towards the FRG analysis of more involved models, investigating how different groups and more involved forms of interaction kernels affect the results, and especially towards models with a more complete quantum geometric interpretation, and stronger links with simplicial quantum gravity and loop quantum gravity. The road ahead is long but promising.

\

\section*{Acknowledgements}
We thank Dario Benedetti for many helpful comments.

\appendix

\renewcommand{\theequation}{\Alph{section}.\arabic{equation}}
\setcounter{equation}{0}

\section{ Evaluation of $\beta$-functions  in rank $d$}
\label{app:betaTGFT}

In this appendix, we provide the detailed calculation of the $\beta$ equations and emphasise its particularities. Note that, this computation
of the $\beta$-functions is performed in the regularised framework and only, at the end, we take the thermodynamic limit. The system of
equations that we obtain is an autonomous system in a continuous non-compact space.

\noindent{\bf Notations.} 
Given the regularization prescription introduced in section \ref{IRthermo}, we set the notation
$\delta_{D^*}(\textbf{p},\textbf{q})=\delta(\textbf{p}-\textbf{q})$ not to be confused with the continuous Dirac delta that
we do not use in this appendix.
We also define $\ca{D}$ to be the one dimensional lattice, that is, the domain of a single component of objects in $D^*$.
We have $D^*=\ca{D}^{\times d}$ so that:
\begin{eqnarray}
 l\sum_{p_i}=\int_{\ca{D}}dp_i\,.
\end{eqnarray}
A change of
notation helps during the calculation: 
\begin{align*}
\textbf{q}=(q_1,q_2, \dots, q_d)\quad\Rightarrow
  \quad q_1:=q_1\,;\qquad \textbf{q}^{(d-1)}_1:=(q_2,q_3,\dots, q_d)\,;\quad q^{(d-1)}_1:=\sqrt{q_2^2+q_3^2 +\dots, q_d^2}\,, 
\end{align*}
for a generic $d$-dimensional momentum $\textbf{q}$.
When there is no possible confusion, we will simply
forget the subscript 1 of $\textbf{q}^{(d-1)}_1$ and $q^{(d-1)}_1$,
and use $\textbf{q}^{(d-1)}$ and $q^{(d-1)}$, respectively.

Let us recall the second variation of the effective action \eqref{eqq}
in these new notations: 
\begin{align*}
\Gamma_k^{(2)}
&=(Z_k\sum_sp_s^2+\mu_k)\delta(\textbf{p}-\textbf{p'})\crcr
&+\lambda_k\biggl[\int_{\ca{D}^{\times d-1}} dq_2\dots dq_d\;
   \varphi_{p'_1q_2\dots q_d}\ov{\varphi}_{p_1q_2\dots q_d}
   \prod_{i=2}^d\delta(p_i-p_i')\crcr
&+\int_{\ca{D}} dq_1\;\varphi_{q_1p'_2\dots p'_d}\ov{\varphi}_{q_1p_2\dots p_d}\delta(p_1-p'_1)
+\Sym\biggr]\crcr
&=(Z_k\sum_sp_s^2+\mu_k)\delta(\textbf{p}-\textbf{p'})+F_k(\textbf{p},\textbf{p'})\,
\end{align*}
and choose a regulator of the  form \eqref{reR} where $\theta$ is now replaced by $\Theta(f(\textbf{p}))$  the discrete step function. This implies:
\begin{align*}
\de_tR_k=\delta(\textbf{p}-\textbf{p'})\Theta(k^2-\sum_sp_s^2)[\de_tZ_k(k^2-\sum_sp_s^2)+Z_k2k^2]\,.
\end{align*}
Defining $P_k(\textbf{p},\textbf{p'})$ like \eqref{Pterm}, with 
appropriate replacements, we expand and truncate the Wetterich equation as \eqref{wettexpansionPHI4}.  The zeroth order of the previous expansion is the vacuum term and does not provide us any useful information. On the other hand, the first and the second order will provide us with the flow
of the kinetic ($\varphi^2$-) and interaction ($\varphi^4$-) couplings, respectively, namely, the 
$\beta$-functions for the couplings $\mu_k$, $Z_k$ and $\lambda_k$.

\subsection{$\varphi^2$-terms}
\label{app:tgftphi2}

To compute the flow of couplings of the quadratic terms of $\Gamma_k$,
in other words, the $\beta$-functions for $\mu_k$ and $Z_k$, we  focus on the first order of \eqref{wettexpansionPHI4}. 
To have more compact notations, let us introduce the first convolution appearing in the expansion:
\begin{align*}
&\tilde{\de}_tR_k(\textbf{p},\textbf{p}'')=
   \int_{D^*} d\textbf{p}'\;\de_tR_k(\textbf{p},\textbf{p}')(P_k)^{-1}(\textbf{p}',\textbf{p}'')\crcr
&=\int_{D^*} d\textbf{p}'\;\delta(\textbf{p}-\textbf{p}')\delta(\textbf{p}'-\textbf{p}'')\Theta(k^2-\sum_sp_s^2)
   \textstyle{\frac{\de_tZ_k(k^2-\sum_sp_s^2)+2k^2Z_k}{Z_k(k^2-\sum_s {p'}_s^2)
   \Theta(k^2-\sum_s {p'}_s^2)+Z_k\sum_s {p'}_s^2+\mu_k}}\crcr
&=\delta(\textbf{p}-\textbf{p}'')\Theta(k^2-\sum_sp_s^2)
   \frac{\de_t Z_k(k^2-\sum_sp_s^2)+2k^2Z_k}{(Z_kk^2+\mu_k)}\,,
\end{align*}
where we used the fact that, after integration, the two $\Theta$'s appearing in the expression are redundant. 

Thus, calling $(I)_W$ the first order of the Wetterich equation,
we write
\begin{align*}
-(I)_W&=\ov\Tr[\tilde{\de}_tR_k\cdot F_k\cdot(P_k)^{-1}]=\int_{{D^*}^{\times 2}} d\textbf{p}d\textbf{p}'\;\tilde{\de}_tR_k(\textbf{p},\textbf{p}')
    \int_{D^*} d\textbf{q}\;F_k(\textbf{p}',\textbf{q})(P_k)^{-1}(\textbf{q},\textbf{p})\crcr
&=\int_{D^*} d\textbf{p}\;\Theta(k^2-\sum_sp_s^2)
     \frac{\de_t Z_k(k^2-\sum_sp_s^2)+2k^2Z_k}{(Z_k k^2+\mu_k)^2}   F_k(\textbf{p},\textbf{p})\,. 
\end{align*}
To simplify the computation, we split the integral in two pieces, namely:
\bea
&&
A=\frac{\de_t Z_k}{(Z_k k^2+\mu_k)^2}
    \int_{D^*} d\textbf{p}\;\Theta(k^2-\sum_sp_s^2)\biggl(\sum_sp_s^2\biggr)F_k(\textbf{p},\textbf{p}) \,,\cr\cr\cr
&&
B=\frac{k^2(2+\de_t)Z_k}{(Z_kk^2+\mu_k)^2}\int_{D^*} d\textbf{p}\;\Theta(k^2-\sum_sp_s^2)\,F_k(\textbf{p},\textbf{p})\,,
\eea
having $(I)_W=A-B$.
Let us treat the first term and recall that $\delta_{\ca{D}}(0)=\delta(0)=\frac{1}{l}$:
\begin{align*}
A&=\frac{\lambda_k\,\de_t Z_k}{(Z_k k^2+\mu_k)^2}
    \int_{D^*} d\textbf{p}\;\Theta(k^2-\sum_sp_s^2)\biggl(\sum_sp_s^2\biggr)\crcr
&\times\biggl[\frac{1}{l^{d-1}}\int_{\ca{D}^{\times d-1}} dq_2\dots dq_d\;|\varphi_{p_1q_2\dots q_d}|^2
    +\frac{1}{l}\int_{\ca{D}} dq_1\;|\varphi_{q_1p_2\dots p_d}|^2
+\Sym\biggr]\crcr
&=\frac{\lambda_k\,\de_t Z_k}{(Z_k k^2+\mu_k)^2}\times\crcr
&\Bigg\{\frac{1}{l^{d-1}}
    \int_{D^*} dp_1dq_2\dots dq_d\;|\varphi_{p_1q_2\dots q_d}|^2\int_{\ca{D}^{\times d-1}} dp_2\dots dp_d\;
    \Theta[(k^2-p_1^2)-\Sigma_{i=2}^d p_i^2]\Big[\Sigma_{i=2}^d p_i^2+p_1^2\Big]\crcr
&+\frac{1}{l}  \int_{D^*} dq_1dp_2\dots dp_d\;|\varphi_{q_1p_2\dots p_d}|^2\int_{\ca{D}} dp_1\;
    \Theta[(k^2-\Sigma_{i=2}^d p_i^2)-p_1^2]\Big[\Sigma_{i=2}^d p_i^2+p_1^2\Big]\Bigg\}\\
&+\Sym  \,. 
\end{align*}
Now we perform the continuum limit $l\to\infty$ and this corresponds  to:
\begin{eqnarray}
\int_{\ca{D}}\quad\longrightarrow\quad\int_{\R}\,,\qquad
\Theta\quad\longrightarrow\quad\theta\,.
\end{eqnarray}

The negative powers of $l$ appearing in the expressions keep track of the former IR divergences of the continuous model.
Extracting an $l$ dependence from the couplings, we will address them at the end.
In order to simplify the notation, we drop the limit symbol $\lim_{l\to \infty}$ and get
\begin{align*}
&A=\frac{\lambda_k\,\de_t Z_k}{(Z_k k^2+\mu_k)^2}\crcr
&\times\Bigg\{
\frac{1}{l}
    \int_{\R^d} dq_1 dp_2\dots dp_d\;\theta(k^2-\Sigma_{i=2}^d p_i^2)\;|\varphi_{q_1p_2\dots p_d}|^2
    \int_{-\sqrt{k^2-\Sigma_{i=2}^d p_i^2}}^{\sqrt{k^2-\Sigma_{i=2}^d p_i^2}}dp_1\;
    [\Sigma_{i=2}^d p_i^2+p_1^2]\crcr
&+\frac{1}{l^{d-1}}
    \int_{\R^d}  dp_1dq_2\dots dq_d\;\theta(k^2-p_1^2)\;
|\varphi_{p_1q_2\dots q_d}|^2\int d\Omega_{d-1}
    \int_0^{\sqrt{k^2-p_1^2}} dr\;r^{d-2}[r^2+p_1^2]\Bigg\}\crcr
&+\Sym\crcr
&=\frac{\lambda_k\,\de_t Z_k}{(Z_k k^2+\mu_k)^2}\Bigg\{
\frac{1}{l}
    \int_{\R^d} dq_1dp_2\dots dp_d\, \theta(k^2-\Sigma_{i=2}^d p_i^2)\crcr
&\times \biggl[2(\Sigma_{i=2}^d p_i^2)\sqrt{k^2-\Sigma_{i=2}^d p_i^2}
    +\frac{2}{3}(k^2-\Sigma_{i=2}^d p_i^2)^{3/2}\biggr]\;|\varphi_{q_1p_2\dots p_d}|^2\crcr
&+\frac{1}{l^{d-1}}
    \int_{\R^d}  dp_1 dq_2\dots dq_d\,\theta(k^2-p_1^2)
    \biggl[\frac{(k^2-p_1^2)^{\frac{d+1}{2}}}{d+1}+\frac{p_1^2}{d-1}(k^2-p_1^2)^{\frac{d-1}{2}}\biggr]
    \Omega_{d-1}\;|\varphi_{p_1q_2\dots q_d}|^2\Bigg\}\\
&+\Sym  \,,
\end{align*}
where in the first passage we changed variable to the $d-1$ dimensional spherical coordinates and introduced the following notation:
\begin{equation}
\Omega_d=\int d\Omega_d=
   \prod_{i=1}^{d-2}\Big[\int_0^{\pi}d\alpha_i\sin^{d-1-i}(\alpha_i)\Big]\int_0^{2\pi}d\alpha_{d-1}
   =\frac{2\pi^{d/2}}{\Gamma_E(\frac{d}{2})}\,,
\end{equation}
with $\Gamma_E$ the Euler gamma function.
Expanding the term $B$, we find:
\bea
&&
B=\lambda_k\frac{k^2(2+\de_t)Z_k}{(Z_kk^2+\mu_k)^2}\crcr
&&\times 
\int_{D^*} d\textbf{p}\;\Theta(k^2-\sum_sp_s^2)
   \biggl[\frac{1}{l^{d-1}}\int_{\ca{D}^{\times d-1}} dq_2\dots dq_d\;|\varphi_{p_1q_2\dots q_d}|^2
   +\frac{1}{l}\int_{\ca{D}} dq_1\;|\varphi_{q_1p_2\dots p_d}|^2\biggr]\crcr
&&+\Sym\,,
\eea 
which, in the limit, gives
\begin{align*}
B&=\lambda_k\frac{k^2(2+\de_t)Z_k}{(Z_kk^2+\mu_k)^2}
\Bigg\{ \frac{1}{l^{d-1}}
   \int_{\R^d} dp_1dq_2\dots dq_d\;\theta(k^2-p_1^2)\;|\varphi_{p_1q_2\dots q_d}|^2\,\Omega_{d-1}
 \int_0^{\sqrt{k^2-p_1^2}}dr\;r^{d-2}\crcr
&+
\frac{1}{l}
   \int_{\R^d} dq_1dp_2\dots dp_d\;\theta(k^2-\Sigma_{s=2}^d p_s^2)\;
|\varphi_{q_1p_2\dots p_d}|^2
   \int_{-\sqrt{k^2-\Sigma_{s=2}^d p_s^2}}^{\sqrt{k^2-\Sigma_{s=2}^d p_s^2}}dp_1\Bigg\}\crcr
&+\Sym\crcr
&=\lambda_k\frac{k^2(2+\de_t)Z_k}{(Z_kk^2+\mu_k)^2}
\Bigg\{ \frac{1}{l^{d-1}}
   \int_{\R^d} dp_1dq_2\dots dq_d\,\theta(k^2-p_1^2)\;
 \Omega_{d-1}\; \frac{(k^2-p_1^2)^{\frac{d-1}{2}}}{d-1}\; 
|\varphi_{p_1q_2\dots q_d}|^2\;\crcr
&+
\frac{2}{l}
   \int_{\R^d} dq_1dp_2\dots dp_d\;\theta(k^2-\Sigma_{s=2}^d p_s^2)\;
   \sqrt{k^2-\Sigma_{s=2}^d p_s^2}\; |\varphi_{q_1p_2\dots p_d}|^2\Bigg\}\crcr
&+\Sym\,.
\end{align*}
\noindent{\bf $\beta$-functions.}
To find the $\beta$-functions of the coupling constants, we rely on the fact that the l.h.s. of \eqref{wettexpansionPHI4} is of
the form:
\begin{align*}
\de_t\Gamma_{kin}=\int d\textbf{p}\,|\varphi(\textbf{p})|^2\, \biggl(\beta(Z_k)\sum_sp_s^2+\beta(\mu_k)\biggr)\,.
\end{align*}

In fact, this allows us to identify the $\beta$-functions with the coefficients of an expansion in powers of the field momenta of the integrands in $A$ and $B$, up to an $o(p^3)$. Respectively,
the terms with momenta of order $p_i^2$ convoluted with the fields
$\varphi_{\dots, p_i, \dots}$ will contribute to the flow of the wave function renormalisation, while the zeroth order will
be proportional to the scaling of the mass. All remaining terms, falling out of the truncation, must be discarded. Hence, we
have, for $d\geq 3$,
\begin{align*}
A&\simeq \frac{\lambda_k\de_t Z_k}{(Z_k k^2+\mu_k)^2}\Bigg\{\frac{1}{l}
    \int dq_1dp_2\dots dp_d\;
    \Big[\frac{2}{3}k^3+k\Big(\Sigma_{s=2}^dp_s^2\Big)\Big]\;|\varphi_{q_1p_2\dots p_d}|^2\crcr
&+\frac{1}{l^{d-1}}
    \int dp_1dq_2\dots dq_d
    \;\Omega_{d-1}\Big[\frac{k^{d+1}}{d+1}+\Big(\frac{1}{d-1}
    -\frac{1}{2}\Big)k^{d-1}p_1^2\Big] \;|\varphi_{p_1q_2\dots q_d}|^2\Bigg\}\crcr
&+\Sym\,.
\end{align*}

For the $B$ terms, one finds: 
\bea
B&\simeq &
\lambda_k\frac{k^2(2+\de_t)Z_k}{(Z_kk^2+\mu_k)^2}
\Bigg\{ \frac{1}{l^{d-1}}
   \int dp_1dq_2\dots dq_d\Omega_{d-1}
   \Big[\frac{k^{d-1}}{d-1}-\frac{k^{d-3}}{2}p_1^2\Big]\;|\varphi_{p_1q_2\dots q_d}|^2\crcr
&&+\frac{2}{l}
   \int dq_1dp_2\dots dp_d
   \Big[k-\frac{1}{2k}\Big(\Sigma_{s=2}^dp_s^2\Big)\Big]\;|\varphi_{q_1p_2\dots p_d}|^2\Bigg\}\crcr
&&+\Sym\,.
\eea
Now, we concentrate on the coloured symmetric terms.
Note that the procedure and result of the above integrals will not change
for each coloured term in sym$\{\cdot\}$, up to a simple relabelling. Thus, collecting all terms, we  obtain an expression of the form:
\bea
\de_t\Gamma_{kin}&=&\int dp_1\dots dp_d\,|\varphi_{p_1\dots p_d}|^2
   \sum_{j=1}^d\Big[f(k)+g(k)p_j^2+h(k)\Big(\Sigma_{i=1}^{j-1}p_i^2+\Sigma_{i=j+1}^d p_i^2\Big)\Big]\crcr
&=&\int dp_1\dots dp_d\,|\varphi_{p_1\dots p_d}|^2
   \Big\{d\,f(k)+\Big[g(k)+(d-1)h(k)\Big]\sum_{i=1}^dp_i^2\Big\}\,.
\eea
This, by comparison between the two sides of the equation, leads to the following dimensionful $\beta$-functions for the parameters $Z_k$ and $\mu_k$:
\bea
\beta(Z_k)&=&\frac{\lambda_k}{(Z_kk^2+\mu_k)^2}\Big\{\de_tZ_k\Big[2(d-1)\frac{k}{l}
+ \frac{\pi^{\frac{d-1}{2}}}{\Gamma_E\Big(\frac{d+1}{2}\Big)}\frac{k^{d-1}}{l^{d-1}}\Big]
   +2Z_k\Big[(d-1)\frac{k}{l}
   +\frac{\pi^{\frac{d-1}{2}}}{\Gamma_E\Big(\frac{d-1}{2}\Big)}\frac{k^{d-1}}{l^{d-1}}\Big]\Big\}\crcr
\beta(\mu_k)&=&-\frac{d\,\lambda_k}{(Z_kk^2+\mu_k)^2}\Big\{\de_tZ_k\Big[\frac{4}{3}\frac{k^3}{l}
   +\frac{\pi^{\frac{d-1}{2}}}{\Gamma_E\Big(\frac{d+3}{2}\Big)}\frac{k^{d+1}}{ l^{d-1}}\Big]
   +2Z_k\Big[2\frac{k^3}{l}
   +\frac{\pi^{\frac{d-1}{2}}}{\Gamma_E\Big(\frac{d+1}{2}\Big)}\frac{k^{d+1}}{l^{d-1}}\Big]\Big\}\,.
\eea

Already at this level, one realises that each $\beta$-function
does not have homogeneous scaling in $k$ and dimensions in $l$.
This feature clearly comes from the pattern of the convolution
of the interaction which is specific to TGFTs.

\subsection{$\varphi^4$-terms}
\label{app:tgftphi4}

The second order $(II)_W$ of \eqref{wettexpansionPHI4} will provide the $\beta$-function for $\lambda_k$, 
which completes the set of $\beta$-functions of the model. Defining,
$R'_k$ and $P'_k$ such that 
\begin{align}
&
    R_k(\textbf{p},\textbf{p'})=
R'_k(\textbf{p})\Theta(k^2-\sum_sp_s^2)\delta(\textbf{p}-\textbf{p}')
\,,\crcr
&
    P_k(\textbf{p},\textbf{p'})=P'_k(\textbf{p})\delta(\textbf{p}-\textbf{p}')\,,
\end{align}
the terms of interest take the form:
\begin{align}
\label{secondorder}
&(II)_W=\Tr[\de_tR_k\cdot(P_k)^{-1}\cdot F_k\cdot(P_k)^{-1}\cdot F_k\cdot(P_k)^{-1}]\crcr
&
=\int_{{D^*}^{\times 5}} d\textbf{p}d\textbf{p}'d\textbf{p}''d\textbf{q}d\textbf{q}'\;
   \de_tR'_k(\textbf{p})\Theta(k^2-\sum_sp_s^2)\delta(\textbf{p}-\textbf{p}')
   (P'_k)^{-1}(\textbf{p}')\delta(\textbf{p}'-\textbf{p}'')\crcr
& \times F_k(\textbf{p}'',\textbf{q})
   (P'_k)^{-1}(\textbf{q})\delta(\textbf{q}-\textbf{q}')
   F_k(\textbf{q}',\textbf{p})
   (P'_k)^{-1}(\textbf{p})\crcr
&=\int_{D^*} d\textbf{p}\;
   \de_tR'_k(\textbf{p})\Theta(k^2-\sum_sp_s^2)
   (P'_k)^{-1}(\textbf{p})
 \int_{D^*} d\textbf{q}\; F_k(\textbf{p},\textbf{q})
   (P'_k)^{-1}(\textbf{q})
   F_k(\textbf{q},\textbf{p})
   (P'_k)^{-1}(\textbf{p})\,.
\end{align}
We  focus on the intermediate convolution $F_k\cdot P_k^{-1}\cdot F_k$  which expands as: 
\begin{align*}
&(F_k\cdot P_k^{-1}\cdot F_k)(\textbf{p},\textbf{p})=\lambda_k^2
   \int_{D^*} d\textbf{q}F(\textbf{p},\textbf{q})(P'_k)^{-1}(\textbf{q})
   F_k(\textbf{q},\textbf{p})\crcr
&=\lambda_k^2\int_{D^*} dq_1\dots dq_d
   \biggl[\int_{\ca{D}} dm_1\, \varphi_{m_1p_2\dots p_d}\ov{\varphi}_{m_1q_2\dots q_d}\delta(p_1-q_1)\crcr
&+\int_{\ca{D}^{\times d-1}} dm_2\dots dm_d\, \varphi_{p_1m_2\dots m_d}\ov{\varphi}_{q_1m_2\dots m_d}
   \prod_{i=2}^d\delta(p_i-q_i)
+\Sym\biggr]\crcr
&\quad(P'_k)^{-1}(\textbf{q})
   \biggl[\int_{\ca{D}} dm'_1\, \varphi_{m'_1q_2\dots q_d}\ov{\varphi}_{m'_1p_2\dots p_d}\delta(p_1-q_1)\crcr
&+\int_{\ca{D}^{\times d-1}} dm'_2\dots dm'_d\, \varphi_{q_1m'_2\dots m'_d}\ov{\varphi}_{p_1m'_2\dots m'_d}
   \prod_{i=2}^d\delta(p_i-q_i)
   +\Sym\biggr]\,.
\end{align*}
At this level, the product of coloured symmetric terms generates a 
list of terms (among which cross terms) that we must all carefully analyse. 
First, we deal with the case when the product involves two terms of the
same colour, then we will treat the cross-coloured case. Below, we further specialise the study to the product of 
terms of colour 1 and, then on the cross term 1-2 in the above expansion. 
We refer to the first type of term as $(F_k\cdot P_k^{-1}\cdot F_k)(\textbf{p},\textbf{p})|_{1,1}$ and to the overall contribution after tracing over remaining indices as $(II)_{W}|_{1,1}$ (respectively, the symbol $|_{1,2}$ will stand for the cross term product of the colours 1 and 2). 
This evaluation is, of course, without loss of generality because 
one can quickly infer the result for all remaining products. All these contributions, at the end, 
must be summed.

We have  
\begin{align*}
&(F_k\cdot P_k^{-1}\cdot F_k)(\textbf{p},\textbf{p})|_{1,1}=\crcr
& 
\lambda_k^2\int_{D^*} dq_1\dots dq_d\;
   \int_{\ca{D}} dm_1\;\varphi_{m_1p_2\dots p_d}\ov{\varphi}_{m_1q_2\dots q_d}\delta(p_1-q_1)
   (P'_k)^{-1}(\textbf{q})\crcr
&\times\int_{\ca{D}^{\times d-1}} dm'_2\dots dm'_d\;\varphi_{q_1m'_2\dots m'_d}\ov{\varphi}_{p_1m'_2\dots m'_d}
   \prod_{i=2}^d\delta(p_i-q_i)\crcr
&+\lambda_k^2\int_{D^*} dq_1\dots dq_d\;
   \int_{\ca{D}^{\times d-1}} dm_2\dots dm_d\;\varphi_{p_1m_2\dots m_d}\ov{\varphi}_{q_1m_2\dots m_d}
   \delta(p_i-q_i)
   (P'_k)^{-1}(\textbf{q})\crcr
&\times \int_{\ca{D}} dm'_1\;\varphi_{m'_1q_2\dots q_d}\ov{\varphi}_{m'_1p_2\dots p_d}\delta(p_1-q_1)\crcr
&+\lambda_k^2\int_{D^*} dq_1\dots dq_d\;
   \int_{\ca{D}} dm_1\;\varphi_{m_1p_2\dots p_d}\ov{\varphi}_{m_1q_2\dots q_d}\delta(p_1-q_1)
   (P'_k)^{-1}(\textbf{q})\crcr
&\times\int_{\ca{D}} dm'_1\;\varphi_{m'_1q_2\dots q_d}\ov{\varphi}_{m'_1p_2\dots p_d}\delta(p_1-q_1)\crcr
&+\lambda_k^2\int_{D^*} dq_1\dots dq_d\;
   \int_{\ca{D}^{\times d-1}} dm_2\dots dm_d\;\varphi_{p_1m_2\dots m_d}\ov{\varphi}_{q_1m_2\dots m_d}
   \prod_{i=2}^d\delta(p_i-q_i)
   (P'_k)^{-1}(\textbf{q})\crcr
&\times\int_{\ca{D}^{\times d-1}} dm'_2\dots dm'_d\;\varphi_{q_1m'_2\dots m'_d}\ov{\varphi}_{p_1m'_2\dots m'_d}
   \prod_{i=2}^d\delta(p_i-q_i)\,.
\end{align*}
The first two terms, once that the $\delta$'s in $\textbf{q}$ are integrated out, become proportional to the product of two square modulus of the fields, thus they represent disconnected interactions. They can
be  discarded for the same reasons invoked above.  As a remainder, we get:
\bea
\label{FQuadro}
&&
(F_k\cdot P_k^{-1}\cdot F_k)(\textbf{p},\textbf{p})|_{1,1}\simeq \\
&&\frac{\lambda_k^2}{l}\int_{\ca{D}^{\times d-1}} dq_2\dots dq_d\;\int_{\ca{D}^{\times 2}} dm_1dm'_1
   \varphi_{m_1p_2\dots p_d}\ov{\varphi}_{m_1q_2\dots q_d}
   \varphi_{m'_1q_2\dots q_d}\ov{\varphi}_{m'_1p_2\dots p_d}
(P'_k)^{-1}(p_1,q_2,\dots,q_d)\crcr
&&+\frac{\lambda_k^2}{l^{d-1}}\int_{\ca{D}}dq_1\int_{\ca{D}^{\times 2d-2}} dm_2\dots dm_ddm'_2\dots dm'_d\;
   \varphi_{p_1m_2\dots m_d}\ov{\varphi}_{q_1m_2\dots m_d}
\varphi_{q_1m'_2\dots m'_d}\ov{\varphi}_{p_1m'_2\dots m'_d}
(P'_k)^{-1}(q_1,p_2,\dots,p_d)\,.
\nonumber
\eea
Then, plugging back \eqref{FQuadro} in $(II)_W$ and concentrating
on the contribution of this term, one finds:
\begin{align*}
&(II)_W|_{1,1}=\lambda_k^2\int_{D^*} 
     d\textbf{p}\, \Theta(k^2-\sum_sp_s^2)\textstyle{\frac{[\de_t Z_k(k^2-\sum_sp_s^2)+2k^2Z_k]}{(Z_kk^2+\mu_k)^2}}\crcr
&\quad\biggl\{\frac{1}{l}\int_{\ca{D}^{\times d-1}} dq_2\dots dq_d\;\int_{\ca{D}^{\times 2}} dm_1dm'_1\;
     \varphi_{m_1p_2\dots p_d}\ov{\varphi}_{m_1q_2\dots q_d}
     \varphi_{m'_1q_2\dots q_d}\ov{\varphi}_{m'_1p_2\dots p_d}\crcr
&\quad\biggl[Z_k(k^2-p_1^2-\Sigma_{i=2}^dq_i^2)\Theta(k^2-p_1^2-\Sigma_{i=2}^dq_i^2)
     +Z_k(p_1^2+\Sigma_{i=2}^dq_i^2)+\mu_k\biggr]^{-1}\crcr
&+\frac{1}{l^{d-1}}\int_{\ca{D}} dq_1\;\int_{\ca{D}^{\times 2d-2}} dm_2\dots dm_ddm'_2\dots dm'_d\;
   \varphi_{p_1m_2\dots m_d}\ov{\varphi}_{q_1m_2\dots m_d}
   \varphi_{q_1m'_2\dots m'_d}\ov{\varphi}_{p_1m'_2\dots m'_d}\crcr
&\quad\biggl[Z_k(k^2-q_1^2-\Sigma_{i=2}^dp_i^2)\Theta(k^2-q_1^2-\Sigma_{i=2}^dp_i^2)
     +Z_k(q_1^2+\Sigma_{i=2}^dp_i^2)+\mu_k\biggr]^{-1}\biggr\}\,.
\end{align*}
 With the same principle used for evaluation of the $\beta$-functions of $Z_k$ and $\mu_k$, any explicit dependence on the $2d$ momenta involved in the four fields in the spectral sums of \eqref{secondorder} must be discarded.
In other words, any term of the form $p_i^{\alpha} \varphi _{\dots p_i \dots}\bar\varphi_{\dots p_i \dots}\cdot( \varphi \bar\varphi) $ falls out of the truncation. 
After taking the limit 
(again we drop the symbol $\lim_{l\to 0}$), we expand the expression at  zeroth order and get: 
\begin{align*}
&(II)_W|_{1,1}\simeq
    \frac{\lambda_k^2}{l}\int_{\R^{2d}} dm_1dm'_1dp_2\dots dp_ddq_2\dots dq_d\;
    \varphi_{m_1p_2\dots p_d}\ov{\varphi}_{m_1q_2\dots q_d}
    \varphi_{m'_1q_2\dots q_d}\ov{\varphi}_{m'_1p_2\dots p_d}\crcr
   &\times\int_{\R} dp_1\;
    \textstyle{\frac{[\de_t Z_k(k^2-p_1^2)+2k^2Z_k]}{(Z_kk^2+\mu_k)^2}}
    \textstyle{\frac{\theta(k^2-p_1^2)}{Z_k(k^2-p_1^2)\theta(k^2-p_1^2)+Z_kp_1^2+\mu_k}}\crcr
&+\frac{\lambda_k^2}{l^{d-1}}\int_{\R^{2d}} dp_1dq_1dm_2\dots dm_ddm'_2\dots dm'_d\;
    \varphi_{p_1m_2\dots m_d}\ov{\varphi}_{q_1m_2\dots m_d}
    \varphi_{q_1m'_2\dots m'_d}\ov{\varphi}_{p_1m'_2\dots m'_d}\crcr
  &\times\int_{\R^{d-1}} dp_2\dots dp_d\;
    \textstyle{\frac{[\de_t Z_k(k^2-\Sigma_{i=2}^dp_i^2)+2k^2Z_k]}{(Z_kk^2+\mu_k)^2}}
\textstyle{\frac{\theta(k^2-\Sigma_{i=2}^dp_i^2)}
    {Z_k(k^2-\Sigma_{i=2}^dp_i^2)\theta(k^2-\Sigma_{i=2}^dp_i^2)+Z_k(\Sigma_{i=2}^dp_i^2)+\mu_k}}\,.
\end{align*}
The $\theta$'s turn out to be redundant in both the terms and we can simplify their contributions. Call 
$\cV_i$ the vertex of colour $i$ of the effective interaction. 
Rather than using the explicit form of that vertex,
we will simply use $\cV_i$ in the following, when no confusion might arise.

We split the previous terms in two pieces:
\begin{align*}
&(II)'_W|_{1,1}=\crcr
& {\textstyle\frac{1}{l}\frac{\lambda_k^2k^2(2+\de_t) Z_k}{(Z_kk^2+\mu_k)^3}} 
     \int dq_2\dots dq_ddp_2\dots dp_ddm_1dm'_1\;
     \varphi_{m_1p_2\dots p_d}\ov{\varphi}_{m_1q_2\dots q_d}
     \varphi_{m'_1q_2\dots q_d}\ov{\varphi}_{m'_1p_2\dots p_d} 
\int dp_1\;\theta(k^2-p_1^2) \crcr
&-{\textstyle{\frac{1}{l}\frac{\lambda_k^2\de_t Z_k}{(Z_kk^2+\mu_k)^3}}}
     \int dq_2\dots dq_ddp_2\dots dp_ddm_1dm'_1\;
     \varphi_{m_1p_2\dots p_d}\ov{\varphi}_{m_1q_2\dots q_d}
     \varphi_{m'_1q_2\dots q_d}\ov{\varphi}_{m'_1p_2\dots p_d}
\int dp_1\;p_1^2\theta(k^2-p_1^2)\crcr
&=2\frac{\lambda_k^2 k^3}{l}\biggl[\frac{(2+\de_t) Z_k}{(Z_kk^2+\mu_k)^3}
    -\frac{1}{3}\frac{\de_t Z_k}{(Z_kk^2+\mu_k)^3}\biggr] \cV_1 
= \frac{2 \lambda_k^2 k^3}{l \, (Z_kk^2+\mu_k)^3}
 \Big[ 2Z_k +\frac{2}{3}\de_t Z_k  \Big] \cV_1\,.
\end{align*}
The second integral can be computed as: 
\begin{align*}
&(II)''_W|_{1,1}=\crcr
& 
{\textstyle{\frac{1}{l^{d-1}}\frac{ \lambda_k^2 k^2(2+\de_t) Z_k}{(Z_kk^2+\mu_k)^3} } }
    \int dp_1dq_1dm_2\dots dm_ddm'_2\dots dm'_d\;
    \varphi_{p_1m_2\dots m_d}\ov{\varphi}_{q_1m_2\dots m_d}
    \varphi_{q_1m'_2\dots m'_d}\ov{\varphi}_{p_1m'_2\dots m'_d}\crcr
& \times  \int dp_2\dots dp_d\;\theta(k^2-\Sigma_{i=2}^dp_i^2)\crcr
&-{\textstyle{\frac{1}{l^{d-1}}\frac{\lambda_k^2\de_t Z_k}{(Z_kk^2+\mu_k)^3}}}
    \int dp_1dq_1dm_2\dots dm_ddm'_2\dots dm'_d\;
    \varphi_{p_1m_2\dots m_d}\ov{\varphi}_{q_1m_2\dots m_d}
    \varphi_{q_1m'_2\dots m'_d}\ov{\varphi}_{p_1m'_2\dots m'_d}\crcr
&\times \int dp_2\dots dp_d\;\Big(\Sigma_{i=2}^dp_i^2\Big)\theta(k^2-\Sigma_{i=2}^dp_i^2)\displaybreak\crcr
&= \int dp_1dq_1dm_2\dots dm_ddm'_2\dots dm'_d\;
    \varphi_{p_1m_2\dots m_d}\ov{\varphi}_{q_1m_2\dots m_d}
    \varphi_{q_1m'_2\dots m'_d}\ov{\varphi}_{p_1m'_2\dots m'_d}\crcr
&\times\biggl[{\textstyle{\frac{1}{l^{d-1}}\frac{ \lambda_k^2 k^2(2+\de_t) Z_k}{(Z_kk^2+\mu_k)^3} } }
    \int d\Omega_{d-1}\int_0^k dr r^{d-2}
    -{\textstyle{\frac{1}{l^{d-1}}\frac{\lambda_k^2\de_t Z_k}{(Z_kk^2+\mu_k)^3}}}
    \int d\Omega_{d-1}\int_0^k dr r^d\biggr]\crcr
&=\frac{\lambda_k^2}{l^{d-1} (Z_kk^2+\mu_k)^3}
    \biggl[\frac{2k^{d+1}(2+\de_t)Z_k\pi^{\frac{d-1}{2}}}{(d-1)\Gamma_E\Big(\frac{d-1}{2}\Big)}
    -\frac{2\pi^{\frac{d-1}{2}}k^{d+1}\de_tZ_k}{(d+1)\Gamma_E\Big(\frac{d-1}{2}\Big)}\biggl]\cV_1 \crcr
&=\frac{\lambda_k^2k^{d+1}\pi^{\frac{d-1}{2}}}{l^{d-1} (Z_kk^2+\mu_k)^3}
    \biggl[\frac{\de_tZ_k}{\Gamma_E\Big(\frac{d+3}{2}\Big)}
    +\frac{2Z_k}{\Gamma_E\Big(\frac{d+1}{2}\Big)}\biggr]\cV_1 \,. 
\end{align*}
A simple check of the dimensions of these terms and
the dimension of the interaction term of the effective action can be 
given as 
\begin{align*}
[(II)'_W]=[(II)''_W]=2[\lambda]-4+2d+4[\varphi]\,,
\end{align*}
which, considering that $[\varphi]=-\frac{d+2}{2}$, fixes $[\lambda]=4$ as expected.

Let us now focus on the cross term given by 
the product of the contribution of colour 1 and 2:
\begin{align*}
&(II)_W|_{1,2}=\lambda_k^2\int_{{D^*}^{\times 2}} d\textbf{p}d\textbf{j}\;
   \textstyle{\frac{\Theta(k^2-\sum_sp_s^2)}{(Z_kk^2 + \mu_k)^2}
   \frac{\de_tZ_k(k^2-\sum_sp_s^2)+2k^2Z_k}
   {\Theta(k^2-\sum_sj_s^2)Z_k(k^2-\sum_sj_s^2)+Z_k\sum_sj_s^2+\mu_k}}\crcr
&\biggl[
   \int_{\ca{D}^{\times 2}} dm_1dn_2\;
   \varphi_{m_1j_2\dots j_d}\ov{\varphi}_{m_1p_2\dots p_d}
   \varphi_{p_1n_2p_3\dots p_d}\ov{\varphi}_{j_1n_2j_3\dots j_d}
   \delta(p_1-j_1)\delta(p_2-j_2)\crcr
&+\int_{\ca{D}^{2d-2}} dm_2\dots dm_ddn_1dn_3\dots dn_d\;
   \varphi_{j_1m_2\dots m_d}\ov{\varphi}_{p_1m_2\dots m_d}
   \varphi_{n_1p_2n_3\dots n_d}\ov{\varphi}_{n_1j_2n_3\dots n_d}\crcr
&\times\delta(p_1-j_1)\delta(p_2-j_2)\prod_{i=3}^d\delta^2(p_i-j_i)\crcr
&+\int_{D^*} dm_1dn_1dn_3\dots dn_d\;
   \varphi_{m_1j_2\dots j_d}\ov{\varphi}_{m_1p_2\dots p_d}
   \varphi_{n_1p_2n_3\dots n_d}\ov{\varphi}_{n_1j_2n_3\dots n_d}
   \delta^2(p_1-j_1)\prod_{i=3}^d\delta(p_i-j_i)\crcr
&+\int_{D^*} dm_2\dots dm_ddn_2\;
   \varphi_{j_1m_2\dots m_d}\ov{\varphi}_{p_1m_2\dots m_d}
   \varphi_{p_1n_2p_3\dots p_d}\ov{\varphi}_{j_1n_2j_3\dots j_d}
   \delta^2(p_2-j_2)\prod_{i=3}^d\delta(p_i-j_i)\biggr]\,.
\end{align*}
If we integrate the deltas over the $j$ variables, the second term is again a disconnected 4-point function that we neglect.  In rank $d>3$, the first term falls out of the truncation: it generates a ''matrix-like'' convolution with two momenta distinguished from the other $d-2$ labels.
However at the  boundary value $d=3$, it will contribute to the flow.
We find: 
\begin{align*}
&(II)_W|_{1,2}=\delta_{d,3}\; 
{\textstyle{\frac{\lambda^2_k}{(Z_kk^2+\mu_k)^2}}}
\int dp_1dp_2 dp_3 dm_1 dn_2 dj_3\, 
  \varphi_{m_1p_2j_3}\ov{\varphi}_{m_1p_2 p_3}
    \varphi_{p_1n_2p_3}\ov{\varphi}_{p_1n_2j_3} 
\crcr
&\times\frac{\Theta(k^2-\sum_sp_s^2)[\de_tZ_k(k^2-\sum_sp_s^2)+2k^2Z_k]}
   {\Theta(k^2-p_1^2-p_2^2-j_3^2)Z_k
   (k^2-p_1^2-p_2^2-j_3^2)+Z_k(p_1^2+p_2^2
+j_3^2)+\mu_k}\crcr
\crcr
& +
{\textstyle{\frac{\lambda^2_k}{(Z_kk^2+\mu_k)^2}\frac{1}{l}
    \int_{D^*\times\ca{D}}}}{\scriptstyle dp_1\dots dp_ddj_2dm_1dn_1dn_3\dots dn_d}\;
    \varphi_{m_1j_2p_3\dots p_d}\ov{\varphi}_{m_1p_2\dots p_d}
    \varphi_{n_1p_2n_3\dots n_d}\ov{\varphi}_{n_1j_2n_3\dots n_d}\crcr
&\times\frac{\Theta(k^2-\sum_sp_s^2)[\de_tZ_k(k^2-\sum_sp_s^2)+2k^2Z_k]}
   {\Theta(k^2-p_1^2-j_2^2-\Sigma_{i=3}^dp_i^2)Z_k
   (k^2-p_1^2-j_2^2-\Sigma_{i=3}^dp_i^2)+Z_k(p_1^2+j_2^2+\Sigma_{i=3}^dp_i^2)+\mu_k}\crcr
&+{\textstyle{\frac{\lambda^2_k}{(Z_kk^2+\mu_k)^2}\frac{1}{l}
    \int_{D^*\times\ca{D}}}}{\scriptstyle dp_1\dots dp_ddj_1dm_2\dots dm_ddn_2}\;
    \varphi_{j_1m_2\dots m_d}\ov{\varphi}_{p_1m_2\dots m_d}
    \varphi_{p_1n_2p_3\dots p_d}\ov{\varphi}_{j_1n_2p_3\dots p_d}\crcr
&\times\frac{\Theta(k^2-\sum_sp_s^2)[\de_tZ_k(k^2-\sum_sp_s^2)+2k^2Z_k]}
   {\Theta(k^2-j_1^2-\Sigma_{i=2}^dp_i^2)Z_k
   (k^2-j_1^2-\Sigma_{i=2}^dp_i^2)+Z_k(j_1^2+\Sigma_{i=2}^dp_i^2)+\mu_k}\,.
\end{align*}
In the continuum limit, the previous integrals can be evaluated at 0-momentum truncation and the $\Theta$ in the denominator,
put to 1. 
One realises that the first term is proportional to $\delta_{d,3}\cV_3$, 
while the second and third terms are (1,2)-coloured symmetric contributions and are proportional to $\cV_2$ and $\cV_1$, respectively. 
Casting away the $p_i^{2}\varphi^4_{p_i}$-terms, one infers   
\bea
(II)_W|_{1,2}&\simeq&
  \delta_{d,3}
\frac{\lambda^2_k k^2(2+ \de_t)Z_k}{ (Z_kk^2+\mu_k)^3 }\cV_3 \crcr
&+& 
\frac{\lambda_k^2}{(Z_kk^2+\mu_k)^3}\frac{1}{l} \cV_2
\int dp_1\theta(k^2-p_1^2)[\de_tZ_k(k^2-p_1^2)+2k^2Z_k]
+ {\textrm{sym}}\{1\to 2\}.
\eea
Performing the integrals over the external momenta:
\bea
(II)_W|_{1,2}=\frac{\lambda^2_k k^2}{(Z_kk^2+\mu_k)^3 }
\Big\{ 
  \delta_{d,3} (2+ \de_t)Z_k \cV_3
+  
\frac{k}{ l}
    \biggl[-\frac{2}{3}\de_tZ_k+2(2+\de_t)Z_k\biggr]\Big(\cV_2 + \cV_1\Big) \Big\}.
\label{II2ij}
\eea
We are in position to sum all contributions. Taking into account 
the colour symmetry of the vertices, the coefficients obtained from $(II)_W|_{i,i}$ contributes once for each colour $i$, while the terms coming from the cross terms, i.e. $(II)_W|_{i,j\ne i}$, will appear once for each couple of colours $(i,j)$, $j\ne i$.
Thus the later terms gain a factor $2(d-1)$. Especially, the term $\delta_{d,3}\cV_3$
in \eqref{II2ij} and the like, at $d=3$, acquires a factor of 2. Performing these operations,  the $\beta$-function for $\lambda_k$ reads:
\bea
\beta(\lambda_k)&=&\frac{2\lambda_k^2}{(Z_kk^2+\mu_k)^3}
    \Big\{\de_tZ_k\Big[
\frac{\pi^{\frac{d-1}{2}}}{\Gamma_E\Big(\frac{d+3}{2}\Big)}\frac{k^{d+1}}{l^{d-1}}
+ 
\frac{4(2d-1)}{3}\frac{k^3}{l} + 2\delta_{d,3} k^2
       \Big]\crcr
&+&
2Z_k\Big[\frac{\pi^{\frac{d-1}{2}}}{\Gamma_E\Big(\frac{d+1}{2}\Big)}\frac{k^{d+1}}{l^{d-1}}
+ 2(2d-1)\frac{k^3}{l}
   + 2\delta_{d,3} k^2    \Big]\Big\}\,.
\eea

\noindent{\bf Dimensionful $\beta$-functions.}
We write the full set of dimensionful $\beta$-functions for the model as: 
\bea
&&\hspace{-0.8cm}
\left\{
\begin{aligned}
\beta(Z_k)&=\frac{\lambda_k}{(Z_kk^2+\mu_k)^2}\Big\{\de_tZ_k\Big[2(d-1)\frac{k}{l}
   -\frac{\pi^{\frac{d-1}{2}}}{\Gamma_E\Big(\frac{d+1}{2}\Big)}\frac{k^{d-1}}{l^{d-1}}\Big]
   +2Z_k\Big[(d-1)\frac{k}{l}
   +\frac{\pi^{\frac{d-1}{2}}}{\Gamma_E\Big(\frac{d-1}{2}\Big)}\frac{k^{d-1}}{l^{d-1}}\Big]\Big\}\crcr
\beta(\mu_k)&=-\frac{d\,\lambda_k}{(Z_kk^2+\mu_k)^2}\Big\{\de_tZ_k\Big[\frac{4}{3}\frac{k^3}{l}
   +\frac{\pi^{\frac{d-1}{2}}}{\Gamma_E\Big(\frac{d+3}{2}\Big)}\frac{k^{d+1}}{l^{d+1}}\Big]
   +2Z_k\Big[2\frac{k^3}{l}
   +\frac{\pi^{\frac{d-1}{2}}}{\Gamma_E\Big(\frac{d+1}{2}\Big)}\frac{k^{d+1}}{l^{d-1}}\Big]\Big\}\\
\beta(\lambda_k)&=\frac{2\lambda_k^2}{(Z_kk^2+\mu_k)^3}
    \Big\{\de_tZ_k\Big[
\frac{\pi^{\frac{d-1}{2}}}{\Gamma_E\Big(\frac{d+3}{2}\Big)}\frac{k^{d+1}}{l^{d-1}}
+ 
\frac{4(2d-1)}{3}\frac{k^3}{l} + 2\delta_{d,3} k^2
       \Big]\crcr
&+
2Z_k\Big[\frac{\pi^{\frac{d-1}{2}}}{\Gamma_E\Big(\frac{d+1}{2}\Big)}\frac{k^{d+1}}{l^{d-1}}
+ 2(2d-1)\frac{k^3}{l}
   + 2\delta_{d,3} k^2    \Big]\Big\}
\end{aligned}
\right. \crcr
&&
\eea
which is reported in section \ref{betastgft}, \eqref{betadimensionful}.

\section{ Evaluation of $\beta$-functions in the gauge invariant case}
\label{app:gau}

The computation of the dimensionful $\beta$-functions for the gauge projected model follows roughly the same steps of the calculations of 
the model without constraints. However, due to the presence of the extra delta's of the gauge projection, the analysis requires, at some point, a different technique. 
In this appendix, we provide details of the procedure for obtaining the system of the dimensionful RG equations, namely \eqref{dimensfulGaug}
of section \ref{betfuncGaug}, and underline the differences
with the previous calculus.

We start by expanding equation \eqref{wettgaugeTr} of section \ref{sec5.2} and focus, first on the $\varphi^2$-terms and
then calculate higher order terms.

\subsection{$\varphi^2$-terms}

Referring to the conventions introduced at the beginning of section \ref{sec5.2}, say \eqref{Rgaug}--\eqref{Pgaug}, for the scaling of the kinetic term, we have:
\begin{align}\label{itm}
&(I^g)_W=-\Tr[\de_tR_k\cdot(P_k)^{-1}\cdot F_k\cdot(P_k)^{-1}]\crcr
&=-\lambda_k\int_{{D^*}}d\textbf{p}\;
    \Theta(k^2-\Sigma_sp_s^2)\frac{[\de_tZ_k(k^2-\Sigma_sp_s^2)+2k^2Z_k]}{(Z_kk^2+\mu_k)^2}
    \frac{\delta(\Sigma p)}{\delta^2(\Sigma p)}\crcr
&\times\biggl[\frac{1}{l^{d-1}}\int_{\ca{D}^{\times d-1}}dm_2\dots dm_d\;
    |\varphi_{p_1m_2\dots m_d}|^2\delta^2(\Sigma p)\delta^2(p_1+m_2+\dots+m_d)\crcr
&+\frac{1}{l}\int_{\ca{D}}dm_1\;|\varphi_{m_1p_2\dots p_d}|^2
    \delta^2(\Sigma p)\delta^2(m_1+p_2+\dots+p_d)+\Sym\biggr]\,. 
\end{align}
In the same perspective, the square delta's can be reduced as $\delta^2(p)=\delta(p)\delta(0)=\frac{1}{l}\delta(p)$. The second integral in the above expression can be directly computed by integrating over $p_1$ the $\delta(\sum p)$ as
\bea\label{i'g}
(I^g)'_W&=&-\frac{\lambda_k}{l^2}\int_{{D^*}}d\textbf{p}\;
    \Theta(k^2-\Sigma_sp_s^2)\frac{[\de_tZ_k(k^2-\Sigma_sp_s^2)+2k^2Z_k]}{(Z_kk^2+\mu_k)^2}
   \delta(\Sigma p)\crcr
&\times& \int_{\ca{D}}dm_1\;|\varphi_{m_1p_2\dots p_d}|^2
  \delta(m_1+p_2+\dots+p_d) +\Sym\cr\cr
&=&  -\frac{\lambda_k}{l^2(Z_kk^2+\mu_k)^2}
\int_{D^*}dm_1dp_2 \dots dp_d\;\;|\varphi_{m_1p_2\dots p_d}|^2
 \delta(m_1+p_2+\dots+p_d) \crcr
&& \times 
\int_{{\ca{D}}}dp_1\;
    \Theta(k^2-\Sigma_sp_s^2)
[\de_tZ_k(k^2-\Sigma_sp_s^2)+2k^2Z_k]
   \delta(\Sigma p) 
+ \Sym \cr\cr
&=& 
 -\frac{\lambda_k}{l^2 (Z_kk^2+\mu_k)^2}
\int_{D^*}d\textbf{p} \, |\varphi_{p_1p_2\dots p_d}|^2\, 
  \delta(p_1+p_2+\dots+p_d) 
\Big[ d k^2(2+\de_t)Z_k - d \de_t Z_k\Sigma_{s=1}^{d}p_s^2 \Big]\cr\cr
&=& 
 -\frac{d\lambda_k}{l^2 (Z_kk^2+\mu_k)^2}
\int_{D^*}d\textbf{p} \, |\varphi_{p_1p_2\dots p_d}|^2\, 
  \delta(p_1+p_2+\dots+p_d) 
\Big[  2 k^2Z_k   + \de_tZ_k[  k^2 - \Sigma_{s=1}^{d}p_s^2] \Big]. 
\eea

We discuss now the first term in the brackets in \eqref{itm}
that we denote
\bea\label{i''g}
&&(I^g)''_W=-\frac{\lambda_k}{l^d(Z_kk^2+\mu_k)^2}
    \int_{D^*} dp_1dm_2\dots dm_d\;
    |\varphi_{p_1m_2\dots m_d}|^2\delta(p_1+m_2+\dots+m_d)\crcr
&&
\times \int_{\ca{D}^{\times d-1}}dp_2 \dots dp_d\;
    \Theta(k^2-\Sigma_sp_s^2)
[\de_tZ_k(k^2-\Sigma_sp_s^2)+2k^2Z_k]\delta(\sum p)
+\Sym 
\eea
Because of the combinatorial pattern chosen for the interaction, the case $d=3$ represents again a special situation that
we deal with by direct evaluation. We integrate over the third variable, imposing the constraint $p_3=-(p_1+p_2)$. The resulting domain of integration of $p_2$ is known, in the continuous limit, as the $\theta$ distribution is non-zero when $-2p_2^2-2p_2p_1+(k^2-2p_1^2)\geq 0$. The boundary of this inequality, solved in $p_2$, is given by the roots 
\bea
\label{solutionsp2}
p_2^{\pm}=\frac{1}{2}\Big(-p_1\pm\sqrt{2k^2-3p_1^2}\Big)\,. 
\eea
The non-zero values of the Heaviside distribution hold when $p_2\in [p_2^-,p_2^+]$. There is still a residual constraint over $p_1$ which has to be imposed in order to keep real the square root appearing in \eqref{solutionsp2}, that is,
$3p_1^2\leq 2k^2$. Thus, \eqref{i''g} becomes
\begin{align}
&(I^g)''_{W;d=3}=-\frac{\lambda_k}{l^3(Z_kk^2+\mu_k)^2}\int dp_1dm_2dm_3\;
 |\varphi_{p_1m_2m_3}|^2\delta(p_1+m_2+m_3)\crcr
&\times\theta(2k^2-3p_1^2)\int_{\frac{1}{2}(-p_1-\sqrt{2k^2-3p_1^2})}^{\frac{1}{2}(-p_1+\sqrt{2k^2-3p_1^2})}dp_2\;
      \{\de_tZ_k[k^2-2(p_2^2+p_1^2+p_2p_1)]+2k^2Z_k\}\crcr
&+\Sym\crcr
&=-\frac{\lambda_k}{l^3(Z_kk^2+\mu_k)^2}\int dp_1dm_2dm_3\;
   |\varphi_{p_1m_2m_3}|^2\delta(p_1+m_2+m_3)\theta(2k^2-3p_1^2)\crcr
&\times\Big\{k^2\sqrt{2k^2-3p_1^2}(2+\de_t)Z_k-\frac{3}{2}\sqrt{2k^2-3p_1^2}\de_tZ_kp_1^2
    -\frac{1}{6}(2k^2-3p_1^2)^{3/2}\de_tZ_k\Big\}\crcr
&+\Sym\,.
\end{align}
Expanding the last result up to the third order in momenta, one obtains
\bea
&&
(I^g)''_{W;d=3}\simeq
-\frac{\lambda_k}{l^3(Z_kk^2+\mu_k)^2}
\int dp_1dm_2dm_3\;
    |\varphi_{p_1m_2m_3}|^2\delta(p_1+m_2+m_3)\cr\cr
&&\times\Big[k^3\Big(\sqrt{2}-\frac{\sqrt{8}}{6}\Big)\de_tZ_k+2\sqrt{2}k^3Z_k
   -\frac{3}{\sqrt{2}}k(1+\de_t)Z_kp_1^2\Big]+\Sym \cr\cr
&&\simeq 
-\frac{\lambda_k}{l^3(Z_kk^2+\mu_k)^2}
\int_{D^*} d\textbf{p}\;
    |\varphi_{p_1p_2p_3}|^2\delta(p_1+p_2+p_3)
\Big[2\sqrt{2}dk^3(\frac{1}{3}\de_t+1)Z_k
   -\frac{3}{\sqrt{2}}k(1+\de_t)Z_k(\sum_{s=1}^d p_s^2)\Big],
\eea 
where in the last line we include the symmetry factors. 
From this point, and combining it with \eqref{i'g} restricted
at $d=3$, 
 we write the $\beta$-functions for the couplings $\mu_k$ and $Z_k$ as:
\begin{align}\label{betaGZmud3}
\beta_{d=3}(Z_k)&=\frac{\lambda_k}{(Z_kk^2+\mu_k)^2}\Big[\frac{3}{\sqrt{2}}\frac{k}{l^3}(1+\de_t)Z_k
  +\frac{3}{l^2}\de_tZ_k\Big]\,;\crcr
\beta_{d=3}(\mu_k)&=-\frac{3\lambda_k}{(Z_kk^2+\mu_k)^2}\Big[2\sqrt{2}\frac{k^3}{l^3}\Big(1
   +\frac{1}{3}\de_t\Big)Z_k+\frac{k^2}{l^2}(2+\de_t)Z_k\Big]\,.
\end{align}
 
At rank $d\geq 4$, the term \eqref{i''g} has  more integrations to perform and becomes simpler if expressed in spherical coordinates. Considering that the coordinate $p_1$ is convoluted with the field, we will change basis from $(p_2,\dots,p_d)$ to $(r,\Omega_{d-1})$. The $\delta(\sum p)$ defines the hyperplane orthogonal to a vector $\textbf{N}$ of norm $\norm{\textbf{N}}=\sqrt{d}$ and components (in Cartesian coordinates) $\textbf{N}=(1,1,\dots,1)$. We will call $\textbf{n}$ the projection of this vector on the subspace orthogonal to $p_1$ and $\textbf{P}$ the generic vector on this subspace. In this setting the Dirac delta function becomes 
\bea
\delta(p_1+\langle\textbf{P},\textbf{n}\rangle)=\delta(p_1+r\sqrt{d-1}\cos\vartheta)=
   \frac{\delta\Big(\frac{p_1}{r\sqrt{d-1}}+\cos\vartheta\Big)}{r\sqrt{d-1}}\,,
\eea
where $\vartheta$ represents the angle between $\textbf{P}$ and $\textbf{n}$. Considering that the scalar product, as the rest of the integrand, is rotational invariant on the $(d-1)$ dimensional space, we can set $\vartheta$ to be one of the angles appearing in the spherical measure. After the change of coordinates, equation \eqref{i''g} reads
\bea
\label{PhiQuadroGaugeSpherical}
&&(I^g)''_{W;d>3}=-\frac{\lambda_k}{l^d(Z_kk^2+\mu_k)^2}\int dp_1dm_2\dots dm_d \int dr\, 
    d\Omega_{d-2}\;\int_0^{\pi}d\vartheta\;r^{d-2}\sin^{d-3}\vartheta
    {\textstyle{\frac{\delta(\frac{p_1}{r\sqrt{d-1}}+\cos\vartheta)}{r\sqrt{d-1}}}}\crcr
&\times&|\varphi_{p_1m_2\dots m_d}|^2\delta(p_1+m_2+\dots+m_d)
    \theta(k^2-p_1^2-r^2)[\de_tZ_k(k^2-p_1^2-r^2)+2k^2Z_k]\crcr
&+&\Sym\,.
\eea
We focus on the integral over $\vartheta$ and change variable from $\vartheta$ to $X=\cos\vartheta$ and get,  for $d>3$,
\beq
\label{IntegraleCoseno}
\int_0^{\pi}d\vartheta\;\sin^{d-3}\vartheta\, \delta\Big(\frac{p_1}{r\sqrt{d-1}}+\cos\vartheta\Big)
=\int_{-1}^1dX\;(1-X^2)^{\frac{d-4}{2}}\delta\Big(\frac{p_1}{r\sqrt{d-1}}+X\Big)
=\Big[1-\frac{p_1^2}{r^2(d-1)}\Big]^{\frac{d-4}{2}}\,.
\eeq
Substituting \eqref{IntegraleCoseno} in \eqref{PhiQuadroGaugeSpherical}, we get:
 \bea
&&(I^g)''_{W;d>3}=-\frac{\lambda_k}{l^d(Z_kk^2+\mu_k)^2}\frac{\Omega_{d-2}}{\sqrt{d-1}}
    \int dp_1dm_2\dots dm_d\;|\varphi_{p_1m_2\dots m_d}|^2\delta(p_1+m_2+\dots+m_d)\crcr
&&\times \theta(k^2-p_1^2)\Bigg[ \theta(d-5)\int_0^{\sqrt{k^2-p_1^2}}dr\;r^{d-3}\Big[1-\frac{p_1^2}{(d-1)r^2}\Big]^{\frac{d}{2}-2}
    [\de_tZ_k(k^2-p_1^2-r^2)+2k^2Z_k]\crcr
&&
+ \delta_{d,4}\int_0^{\sqrt{k^2-p_1^2}}dr\;r\,
    [\de_tZ_k(k^2-p_1^2-r^2)+2k^2Z_k]\Bigg] 
+\Sym \,.
\eea 
%
Expanding the result of the integral over $\vartheta$ at the second order in $p_1$, we obtain an integral over $r$ of the form: 
\bea
\label{expr:interm:gauge0}
&&(I^g)''_{W;d>3}\simeq-\frac{\lambda_k}{l^d(Z_kk^2+\mu_k)^2}\frac{\Omega_{d-2}}{\sqrt{d-1}}
    \int dp_1dm_2\dots dm_d\;|\varphi_{p_1m_2\dots m_d}|^2\delta(p_1+m_2+\dots+m_d)\crcr
&&\times \theta(k^2-p_1^2)\Bigg[ \theta(d-5)\int_0^{\sqrt{k^2-p_1^2}}dr\;r^{d-3}\Big[1-\frac{d-4}{2(d-1)r^2}p_1^2\Big]
    [\de_tZ_k(k^2-p_1^2-r^2)+2k^2Z_k]\crcr
&&
+   \delta_{d,4}\int_0^{\sqrt{k^2-p_1^2}}dr\;r\,
    [\de_tZ_k(k^2-p_1^2-r^2)+2k^2Z_k]  \Bigg] 
+\Sym\,. 
\eea
Computing the last integral and expanding the result, we expand the r.h.s. of \eqref{expr:interm:gauge0} to the second order in the momenta convoluted with the fields and this yields: 
\bea
\label{expr:interm:gauge}
&&(I^g)''_{W;d>3}\simeq -\frac{\lambda_k}{l^d(Z_kk^2+\mu_k)^2}\frac{\Omega_{d-2}}{\sqrt{d-1}}
    \int dp_1dm_2\dots dm_d\;|\varphi_{p_1m_2\dots m_d}|^2\delta(p_1+m_2+\dots+m_d)\crcr
&&\times \theta(k^2-p_1^2) \Bigg\{ 
\theta(d-5) \Big[ \frac{2 k^{d}  }{ d-2 }Z_k
+ 
\frac{ 2 k^{d} }{ d(d-2) }\de_t Z_k  
-p_1^2 k^{d-2}\Big[ \frac{d}{(d-1)} Z_k  + \frac{d\,\de_tZ_k}{(d-1)(d-2)} \Big] 
   \Big]\crcr
&&
+ \delta_{d,4}\Big[ \frac12[\frac12\de_tZ_k+2Z_k ]k^4
-p_1^2k^2\Big[\frac12\de_tZ_k+Z_k  \Big] 
\Big] 
\Bigg\} 
+\Sym\,. 
\eea
 We sum \eqref{i'g} and \eqref{expr:interm:gauge} and write at rank $d=4$,
\bea
&&
(I^g)_{W;d=4} \simeq
-\frac{\lambda_k}{(Z_kk^2+\mu_k)^2}
 \int d\textbf{p}\;|\varphi_{p_1p_2\dots p_4}|^2\delta(\sum_sp_s)
\Bigg\{ 
   \crcr
&& \frac{2\pi}{l^4 \sqrt{3}} \Big[
 2[\frac12\de_tZ_k+2Z_k ]k^4
-k^2\Big[\frac12\de_tZ_k+Z_k  \Big] (\sum_sp_s^2)
\Big] 
+ \frac{4}{l^2 }
\Big[  k^2(2+\de_t)Z_k -\de_t Z_k(\sum_{s=1}^{d}p_s^2) \Big]
\Bigg\} 
\eea

and at rank $d>4$, summing  \eqref{i'g} and \eqref{expr:interm:gauge}
gives 
\bea
(I^g)_W&\simeq&
\frac{\lambda_k}{(Z_kk^2+\mu_k)^2}
\int_{D^*}d\textbf{p} \, |\varphi_{p_1p_2\dots p_d}|^2\, 
  \delta(\sum_sp_s)  \Bigg\{ \crcr
&&
-\frac{1}{l^d}\frac{\Omega_{d-2}}{\sqrt{d-1}}
 \Big\{dk^d\Big[\frac{(2+\de_t)Z_k}{d-2}-\frac{\de_tZ_k}{d}\Big]
    -k^{d-2}\Big[ \frac{d\,\de_tZ_k}{(d-1)(d-2)} 
 +\frac{d\,Z_k}{d-1}\Big](\sum_sp_s^2)\Big\}\crcr
&&-\frac{d}{l^2} 
\Big[  k^2(2+\de_t)Z_k -\de_t Z_k(\sum_{s=1}^{d}p_s^2 )\Big]
\Bigg\}\,.
\eea
Hence, we write the $\beta$-functions for the couplings $\mu_k$ and $Z_k$ 
at rank $d=4$, as 
\bea\label{betaGZmud4}
\beta_{d=4}(Z_k)&=&\frac{\lambda_k}{(Z_kk^2+\mu_k)^2}
\Big\{  \frac{2\pi}{\sqrt{3}} \Big[\frac12\de_tZ_k+Z_k  \Big] \frac{k^2}{l^4}
+ \de_t Z_k \frac{4}{l^2 }
\Big\} ;\crcr
\beta_{d=4}(\mu_k)&=&
-\frac{4\lambda_k}{(Z_kk^2+\mu_k)^2}
\Big\{ \frac{\pi}{\sqrt{3}} \Big[
 \frac12\de_tZ_k+2Z_k \Big] \frac{ k^4  }{l^4 } 
+ (2+\de_t)Z_k \frac{ k^2}{l^2 } 
\Big\},
\eea
and for $d>4$, as 
\bea\label{betaGZmu}
\beta_{d>4}(Z_k)&=&\frac{d\lambda_k}{(Z_kk^2+\mu_k)^2}\Big\{\de_tZ_k\Big[
    \frac{\pi^{\frac{d-2}{2}}}{(d-1)^{\frac{3}{2}}\Gamma_E\Big(\frac{d}{2}\Big)}\frac{k^{d-2}}{l^d}
    +\frac{1}{l^2}\Big]+\frac{2\,\pi^{\frac{d-2}{2}}Z_k}
    {(d-1)^{\frac{3}{2}}\Gamma_E\Big(\frac{d-2}{2}\Big)}\frac{k^{d-2}}{l^d}\Big\}\,;\crcr
\beta_{d>4}(\mu_k)&=&-\frac{d\lambda_k}{(Z_kk^2+\mu_k)^2}
    \Big\{\de_tZ_k\Big[\frac{k^d}{l^d}\frac{\pi^{\frac{d-2}{2}}}
    {\sqrt{d-1}\Gamma_E\Big(\frac{d+2}{2}\Big)}+\frac{k^2}{l^2}\Big]
    +2Z_k\Big[\frac{k^d}{l^d}\frac{\pi^{\frac{d-2}{2}}}
    {\sqrt{d-1}\Gamma_E\Big(\frac{d}{2}\Big)}+\frac{k^2}{l^2}\Big]\Big\}\,.
\eea
We note that setting $d=3$ in \eqref{betaGZmu}, we recovers
\eqref{betaGZmud3}. We can therefore extend the last formula
to $d=3$, and will denote them $\beta_{d\ne 4}(Z_k)$
and $\beta_{d\ne 4}(\mu_k)$. The case $d=4$ must
be distinguished from the rest of the ranks, because we observe that
$\beta_{d= 4}(Z_k)$ is not the evaluation of $\beta_{d\ne 4}(Z_k)$
at $d=4$. Note that the mass
equation can be however  recovered from $\beta_{d\ne 4}(\mu_k)$
at $d=4$. 


\subsection{$\varphi^4$-terms}

The next order of the truncation made on the Wetterich equation,
i.e. $(II^g)_W=\Tr[\de_tR_k\cdot(P_k)^{-1}\cdot F_k\cdot(P_k)^{-1}\cdot F_k\cdot(P_k)^{-1}]$, provides the $\beta$-function for the coupling $\lambda_k$. Introducing the notation $\hat\varphi_{\textbf{p}} = \varphi _{\textbf{p}}\delta(\sum p)$ for the gauge invariant field, we write: 
\bea
 &&(II^g)_W
=\lambda_k^2\int_{{D^*}^{\times 2}}d\textbf{p}d\textbf{r}\;
    {\textstyle\frac{\Theta(k^2-\Sigma_sp_s^2)[\de_tZ_k(k^2-\Sigma_sp_s^2)+2k^2Z_k]\delta(\Sigma p)}
    {(Z_kk^2+\mu_k)^2[Z_k \Sigma_s{r}_s^2+\mu_k+
    \Theta(k^2-\Sigma_s{r}_s^2)Z_k(k^2-\Sigma_s{r}_s^2)]\delta(\Sigma r)\delta^2(\Sigma p)}}\crcr
&&\times\biggl[\int_{\ca{D}^{d-1}}dm_2\dots dm_d\;
    \hat\varphi_{r_1m_2\dots m_d}\hat{\ov{\varphi}}_{p_1m_2\dots m_d} \delta(\Sigma p)
   \delta(r_1+p_2+\dots+p_d)
    \prod_{i=2}^d\delta(p_i-r_i)\crcr
&&+\int_{\ca{D}}dm_1\;\hat\varphi_{m_1r_2\dots r_d}\hat{\ov{\varphi}}_{m_1p_2\dots p_d}\delta(\Sigma p)
 \delta(p_1+r_2+\dots+r_d)\delta(p_1-r_1)+\Sym\biggr]\crcr
&&\times \biggl[\int_{\ca{D}^{\times d-1}}dn_2\dots dn_d\;
    \hat\varphi_{p_1n_2\dots n_d}\hat{\ov{\varphi}}_{r_1n_2\dots n_d}\delta(\Sigma r)
    \delta(p_1+r_2+\dots+r_d)
    \prod_{i=2}^d\delta(r_i-p_i)\crcr
&&+\int_{\ca{D}}dn_1\;
\hat\varphi_{n_1p_2\dots p_d}\hat{\ov{\varphi}}_{n_1r_2\dots r_d}\delta(\Sigma r)
   \delta(r_1+p_2+\dots p_d)\delta(r_1-p_1)+\Sym\biggr]\,,
\eea
where the redundant $\Theta$-functions are set to 1.  The combinatorics of the present model is the same studied in the previous appendix, we therefore proceed in the same way by collecting different types of coloured contributions. We first discuss the contribution obtained by the product of colour 1-1: 
\bea
&&(II^g)_W|_{1,1}= \crcr
&&=\lambda_k^2\int_{{D^*}^{\times 2}}d\textbf{p}d\textbf{r}\;
    {\textstyle\frac{\Theta(k^2-\Sigma_sp_s^2)[\de_tZ_k(k^2-\Sigma_sp_s^2)+2k^2Z_k]}
    {(Z_kk^2+\mu_k)^2[Z_k \Sigma_s{r}_s^2 +\mu_k+
    \Theta(k^2-\Sigma_s{r}_s^2)Z_k(k^2-\Sigma_s{r}_s^2)]}}
\crcr
&&\times \biggl[\int_{\ca{D}^{\times 2}}dm_1dn_1\;
  \hat\varphi_{m_1r_2\dots r_d}\hat{\ov{\varphi}}_{m_1p_2\dots p_d}
   \hat\varphi_{n_1p_2\dots p_d}\hat{\ov{\varphi}}_{n_1r_2\dots r_d}
    \delta(p_1+r_2+\dots+r_d)
\delta(r_1+p_2+\dots+p_d)\delta^2(r_1-p_1)\cr\cr
&&+\int_{\ca{D}^{\times 2d-2}}dm_2\dots dm_ddn_2\dots dn_d\;
     \hat{\varphi}_{r_1m_2\dots m_d}\hat{\ov{\varphi}}_{p_1m_2\dots m_d}
     \hat{\varphi}_{p_1n_2\dots n_d}\hat{\ov{\varphi}}_{r_1n_2\dots n_d}
\cr\cr
&&\times \delta(r_1+p_2+\dots+p_d)
    \delta(p_1+r_2+\dots+r_d)
    \prod_{i=2}^d\delta^2(r_i-p_i)\crcr
&&+\textrm{disconnected}\biggr]\,,
\eea
where the  terms denoted by ``disconnected'' describe disconnected interactions which we discard. Integrating over $r_i$,
in the delta functions which are not convoluted with the fields,
and replacing the redundant $\delta$ by $1/l$, one gets:
\bea
 &&(II^g)_W|_{1,1}\simeq \lambda_k^2\int_{{D^*}^{\times 2}}dm_1dp_2\dots dp_ddn_1dr_2\dots dr_d\;
   {\textstyle\frac{\hat{\varphi}_{m_1r_2\dots r_d}\hat{\ov{\varphi}}_{m_1p_2\dots p_d}
   \hat{\varphi}_{n_1p_2\dots p_d}\hat{\ov{\varphi}}_{n_1r_2\dots r_d}}
    {(Z_kk^2+\mu_k)^2}}\cr\cr
&&\times \frac{1}{l}\int_{\ca{D}}dp_1\;{\textstyle\frac{\Theta(k^2-\Sigma_sp_s^2)[\de_tZ_k(k^2-\Sigma_sp_s^2)+2k^2Z_k]}
    {Z_k(p_1^2+\Sigma_{i=2}^dr_i^2)+\mu_k+
    \Theta[k^2-p_1^2-\Sigma_{i=2}^dr_i^2]Z_k[k^2-p_1^2-\Sigma_{i=2}^dr_i^2]}}
\delta(\Sigma p)\delta(p_1+\Sigma_{i=2}^dr_i)
\cr\cr
&&+\lambda_k^2\int_{{D^*}^{\times 2}}dp_1dm_2\dots dm_ddr_1dn_2\dots dn_d\;
    {\textstyle\frac{\hat\varphi_{r_1m_2\dots m_d}
\hat{\ov{\varphi}}_{p_1m_2\dots m_d}\hat\varphi_{p_1n_2\dots n_d}\hat{\ov{\varphi}}_{r_1n_2\dots n_d}}
      {(Z_kk^2+\mu_k)^2}}\cr\cr
&&\times \frac{1}{l^{d-1}}\int_{\ca{D}^{\times (d-1)}}dp_2\dots dp_d\;
    {\textstyle\frac{\Theta(k^2-\Sigma_sp_s^2)[\de_tZ_k(k^2-\Sigma_sp_s^2)+2k^2Z_k]}
    {Z_k(r_1^2+\Sigma_{i=2}^dp_i^2)+\mu_k+
    \Theta[k^2-r_1^2-\Sigma_{i=2}^dp_i^2]
Z_k[k^2-r_1^2-\Sigma_{i=2}^dp_i^2]}}\cr\cr
&& \times
\delta(\Sigma p)\delta(r_1+p_2+\dots+p_d)
\,. 
\eea 
  Once again, the case $d=3$ requires a special care 
during the evaluation of the above integrals. 
For $d=3$, we have by direct evaluation: 
\bea
 &&(II^g)_{W;d=3}|_{1,1}\simeq \lambda_k^2\int_{{D^*}^{\times 2}}dm_1dp_2 dp_3dn_1dr_2dr_3\;
   {\textstyle\frac{\hat{\varphi}_{m_1r_2r_3}\hat{\ov{\varphi}}_{m_1p_2
p_3}
   \hat{\varphi}_{n_1p_2p_3}\hat{\ov{\varphi}}_{n_1r_2 r_3}}
    {(Z_kk^2+\mu_k)^2}}\cr\cr
&&\times \frac{1}{l}\int_{\ca{D}}dp_1\;{\textstyle\frac{\Theta(k^2-\Sigma_sp_s^2)[\de_tZ_k(k^2-\Sigma_sp_s^2)+2k^2Z_k]}
    {Z_k(p_1^2+\Sigma_{i=2}^3r_i^2)+\mu_k+
    \Theta[k^2-p_1^2-\Sigma_{i=2}^3r_i^2]Z_k[k^2-p_1^2-\Sigma_{i=2}^3r_i^2]}}
\delta(\Sigma p)\delta(p_1+r_2+r_3)
\cr\cr
&&+\lambda_k^2\int_{{D^*}^{\times 2}}dp_1dm_2dm_3dr_1dn_2dn_3\;
    {\textstyle\frac{\hat\varphi_{r_1m_2 m_3}
\hat{\ov{\varphi}}_{p_1m_2 m_3}\hat\varphi_{p_1n_2 n_3}\hat{\ov{\varphi}}_{r_1n_2n_3}}
      {(Z_kk^2+\mu_k)^2}}\cr\cr
&&\times \frac{1}{l^{2}}\int_{\ca{D}^{\times 2}}dp_2 dp_3\;
    {\textstyle\frac{\Theta(k^2-\Sigma_sp_s^2)[\de_tZ_k(k^2-\Sigma_sp_s^2)+2k^2Z_k]}
    {Z_k(r_1^2+\Sigma_{i=2}^3p_i^2)+\mu_k+
    \Theta[k^2-r_1^2-\Sigma_{i=2}^3p_i^2]
Z_k[k^2-r_1^2-\Sigma_{i=2}^3p_i^2]}}
\delta(\Sigma p)\delta(r_1+p_2+p_3)
\,. 
\eea 
We integrate over $p_1$ the first term and over $p_3$ the second
term, replace redundant deltas by appropriate factors $1/l$ and then put to 0 all momentum variables involved in the field convolutions, to get
\bea
 &&(II^g)_{W;d=3}|_{1,1}\simeq \frac{\lambda_k^2}{(Z_kk^2+\mu_k)^3}\frac{k^2(2+\de_t)Z_k}{l^2}\ca{V}_1 \cr\cr
&&+\frac{\lambda_k^2}{(Z_kk^2+\mu_k)^3}
\frac{1}{l^{3}} \int_{-\sqrt{k^2/2}}^{\sqrt{k^2/2}} dr \;
  [\de_tZ_k(k^2-2r^2)+2k^2Z_k]\ca{V}_1 \crcr
&&
\simeq 
\frac{\lambda_k^2}{(Z_kk^2+\mu_k)^3}\Big[ 
\frac{k^2(2+\de_t)Z_k
 }{l^2}+
\frac{k^3}{l^{3}} \big[ \sqrt{2}(\de_t+ 2)Z_k
- \frac{\sqrt{2}}{3} \de_tZ_k\big] 
 \Big] \ca{V}_1 
\crcr
&&
\simeq 
\frac{\lambda_k^2}{(Z_kk^2+\mu_k)^3}\Big[ 
\Big[\frac{k^2}{l^2}  + \frac{2\sqrt{2}}{3}\frac{k^3}{l^{3}} \Big] 
\de_tZ_k
+ 
2\Big[ \frac{k^2 }{l^2} +\sqrt{2}\frac{k^3}{l^{3}}  \Big] Z_k 
\Big] \ca{V}_1 \,. 
\label{iiw113}
\eea

At rank $d> 3$, using again the spherical coordinates $(R,\Omega_{d-1})$, and taking the continuum limit,  
we write:
\bea
&&(II^g)_{W;d>3}|_{1,1}\simeq 
\lambda_k^2\int dm_1dp_2\dots dp_ddn_1dr_2\dots dr_d\;
   {\textstyle\frac{\hat\varphi_{m_1r_2\dots r_d}\hat{\ov{\varphi}}_{m_1p_2\dots p_d}
   \hat\varphi_{n_1p_2\dots p_d}\hat{\ov{\varphi}}_{n_1r_2\dots r_d}}
    {l(Z_kk^2+\mu_k)^2}}
\cr\cr
&&\times \int dp_1\;{\textstyle\frac{\theta(k^2-\Sigma_sp_s^2)[\de_tZ_k(k^2-\Sigma_sp_s^2)+2k^2Z_k]}
    {Z_k(p_1^2+\Sigma_{i=2}^dr_i^2)+\mu_k+
    \theta[k^2-p_1^2-\Sigma_{i=2}^dr_i^2]Z_k[k^2-p_1^2-\Sigma_{i=2}^dr_i^2]}}
\delta(\Sigma p)\delta(p_1+\Sigma_{i=2}^dr_i)
\cr\cr\cr
&&+\lambda_k^2\int dp_1dm_2\dots dm_ddr_1dn_2\dots dn_d\;
    {\textstyle\frac{\hat\varphi_{r_1m_2\dots m_d}
\hat{\ov{\varphi}}_{p_1m_2\dots m_d}\hat\varphi_{p_1n_2\dots n_d}\hat{\ov{\varphi}}_{r_1n_2\dots n_d}}
      {l^{d-1}(Z_kk^2+\mu_k)^2}}\cr\cr
&&\int dR\;\int d\Omega_{d-1}\;\frac{R^{d-2}\theta(k^2-p_1^2-R^2)[\de_tZ_k(k^2-p_1^2-R^2)+2k^2Z_k]}
     {Z_k(r_1^2+R^2)+\mu_k+\theta(k^2-r_1^2-R^2)Z_k[k^2-r_1^2-R^2]}\cr\cr
&&\times\delta(p_1+R\sqrt{d-1}\cos\vartheta)
\delta(r_1+R\sqrt{d-1}\cos\vartheta)
\cr\cr\cr
&&=
\lambda_k^2\int dm_1dp_2\dots dp_ddn_1dr_2\dots dr_d\;
   {\textstyle\frac{\hat\varphi_{m_1r_2\dots r_d}\hat{\ov{\varphi}}_{m_1p_2\dots p_d}
   \hat\varphi_{n_1p_2\dots p_d}\hat{\ov{\varphi}}_{n_1r_2\dots r_d}}
{l(Z_kk^2+\mu_k)^2}}\cr\cr
&&\times \int dp_1\;{\textstyle\frac{\theta(k^2-\Sigma_sp_s^2)[\de_tZ_k(k^2-\Sigma_sp_s^2)+2k^2Z_k]}
    {Z_k(p_1^2+\Sigma_{i=2}^dr_i^2)+\mu_k+
    \theta[k^2-p_1^2-\Sigma_{i=2}^dr_i^2]Z_k[k^2-p_1^2-\Sigma_{i=2}^dr_i^2]}}
\delta(\Sigma p)\delta(p_1+\Sigma_{i=2}^dr_i)
\cr\cr
&&+\lambda_k^2\int dp_1dm_2\dots dm_ddr_1dn_2\dots dn_d\;
    {\textstyle\frac{\hat\varphi_{r_1m_2\dots m_d}
\hat{\ov{\varphi}}_{p_1m_2\dots m_d}\hat\varphi_{p_1n_2\dots n_d}\hat{\ov{\varphi}}_{r_1n_2\dots n_d}}
{l^{d-1}(Z_kk^2+\mu_k)^2}}\cr\cr
&&\times \int dR\;\int d\Omega_{d-2}\;\int_0^{\pi}
   \frac{R^{d-2}}{R^2(d-1)}d\vartheta\sin^{d-3}\vartheta
   \delta\Big(\frac{p_1}{R\sqrt{d-1}}+\cos\vartheta\Big)\delta\Big(\frac{r_1}{R\sqrt{d-1}}+\cos\vartheta\Big)\cr\cr
&&\times \frac{\theta(k^2-p_1^2-R^2)[\de_tZ_k(k^2-p_1^2-R^2)+2k^2Z_k]}
{Z_k(r_1^2+R^2)+\mu_k+\theta(k^2-r_1^2-R^2)Z_k(k^2-r_1^2-R^2)}\cr\cr\cr
&&=\lambda_k^2\int dm_1dp_2\dots dp_ddn_1dr_2\dots dr_d\;
   {\textstyle\frac{\hat\varphi_{m_1r_2\dots r_d}\hat{\ov{\varphi}}_{m_1p_2\dots p_d}
   \hat\varphi_{n_1p_2\dots p_d}\hat{\ov{\varphi}}_{n_1r_2\dots r_d}}
{l(Z_kk^2+\mu_k)^2}}\cr\cr
&&\times\frac{\theta\Big[k^2-2\Big(\Sigma_{i=2}^dp_i^2+\Sigma_{1<i< j}p_ip_j\Big)\Big]
    \Big\{\de_tZ_k\Big[k^2-2\Big(\Sigma_{i=2}^dp_i^2+\Sigma_{1<i< j}p_ip_j\Big)\Big]+2k^2Z_k\Big\}
    \delta(\Sigma_{i=2}^d(r_i-p_i))}
    {Z_k[\Sigma_{i=2}^dr_i^2+(\Sigma_{i=2}^dp_i)^2]+\mu_k
    +\theta[k^2-\Sigma_{i=2}^dr_i^2-(\Sigma_{i=2}^dp_i)^2]Z_k[k^2-\Sigma_{i=2}^dr_i^2-(\Sigma_{i=2}^dp_i)^2]}
\cr\cr
&&+\lambda_k^2\Omega_{d-2}\int dp_1dm_2\dots dm_ddr_1dn_2\dots dn_d\;
    {\textstyle\frac{\hat\varphi_{r_1m_2\dots m_d}
\hat{\ov{\varphi}}_{p_1m_2\dots m_d}\hat\varphi_{p_1n_2\dots n_d}\hat{\ov{\varphi}}_{r_1n_2\dots n_d}}
{l^{d-1}(Z_kk^2+\mu_k)^2}}\cr\cr
&&\times\int dR\Big[1-\frac{r_1^2}{R^2(d-1)}\Big]^{\frac{d-4}{2}}\frac{R^{d-3}}{\sqrt{d-1}}\delta(p_1-r_1)
    {\textstyle{\frac{\theta(k^2-p_1^2-R^2)[\de_tZ_k(k^2-p_1^2-R^2)+2k^2Z_k]}
{Z_k(r_1^2+R^2)+\mu_k+\theta(k^2-r_1^2-R^2)Z_k(k^2-r_1^2-R^2)}}}\,.
\eea
Considering that all interaction terms which explicitly depend on the momenta involved in their fields fall out of our truncation, 
considering also that the deltas $\delta(p_1-r_1)$
and $ \delta(\Sigma_{i=2}^d(r_i-p_i))$  turn out to be redundant
with the gauge invariance conditions, we can then set to zero the labels  $p_i$ and $r_i$ appearing in the integrals, coefficients of the  gauge projected fields, and get:
\bea
&&(II^g)_{W;d>3}|_{1,1}\simeq \frac{\lambda_k^2k^2(2+\de_t)Z_k}{l^2(Z_kk^2+\mu_k)^3}
   \int_{D^{* \times 2}} dm_1dp_2\dots dp_ddn_1dr_2\dots dr_d\;
   \hat\varphi_{m_1r_2\dots r_d}\hat{\ov{\varphi}}_{m_1p_2\dots p_d}
   \hat\varphi_{n_1p_2\dots p_d}\hat{\ov{\varphi}}_{n_1r_2\dots r_d}
\crcr
&&+\frac{2\pi^{\frac{d-2}{2}}\lambda_k^2}{(Z_kk^2+\mu_k)^3\Gamma_E\Big(\frac{d-2}{2}\Big)\sqrt{d-1}}
    \frac{k^d}{l^d}\Big[\frac{(2+\de_t)Z_k}{d-2}-\frac{\de_tZ_k}{d}\Big]\crcr
&&\times\int_{D^{* \times 2}} dp_1dm_2\dots dm_ddr_1dn_2\dots dn_d\;
    \hat\varphi_{r_1m_2\dots m_d}
\hat{\ov{\varphi}}_{p_1m_2\dots m_d}\hat\varphi_{p_1n_2\dots n_d}\hat{\ov{\varphi}}_{r_1n_2\dots n_d}\crcr
&&=\frac{\lambda_k^2}{(Z_kk^2+\mu_k)^3}\Big\{(2+\de_t)Z_k
    \Big[\frac{\pi^{\frac{d-2}{2}}}{\Gamma_E\Big(\frac{d}{2}\Big)\sqrt{d-1}}\frac{k^d}{l^d}
    +\frac{k^2}{l^2}\Big]
    -\frac{2\pi^{\frac{d-2}{2}}\de_tZ_k}{d\sqrt{d-1}\Gamma_E\Big(\frac{d-2}{2}\Big)}
    \frac{k^d}{l^d}\Big\}\ca{V}_1\,,
\label{IIga11}
\eea
where we used for the coloured vertex the same notation introduced in section \ref{app:tgftphi4}. We note that setting $d=3$ in the last result leads us to
\eqref{iiw113}. Then, we prolong $(II^g)_{W}$ to $d \geq 3$. 

Inspecting the 2-colour cross terms, we focus on the product
of terms 1-2.
Discarding the disconnected interactions and the terms which fall out of the chosen truncation, while paying a special care
on the case $d=3$, one has:
\bea
&&(II^g)_W|_{1,2}\simeq 
    \lambda_k^2\int_{{D^*}^{\times 2}}d\textbf{p}d\textbf{r}\;
    {\textstyle{\frac{\Theta(k^2-\Sigma_sp_s^2)[\de_tZ_k(k^2-\Sigma_sp_s^2)+2k^2Z_k]}
    {(Z_kk^2+\mu_k)^2[Z_k\Sigma_s{r}_s^2+\mu_k+
    \Theta(k^2-\Sigma_s{r}_s^2)Z_k(k^2-\Sigma_s{r}_s^2)]}}}\crcr
&&\times\biggl[\delta_{d,3}
\int_{D^2}dm_1dn_2 \;
    \hat\varphi_{m_1r_2 r_3}\hat{\ov{\varphi}}_{m_1p_2p_3}
    \hat\varphi_{p_1n_2p_3}\hat{\ov{\varphi}}_{r_1n_2r_3} 
 \delta(p_1+r_2+p_3)
\delta(p_1+r_2+r_3)
    \delta(r_1-p_1)\delta(p_2-r_2)  \crcr
\crcr
&&+\int_{D^*}dm_1dn_1dn_3\dots dn_d\;
    \hat\varphi_{m_1r_2\dots r_d}\hat{\ov{\varphi}}_{m_1p_2\dots p_d}
    \hat\varphi_{n_1p_2n_3\dots n_d}\hat{\ov{\varphi}}_{n_1r_2n_3\dots n_d}\crcr
&&\times
\delta(p_1+r_2+\dots+r_d)
\delta(r_1+p_2+r_3+\dots+r_d)
    \delta^2(r_1-p_1)\prod_{i=3}^d\delta(p_i-r_i)\crcr
&&+\int_{\ca{D^*}}dn_2dm_2\dots dm_d\;
    \hat\varphi_{r_1m_2\dots m_d}\hat{\ov{\varphi}}_{p_1m_2\dots m_d}
\hat\varphi_{p_1n_2p_3\dots p_d}\hat{\ov{\varphi}}_{r_1n_2r_3\dots p_d}\crcr
&&\times
\delta(r_1+p_2+\dots+p_d) 
\delta(p_1+r_2+p_3+\dots+p_d)\delta^2(p_2-r_2)\prod_{i=3}^d\delta(p_i-r_i)\biggr]\crcr
&&\simeq 
\frac{\lambda_k^2}{(Z_kk^2+\mu_k)^2} \Big\{
\delta_{d,3}
\int_{D^2}dm_1dn_2 dr_3 dp_1dp_2dp_3\;
    \hat\varphi_{m_1p_2 r_3}\hat{\ov{\varphi}}_{m_1p_2p_3}
    \hat\varphi_{p_1n_2p_3}\hat{\ov{\varphi}}_{p_1n_2r_3}
\delta(p_1+p_2+p_3)
\delta(p_1+p_2+r_3)
\crcr
&&
\times\frac{\Theta(k^2-\Sigma_sp_s^2)[\de_tZ_k(k^2-\Sigma_sp_s^2)+2k^2Z_k]}
    {Z_k(p_1^2+p_2^2+r_3^2)+\mu_k+
    \Theta[k^2-(p_1^2+p_2^2+r_3^2)]
Z_k[k^2-(p_1^2+p_2^2+r_3^2)]}
\crcr
&&
+ 
\frac{1}{l}\int_{{D^*}^{\times 2}}dm_1dp_2\dots dp_ddn_1dr_2dn_3\dots dn_d\;
    \hat\varphi_{m_1r_2p_3\dots p_d}\hat{\ov{\varphi}}_{m_1p_2\dots p_d}
    \hat\varphi_{n_1p_2n_3\dots n_d}\hat{\ov{\varphi}}_{n_1r_2n_3\dots n_d}\crcr
&&\times\int_{\ca{D}}dp_1\;\delta(p_1+r_2+p_3+\dots+p_d)\delta(\Sigma p) 
\crcr
&&\times\frac{\Theta(k^2-\Sigma_sp_s^2)[\de_tZ_k(k^2-\Sigma_sp_s^2)+2k^2Z_k]}
    {Z_k(p_1^2+r_2^2+\Sigma_{i=3}^dp_i^2)+\mu_k+
    \Theta[k^2-(p_1^2+r_2^2+\Sigma_{i=3}^dp_i^2)]Z_k[k^2-(p_1^2+r_2^2+\Sigma_{i=3}^dp_i^2)]}\crcr
&&+\frac{1}{l}\int_{{D^*}^{\times 2}}dr_1dm_2\dots dm_ddp_1dn_2dp_3\dots dp_d\;
    \hat\varphi_{r_1m_2\dots m_d}\hat{\ov{\varphi}}_{p_1m_2\dots m_d}
    \hat\varphi_{p_1n_2p_3\dots p_d}\hat{\ov{\varphi}}_{r_1n_2p_3\dots p_d}\crcr
&&\times\int_{\ca{D}}dp_2\;\delta(r_1+p_2+\dots+p_d)\delta(\Sigma p)
    \crcr
&&\times\frac{\Theta(k^2-\Sigma_sp_s^2)[\de_tZ_k(k^2-\Sigma_sp_s^2)+2k^2Z_k]}
    {Z_k(r_1^2+p_2^2+\dots+p_d^2)+\mu_k+
    \Theta[k^2-(r_1^2+p_2^2+\dots+p_d^2)]Z_k[k^2-(r_1^2+p_2^2+\dots+p_d^2)]}
\Big\}\,.
\eea

Performing the integral over $p_1$ and $p_2$ in the last two terms and evaluating at the 0-momentum we find:
\bea\label{IIga12} 
(II^g)_W|_{1,2} &\simeq&
\delta_{d,3}
\frac{\lambda^2_k}{(Z_kk^2+\mu_k)^3 }\frac{ k^2}{l^2} (2+ \de_t)Z_k
 \cV_3  
+ 
 \frac{\lambda_k^2}{(Z_kk^2+\mu_k)^3} \frac{ k^2}{l^2}(2+\de_t)Z_k\Big[  \cV_2 + \cV_1 \Big] 
\crcr
&\simeq& \frac{\lambda_k^2}{(Z_kk^2+\mu_k)^3} \frac{ k^2}{l^2}(2+\de_t)Z_k\Big[   \delta_{d,3} \cV_3 + \cV_2 + \cV_1 \Big] 
\eea
The combinatorics of the $\varphi^4$ is the same with or without the presence of  (gauge) constraints, the contribution to the coefficients coming from  the color
symmetry is the same as for the previous model.
Collecting all contributions, $(II^g)_W|_{i,i}$ \eqref{IIga11}, $i=1,\dots,d$, and $(II^g)_W|_{i,j}$ \eqref{IIga12}, $i<j$, $i,j=1,\dots,d$, the $\beta$-function for $\lambda_k$, in any rank $d$,  expresses as
\beq\label{betaGl}
\beta(\lambda_k)=
\frac{2\lambda_k^2}{(Z_kk^2+\mu_k)^3}
   \Big\{\de_tZ_k\Big[\frac{2\pi^{\frac{d-2}{2}}}{d\sqrt{d-1}\Gamma\Big(\frac{d}{2}\Big)}
   \frac{k^d}{l^d}+(2d-1)\frac{k^2}{l^2}\Big]
+2Z_k\Big[\frac{\pi^{\frac{d-2}{2}}}{\sqrt{d-1}\Gamma\Big(\frac{d}{2}\Big)}
   \frac{k^d}{l^d}+(2d-1)\frac{k^2}{l^2}\Big]\Big\}\,.
\eeq
\noindent{\bf Dimensionful $\beta$-functions.}
Let us collect all $\beta$-functions. 
At rank $d\ne 4$, we gather \eqref{betaGZmu} and  \eqref{betaGl} for 
the complete system of $\beta$-functions for the gauge 
invariant TGFT model which expresses as: 
\beq
\left\{
\begin{aligned}
\beta_{d>4}(Z_k)&=\frac{d\lambda_k}{(Z_kk^2+\mu_k)^2}\Big\{\de_tZ_k\Big[
    \frac{\pi^{\frac{d-2}{2}}}{(d-1)^{\frac{3}{2}}\Gamma_E\Big(\frac{d}{2}\Big)}\frac{k^{d-2}}{l^d}
    +\frac{1}{l^2}\Big]+\frac{2\,\pi^{\frac{d-2}{2}}Z_k}
    {(d-1)^{\frac{3}{2}}\Gamma_E\Big(\frac{d-2}{2}\Big)}\frac{k^{d-2}}{l^d}\Big\}\crcr
\beta_{d\ne 4}(\mu_k)&=-\frac{d\lambda_k}{(Z_kk^2+\mu_k)^2}
    \Big\{\de_tZ_k\Big[\frac{k^d}{l^d}\frac{\pi^{\frac{d-2}{2}}}
    {\sqrt{d-1}\Gamma_E\Big(\frac{d+2}{2}\Big)}+\frac{k^2}{l^2}\Big]
    +2Z_k\Big[\frac{k^d}{l^d}\frac{\pi^{\frac{d-2}{2}}}
    {\sqrt{d-1}\Gamma_E\Big(\frac{d}{2}\Big)}+\frac{k^2}{l^2}\Big]\Big\}\crcr
\beta_{d\ne 4}(\lambda_k)&=
\frac{2\lambda_k^2}{(Z_kk^2+\mu_k)^3}
   \Big\{\de_tZ_k\Big[\frac{2\pi^{\frac{d-2}{2}}}{d\sqrt{d-1}\Gamma\Big(\frac{d}{2}\Big)}
   \frac{k^d}{l^d}+(2d-1+2\delta_{d,3})\frac{k^2}{l^2}\Big]
+2Z_k\Big[\frac{\pi^{\frac{d-2}{2}}}{\sqrt{d-1}\Gamma\Big(\frac{d}{2}\Big)}
   \frac{k^d}{l^d}+(2d-1+2\delta_{d,3} )\frac{k^2}{l^2}\Big]\Big\}
\end{aligned}
\right. 
\eeq
which is reported in \eqref{dimensfulGaug} in section \ref{betfuncGaug}
and at $d=4$, we obtain the expression 
\beq
\left\{
\begin{aligned}
&
\beta_{d=4}(Z_k)=\frac{\lambda_k}{(Z_kk^2+\mu_k)^2}
\Big\{  \frac{2\pi}{\sqrt{3}} \Big[\frac12\de_tZ_k+Z_k  \Big] \frac{k^2}{l^4}
+ \de_t Z_k \frac{4}{l^2}
\Big\} \crcr
&
\beta_{d=4}(\mu_k)=
-\frac{4\lambda_k}{(Z_kk^2+\mu_k)^2}
\Big\{ \frac{\pi}{\sqrt{3}} \Big[
 \frac12\de_tZ_k+2Z_k \Big] \frac{ k^4  }{l^4 } 
+ (2+\de_t)Z_k \frac{ k^2}{l^2 } 
\Big\} \crcr
&\beta_{d=4}(\lambda_k)=
\frac{2\lambda_k^2}{(Z_kk^2+\mu_k)^3}
   \Big\{\de_tZ_k\Big[\frac{2\pi}{4\sqrt{3}}
   \frac{k^4}{l^4}+7\frac{k^2}{l^2}\Big]
+2Z_k\Big[\frac{\pi}{\sqrt{3}}
   \frac{k^4}{l^4}+7\frac{k^2}{l^2}\Big]\Big\}
\end{aligned}
\right. 
\eeq
as reported in \eqref{dimensfulG4}.


\end{document}